\documentclass[11pt]{article}

\usepackage{fullpage}
\usepackage{authblk}
\usepackage{natbib}
\usepackage{algorithm}
\usepackage{algpseudocode}
\usepackage{amsmath,amsthm,amsfonts}
\usepackage{amsmath}
\usepackage[subnum]{cases}
\usepackage{tcolorbox}
\usepackage{mathrsfs}
\usepackage{amssymb}
\usepackage{color}
\usepackage{mathrsfs}
\usepackage{enumitem}
\usepackage{bm}
\usepackage{multirow}
\usepackage{booktabs}
\usepackage{makecell}
\usepackage{graphicx}
\usepackage{subcaption}
\usepackage{comment}
\usepackage{cases}
\usepackage{appendix}
\usepackage{tikz}
\usetikzlibrary{arrows,shapes}
\usepackage[colorlinks= true, linkcolor=red, citecolor=blue, urlcolor=red]{hyperref}
\usepackage{cleveref}
\usepackage{marginnote}
\usepackage{diagbox} 
\usepackage{mdframed}

\numberwithin{equation}{section}

\theoremstyle{definition}

\usepackage{booktabs}
\usepackage{tikz}
\usepackage{pgfplots}
\usetikzlibrary{calc,intersections}

\usetikzlibrary{shapes.geometric, arrows, positioning}

\tikzset{
	mybox/.style  = {draw, rectangle, minimum width=4cm, minimum height=0.8cm, text centered, text width=4.4cm,   
		font=\normalsize},
	box/.style  = {draw, rectangle, minimum width=2.0cm, minimum height=0.6cm, text centered, text width=3.0cm,   
		font=\normalsize},
	myarrow/.style = {line width=0.2pt, draw=black, -triangle 60, postaction={draw, line width=0.2pt, shorten >=10pt,-}}
}

\tikzstyle{arrow} = [->, >=stealth, -triangle 60]

\allowdisplaybreaks

\makeatletter
\newcommand{\leqnomode}{\tagsleft@true}
\newcommand{\reqnomode}{\tagsleft@false}
\makeatother

\begin{document}

\title{Optimal Disturbances of Blocking: A Barotropic View}

\author[1,3]{Bin Shi\thanks{Corresponding author: \texttt{shibin@lsec.cc.ac.cn} } }
\author[2,4]{Dehai Luo}
\author[2]{Wenqi Zhang}


\affil[1]{Academy of Mathematics and Systems Science, Chinese Academy of Sciences, Beijing 100190, China}
\affil[2]{Institute of Atmospheric Physics, Chinese Academy of Sciences, Beijing 100029, China}

\affil[3]{School of Mathematical Sciences, University of Chinese Academy of Sciences, Beijing 100049, China}
\affil[4]{College of Earth and Planetary Sciences, University of Chinese Academy of Sciences, Beijing 100049, China}


\date\today

\maketitle

\begin{abstract}

In this paper, we explore optimal disturbances of blockings in the equivalent barotropic atmosphere using the conditional nonlinear optimal perturbation (CNOP) approach. Considering the initial blocking amplitude, the optimal disturbance exhibits a solitary wave-like pattern. As the size increases incrementally, the spatial pattern becomes more concentrated, and the nonlinear evolution becomes more pronounced. During the evolution, it only focuses on gradually intensifying the blocking amplitude without any other influence. Additionally, based on the medium-range experiments, the time-delay optimal disturbance appears to lead to larger errors, making it more challenging to predict. Considering the preexisting synoptic-scale eddies, the optimal disturbance displays a sharply concentrated pattern, even more concentrated by increasing the size. However, it is worth noting that the nonlinear evolution undergoes significant changes, compared to disturbances of the initial blocking amplitude. Meanwhile, we find that the optimal disturbance not only strongly impacts the amplitude of blockings but also their shape, making eddy straining and wave breaking more chaotic and predominant,  further influencing the development of weather extremes. This suggests that blockings are more sensitive to perturbations of preexisting synoptic-scale eddies than initial blocking amplitudes. Furthermore, the perturbations of the synoptic-scale eddies are more likely to lead to the development of weather extremes, making them less predictable. In medium-range experiments, it is also found that time-delay disturbances result in larger errors, particularly during the decay period. Finally, we discuss how the variations of westerly wind influence optimal disturbances in spatial patterns and nonlinear evolution as well as their relation to predictability.

\end{abstract}

%

\section{Introduction}
\label{sec: introduction}

Weather extremes have a significant impact on society as they can pose a threat to human life and safety, as well as cause significant economic damage and disruption. For instance, heat waves and extreme droughts can lead to devastating forest fires, damaging agriculture and causing air pollution that poses health risks~\citep{witte2011nasa}. Similarly, cold spells with low temperatures and heavy snowfall can greatly disrupt transportation systems and daily life~\citep{davolio2015exceptional}. Additionally, floods caused by heavy precipitation are another type of high-impact weather event that can result in severe consequences, affecting infrastructure, displacing communities, and causing property damage~\citep{lenggenhager2019dynamical}. Despite the diverse nature of extreme weather events, they do share a common factor --- the prevailing large-scale flow pattern in the troposphere over the North Pacific and North Atlantic Oceans, which is strongly influenced by atmospheric blocking (hereinafter referred to as~\textit{blocking}). Hence, understanding the mechanism behind blocking is crucially important for comprehending and predicting these high-impact weather events~\citep{kautz2022atmospheric}.


Before delving into the mechanism study of blockings, it is important to understand their features. These blockings are characterized as long-lasting, quasi-stationary, and self-preserving in the midlatitudes, as highlighted in~\citep{liu1994definition, nakamura2018atmospheric}. In the unfiltered geopotential height field, blocking flow often manifests as a significant meandering of westerly jet streams, as described in~\citep{berggren1949aerological}. This meandering resembles the westerly winds flowing around and bypassing the obstacle created by the blocking pattern. According to the pioneering study~\citep{rex1950blocking}, a key feature of blocking is the abrupt transition from a zonal (east-west) to a meridional (north-south) flow pattern. This transition often leads to the splitting of the jet stream into two branches around the blocking. Generally, these blockings can be classified into three types, dipole blockings~\citep{hoskins1985use, pelly2003new, weijenborg2012direction, masato2013wave}, Omega blockings~\citep{hakkinen2014greenland, steinfeld2019role}, and amplified ridges~\citep{sousa2018european}, with more examples provided in~\citep{woollings2018blocking}. Earlier studies, such as~\citep{berggren1949aerological, charney1979multiple, tung1979theory, shutts1983propagation, illari1983interpretation, holopainen1987high, mullen1987transient}, suggested that blockings were primarily caused by traveling synoptic-scale eddies and large-scale topography. However,~\citet{ji1983numerical} conducted numerical experiments, indicating that the topographic forcing plays a secondary role compared to traveling synoptic-scale eddies. Furthermore, the observation that dipole blockings mainly occur downstream of the storm track in the Pacific or Atlantic basin supports the idea that synoptic-scale eddies likely contribute to the formation and maintenance of dipole blocking downstream of the storm tracks~\citep{illari1983interpretation, holopainen1987high, mullen1987transient, nakamura1993synoptic}.


%

There have been several classical and theoretical models put forward to explain the mechanisms behind the maintenance of blocking patterns. Three well--known models in this regard are the global theory of multiple flow equilibria proposed in~\citep{charney1979multiple}, the local theories of modon proposed in~\citep{mcwilliams1980application}, and the eddy straining proposed in~\citep{shutts1983propagation}. In the global theory of multiple flow equilibria,~\citet{charney1979multiple} utilized a highly-truncated, nonlinear, barotropic channel model to study the blocking phenomenon from a global perspective. However, observations have indeed shown that most blocking events are primarily a local phenomenon~\citep{dole1983persistent, diao2006new}. This suggests that a local approach appears to be more consistent with synoptic blocking observations. In the past,~\citet{mcwilliams1980application} utilized modon or vortex pair solutions of the equivalent barotropic vorticity equation as nonlinear free modes to describe the observed blocking features over the Atlantic region. However, it was also noted in~\citep{mcwilliams1980application} that the existing condition of the modon solution is not easily satisfied by the observed mean zonal wind. According to~\citep{higgins1994simulated}, the composite field of observed blocking events does not align with the modon or vortex pair structure.~\citet{shutts1983propagation} proposed an eddy straining mechanism that takes into account the time-mean eddy vorticity flux, which aligns with the observed maintenance of blocking and suggests that the eddy straining around the blocking region's two sides plays a crucial role. However,~\citet{shutts1983propagation} also mentioned that blocking is essentially an unsteady phenomenon, meaning that the dynamic life cycle, including onset, growth, maintenance, and decay, is not fully explained by the eddy straining mechanism. Further studies conducted in~\citep{pierrehumbert1984forced, haines1987eddy, holopainen1987high} have expanded on the three well-known theoretical models of blockings. Additionally,~\citet{farrell1996generalized} and~\citet{mak1989local} proposed another viewpoint that suggests barotropic instability as a factor in the occurence of blocking events.

%


According to~\citep{berggren1949aerological}, the concept of spatial scale separation has emerged, highlighting the interaction between different scales in blocking systems. This interaction involves the interply between fast-moving synoptic-scale eddies and quasi-stationary planetary-scale blocking, which can result in the presence of a low-PV (potential vorticity) air mass. The idea of the eddy straining mechanism, as described in~\citep{shutts1983propagation}, further supports this concept by suggesting a ``drip-feeding'' of low-PV air that replaces the original air mass. Additionally, subsequent wave-breaking events, as described by~\citep{hoskins1985use}, serve as another means of replacing the original air mass in this exchange. In contrast to the previous steady-state theorems that primarily focused on the time-mean eddy vorticity flux~\citep{berggren1949aerological, shutts1983propagation, pierrehumbert1984forced, hoskins1985use, haines1987eddy, holopainen1987high}, recent works by~\citet{luo2000planetary, luo2005barotropic} have introduced a dynamic consideration of scale separation in the spatial structure. This approach takes into account the westard propagation of a Rossby wave packet and separates it into a planetary-scale blocking anomaly and preexisting synoptic-scale eddies, based on the background westerly wind. As discussed in~\citep{luo2000planetary, luo2005barotropic}, asymptotic analysis has revealed that the slow-varying amplitude of the planetary-scale blocking anomaly behaves like a solitary wave, governed by the forced nonlinear Schr\"odinger (NLS) equation. Additionally, the preexisting synoptic-scale eddies act as an external force on the blocking anomaly. Furthermore,~\citet{luo2014nonlinear} introduced the eddy-blocking matching (EBM) mechanism and the nonlinear multiscale interaction (NMI) model to provide insights into the dynamics of blocking flows. The EBM mechanism helps explain how synoptic-scale eddies can either enhance or suppress a blocking flow. Surprisingly, the NMI model accurately captures the entire life cycle of blockings, including their onset, growth, maintenance, and decay. Additionally,~\citet{luo2019nonlinear} proposed a theory that elucidates how the meridional gradient of potential vorticity ($PV_y$) influences the dispersive and nonlinear behavior of blocking. This theory finds support in observations of the background westerly jet stream~\citep{luo2019nonlinear}.

Despite the progress made in understanding blocking through various theories, accurately numerical predicting the blocking event in weather forecasts remains a challenge~\citep{zhang2019predictability}. The abrupt onset of block flow, as observed in midlatitudes weather over the past century, contributes to the difficulty~\citep{vautard1990multiple}. This is primarily attributed to the inherent instability of fluid dynamics, regardless of whether it is in normal or nonnormal modes.  In other words, the instability of fluid dynamics act as a barrier that hampers accurate predictions~\citep{pierrehumbert1984local, pierrehumbert1995baroclinic, swanson2001blocking}. The conditional nonlinear optimal perturbation (CNOP) approach, introduced by~\citet{mu2003conditional}, is indeed a valuable method for quantifying fluid instability using nonlinear optimization techniques. Unlike approaches that rely on linear approximation assumptions, the CNOP approach takes into account the full nonlinear effects within the system. By maximizing the objective value while adhering to reasonable physical constraints at a fixed time $T$, the CNOP approach helps identifies the most unstable scenario, or says optimal disturbance.  These optimal disturbances, also referred to as optimal precursors, often serve as important signals that can lead to some unstable fluid phenomenon being studied. Therefore, the CNOP approach has been widely applied in various fields such as fluid dynamics, atmospheric science, and oceanography. It has been used to study phenomena like turbulence in shear flows~\citep{pringle2010using, kerswell2018nonlinear}, disturbance energy in vortex-pairs~\citep{navrose2018optimal}, detecting blocking onset~\citep{mu2008method},  typhoon observations~\citep{mu2009method, qin2012influence}, predictability of El Ni\~no-Southern Oscillation~\citep{duan2009exploring, duan2016initial} and variations in the Kuroshio path~\citep{wang2015new}. Furthermore, recent advancements in statistical machine learning techniques have further improved and accelerated the CNOP approach in practical applications~\citep{shi2023adjoint, shi2023sampling}.

In this paper, we investigate optimal disturbances of blockings using the NMI model, where the goal is to understand their static and dynamic behaviors and their relation to predictability. In~\Cref{sec: nmi-od}, the derivation of the NMI model and the basic CNOP settings are briefly described to obtain the optimal disturbances.~\Cref{sec: cnop-init} provides a theoretical analysis of the optimal disturbance of the initial blocking amplitude,  including spatial patterns, nonlinear evolution, the eastward propagation of blockings, and the time-delay effect. Additionally, a one-by-one comparison is made  in~\Cref{sec: cnop-init} with the optimal disturbance of the preexisting synoptic-scale eddies in terms of spatial patterns, nonlinear evolution, the eastward propagation of blockings, and the time-delay effect.~\Cref{sec: role-westerly} discusses how variations in westerly wind influence optimal disturbances in spatial patterns and nonlinear evolution, as well as their relation to predictability.  Finally,  in~\Cref{sec: discussion}, this paper concludes with a summary and discussion.



\section{The NMI model and optimal disturbances}
\label{sec: nmi-od}

In this section, we provide a brief description of the barotropic NMI model, which has been derived and developed in~\citep{luo2000planetary, luo2005barotropic, luo2014nonlinear, luo2019nonlinear}.  The barotropic NMI model serves as a mathematical framework used to study atmospheric phenomena, particularly those related to blockings. After introducing the barotropic NMI model, we proceed to formalize the objective functions that need to be maximized.  These objective functions play a crucial role in determining the optimal disturbances of both the initial blocking amplitude and the preexisting synoptic-scale eddies. By maximizing these objective functions, our aim is to find the optimal perturbations that contribute to the occurrence and development of blockings. 

\subsection{The NMI Model}
\label{subsec: nmi-model}
In the initial stage, we provide a list of various values for the object parameters in~\Cref{tab: nmi}.  Let $F = (L/R)^2$ be the Froude number, where $R \approx L$ is the radius of Rossby deformation. The meridional gradient of the Coriolis parameter at the given latitude $\varphi_0$ is denoted as $\beta_0$, and the nondimensional parameter is set as $\beta = \beta_0 L^2/U$. Typically, the background westerly wind is observed to have a speed of approximately $ 7\;m/s$~\citep{luo2005barotropic}.  Considering that the dimension of wind speed is $10\;m/s$, we set the nondimensional wind speed as $U = 0.7$.

\begin{table}[htbp!]
\centering
\begin{tabular}{l|ll}
   \toprule
   object                       &parameter                                    & value                             \\
   \midrule
                                &reference latitude                           & $\varphi_0=55^\circ N$           \\
   \midrule 
   horizontal scale             &characteristic length                        & $L \sim 10^6m$                   \\
                                &characteristic wind speed                    & $U \sim 10ms^{-1}$               \\  
   \midrule
   $\beta$-channel              &nondimensional width                         & $L_y = 5$                        \\ 
   \midrule                                                                                                       
   Total Rossby wave packet     &nondimensional zonal wavenumber              & $k_0=1/(6.371\cos\varphi_0)$     \\  
                                &nondimensional wind speed                    & $U=0.7$                          \\
                                &(uniform background westerly)                &                                  \\ 
   \midrule
   blocking dipole              &nondimensional zonal wavenumber              & $k=2k_0$       \\
   \midrule
   preexisting synoptic eddies  &nondimensional zonal wavenumber              & $k_1=9k_0$     \\
                                &nondimensional zonal wavenumber              & $k_2=11k_0$    \\   
                                &zonal location                               & $x_T=1.435$                      \\  
                                &amplitude                                    & $a_0=0.17$                      \\
                                &variance parameter                           & $\mu=1.2$                        \\      
                                &variance parameter                           & $\epsilon=0.24$                  \\    
   \bottomrule
\end{tabular}
\caption{The values of the object parameters in the NMI model. }
\label{tab: nmi}
\end{table}

In the given context, we consider the zonal westerly wind, denoted as $U = U(y)$. Regarding a blocking event, its nondimensional governing equation is expressed as the barotropic quasi-geostrophic equation with $x$-periodic and lateral boundary conditions:
\begin{subequations}
\begin{align}
        &\frac{\partial}{\partial t}\left(\nabla^2\psi_T - F\psi_T \right) + J(\psi_T, \nabla^2\psi_T) + \beta \frac{\partial \psi_T}{\partial x} = 0, \label{eqn: barotropic-qg} \\
x-\text{periodic}:\qquad       & \psi_{T}(-L_x,y,t) =  \psi_{T}(L_x,y,t),  \label{eqn: x-periodic}\\
y-\text{lateral}:\phantom{\mathrel{a}}\qquad       & \left.\frac{1}{2L_x}\int_{-L_x}^{L_x} \frac{\partial \psi_T}{\partial y}dx\right|_{y=0} = -U(0), \quad \left.\frac{1}{2L_x}\int_{-L_x}^{L_x} \frac{\partial \psi_T}{\partial y}dx\right|_{y=L_y} = -U(L_y),   \label{eqn: lateral}
\end{align}
\end{subequations}
where $\psi_T$ is the instantaneous total streamfunction.  Then, we can decompose the total streamfunction $\psi_T$ by scales into three parts as
\begin{equation}
\label{eqn: total}
\psi_T = \overline{\psi} + \psi + \psi',
\end{equation}
where $\overline{\psi}= \overline{\psi}(y) = -\int_{0}^{y} U(y')dy'$ represents the basic westerly flow, which is only dependent on the meridional direction $y$, $\psi = \psi(x,y,t)$ represents the planetary-scale blocking anomaly, and $\psi'=\psi'(x,y,t)$ represents the preexisting synoptic-scale eddies. Based on observations in the mid-latitudes of the northern hemisphere~\citep{colucci1981spectral}, the planetary-scale blocking anomaly $\psi$ in the zonal direction exhibits a single wave with wavenumber $k=2k_0$ (\Cref{tab: nmi}), assuming a corresponding frequency of $\omega$ as discussed in~\citep{charney1979multiple,luo2005barotropic}. In the case of the synoptic-scale eddies in the zonal direction, it is believed that they are a superposition of two single waves with wavenumbers, $k_1 = 9k_0$ and $k_2 = 11k_0$ (\Cref{tab: nmi}). The corresponding frequencies for these waves are $\omega_1$ and $\omega_2$, respectively, as mentioned in~\citep{luo2005barotropic, luo2007dynamics}. Regarding the equivalent barotropic atmosphere, it is widely recognized that the potential vorticity ($PV$) is governed by the equation $PV=f_0 + \beta y - U_y - F\overline{\psi}$,  as described in~\citep{pedlosky1987geophysical}. By substituting the instantaneous total streamfunction~\eqref{eqn: total} into the nondimensional barotropic quasi-geostrophic equation,~\eqref{eqn: barotropic-qg}, we can establish a relationship between the three wavenumbers. This relationship leads to two equations for the planetary-scale blocking anomaly $\psi$ and the preexisting synoptic-scale eddies $\psi'$ as
\begin{subequations}
\begin{align}
& \left( \frac{\partial}{\partial t} + U\frac{\partial}{\partial x} \right)\left(\nabla^2\psi - F\psi \right) + J(\psi, \nabla^2\psi) + PV_y \frac{\partial \psi}{\partial x} = - J(\psi', \nabla^2 \psi')_P, \label{eqn: planetary-scale}\\
& \left( \frac{\partial}{\partial t} + U\frac{\partial}{\partial x} \right)\left(\nabla^2\psi' - F\psi' \right) +  PV_y \frac{\partial \psi'}{\partial x} = - J(\psi', \nabla\psi) - J(\psi, \nabla^2\psi'), \label{eqn: synoptic-scale} 
\end{align}
\end{subequations}
where the meridional gradient of potential vorticity ($PV_y$) satisfies the equation $PV_y = \beta + FU - U_{yy}$, which indicates the $PV_y$ is slow-varying and the subscript $P$ represents the force driven by the synoptic-scale eddies, which is denoted as $-J(\psi', \nabla^2\psi')$ and has the wavenumber $2k_0$.  
Indeed, the relative vorticity, represented as $q'$, can be expressed as $q'=\nabla^2\psi' - F\psi'$. By considering this equation, we can derive that the synoptic-scale eddies satisfy $J(\psi', \nabla^2\psi')_P = \nabla \cdot (\pmb{v'}q')_P$, which represents the planetary-scale component of the divergence of the eddy vorticity flux induced by the preexisting synoptic-scale eddies. It is important to note that the planetary-scale component of the divergence of the eddy vorticity flux $\nabla \cdot (v'q')_P$ is time-dependent. On the contrary, the time-mean $\nabla \cdot \overline{(v'q')}$ in the eddy straining model is time-independent, as discussed in~\citep{shutts1983propagation, haines1987eddy}.


Recall the classical multiscale decomposition, we can decompose the spatial and temporal components using the parameters $\{X_k = \epsilon^k x\}_{k=0}^{\infty}$ and $\{T_k = \epsilon^k t\}_{k=0}^{\infty}$, respectively. Here, $\epsilon$ is a small parameter and $k$ represents the scale. This decomposition allows us to analyze and understand the behavior of the blocking anomaly and the synoptic-scale eddies at different scales. Using this decomposition, we can express the wavefunctions of the planetary-scale blocking anomaly and the synoptic-scale eddies as follows: 
\begin{equation}
\label{eqn: streamfunction}
\psi = \psi(x,y,t;X_1,T_1;X_2,T_2;\cdots), \quad \text{and}\quad \psi' = \psi'(x,y,t;X_1,T_1;X_2,T_2;\cdots).
\end{equation}
Without loss of generality, let us consider the planetary-scale blocking anomaly $\psi$ as an example. By utilizing the multiscale decomposition~\eqref{eqn: streamfunction}, we can express the temporal and spatial derivatives of the streamfunction $\psi$ as follows:
\begin{equation}
\label{eqn: temporal-spatial-derivative}
\frac{d\psi}{d t} = \frac{\partial \psi}{\partial t} + \epsilon \frac{\partial \psi}{\partial T_1} + \epsilon^2\frac{\partial \psi}{\partial T_2} + \cdots, \quad \text{and}\quad \frac{d\psi}{d x} = \frac{\partial \psi}{\partial x} + \epsilon \frac{\partial \psi}{\partial X_1} + \epsilon^2\frac{\partial \psi}{\partial X_2} + \cdots.
\end{equation}
Regarding the streamfunctions of the planetary-scale blocking anomaly and the synoptic-scale eddies~\eqref{eqn: streamfunction}, we can expand them asymptotically  as follows:
\begin{subequations}
\label{eqn: sf-expansion}
\begin{align}
& \psi  = \epsilon \psi_1(x,y,t;X_1,T_1;X_2,T_2;\cdots) + \epsilon^2 \psi_2(y;X_1,T_1;X_2,T_2;\cdots) + \cdots, \label{eqn: planetary-expansion} \\
& \psi' = \epsilon^{\frac32} \psi_1'(x,y,t;X_1,T_1;X_2,T_2;\cdots) + \epsilon^{\frac52} \psi_2'(x,y,t;X_1,T_1;X_2,T_2;\cdots) + \cdots, \label{eqn: synoptic-expansion} 
\end{align}
\end{subequations}
where the fast-varying variable of $\psi_2$ is only meridional or only dependent on $y$, as it represents the blocking's feedback to the zonal-mean westerly wind. Using the derivatives~\eqref{eqn: temporal-spatial-derivative} and the asymptotic expansion~\eqref{eqn: sf-expansion}, we employ Wentzel-Kramers-Brillouin (WKB) method from asymptotic analysis~\citep{nayfeh2008perturbation} to
derive the NMI model as:

\begin{itemize}
\item[(1)] The nondimensional streamfunctions of the blocking wavy anomaly $\psi_1$, the associated zonal-mean anomaly $\psi_2$ and the preexisting synoptic-scale eddies $\psi_1'$ are represented as follows:
\begin{subequations}
\label{eqn: nmi-model}
\begin{align}
&  		 \psi_1 = \frac{1}{\epsilon}\sqrt{\frac{2}{L_y}}\left( B e^{i(kx - \omega t)} + \overline{B} e^{-i(kx-\omega t)} \right)  \sin\left( my - \frac{\pi}{4}\right), \label{eqn: blocking-wave-1} \\
& \psi_2 = - \frac{g |B|^2 \cos(2my)}{\epsilon^2}, \label{eqn: associate-2} \\
& \psi_1'=  \frac{2F_0}{\epsilon^{\frac32}} \big( \cos(k_1x - \omega_1t) - \cos(k_2x - \omega_2t) \big)\sin\left( \frac{my}{2} - \frac{\pi}{8} \right), \label{eqn: synoptic-1} 
\end{align}
\end{subequations}
where $F_0 = a_0  \exp\left[ - \mu \epsilon^2(x + x_T)^2 \right]$\footnote{The external force $F_0$, acting as a filter for the waves, indeed serves as the core ingredient of the preexisting synoptic-scale eddies $\psi'_1$. Therefore, unless specifically mentioned afterward, we use the external force $F_0$ to represent the preexisting synoptic-scale eddies. } and the parameters are calculated as $m = -2\pi/L_y$ and
\[
g = \frac{4m k^2(m^2+k^2+F)^2}{ PV_yL_y\left[(4m^2+F)(m^2+F-k^2) - (m^2+k^2+F)^2 \right]}.
\]

\item[(2)] Both the phase and group velocities of the planetary-scale blocking anomaly and the phase velocities of the synoptic-scale eddies are derived separately as
\begin{subequations}
\label{eqn: velocity}
\begin{align}
& c = \frac{\omega}{k} = U - \frac{PV_y}{ m^2 + k^2 + F},  \label{eqn: p-phase-velocity} \\
& c_g = \frac{\partial \omega}{\partial k} = U - \frac{PV_y(m^2 - k^2 + F)}{(m^2 + k^2 + F)^2}, \label{eqn: p-group-velocity} \\
& c_1=\frac{\omega_{1}}{k_1} = U - \frac{PV_y}{ \frac{m^2}{4} + k_1^2 + F}, \qquad  c_2=\frac{\omega_{2}}{k_2} =  U - \frac{PV_y}{ \frac{m^2}{4} + k_2^2 + F}. \label{eqn: s-phase-velocity}
\end{align}
\end{subequations}

\item[(3)] The blocking amplitude $B$ obeys the 1-dimensional forced NLS equation with the periodic boundary condition as
\begin{equation}
\left\{ \begin{aligned}
& i\left( \frac{\partial B}{\partial t} + c_g \frac{\partial B}{\partial x} \right) + \lambda \frac{\partial^2B}{\partial x^2} + \delta |B|^2 B+ GF_0^2 \exp(-i \Delta \omega t) = 0, \\
& B(0,-L_x) = B(0, L_x), 
         \end{aligned} \right.
 \label{eqn: ampli-nls}
\end{equation}
where $\Delta \omega = \omega_2 - \omega_1 - \omega$ and the parameters are set as
\begin{equation}
\label{eqn: pvy-lambda-delta}
\left\{\begin{aligned}
& \lambda = \frac{PV_yk\left[3(m^2 + F) -k^2 \right]}{(m^2+k^2 +F)^3}, \\
& \delta = \frac{gkm(3m^2-k^2)}{m^2+k^2+F}, \\
& G = -\sqrt{\frac{L_y}{2}}\cdot \frac{ m(k_1+k_2)^2(k_2-k_1)}{4(m^2+k^2+F)}.
\end{aligned}\right.
\end{equation}
\end{itemize}
The detailed derivation of the NMI model is shown in~\Cref{sec: derivation-nmi}.

\subsection{The basic CNOP settings of optimal disturbances}
\label{subsec: cnop}

In the previous derivation of the NMI model, it has been established that the 1-dimensional forced NLS equation, particularly~\cref{eqn: ampli-nls}, is of great importance in understanding the blocking. Concretely, this equation plays a significant role in describing the dynamic behavior of blocking. The motion of the blocking amplitude in the 1-dimensional forced NLS equation~\eqref{eqn: ampli-nls} is determined by three factors: the initial blocking amplitude $B_0$, the preexisting synoptic-scale eddies $F_0$ and the background westerly wind $U$.\footnote{In this study, the blocking's motion is primarily determined by the meridional gradient of potential vorticity ($PV_y$), which is influenced by the westerly wind $U$ and its meridional shear $U_{yy}$. However, for the specific focus of this paper, only the 1-dimensional forced NLS equation is considered, and therefore the meridional shear of the westerly wind $U_{yy}$ is disregarded. As a result, the main factor affecting the blocking's motion is the westerly wind $U$.} Traditionally, the Lyapunov exponent has been used to characterize the nonlinear error growth~\citep{lucarini2020new}. However, it is applicable only to finite-dimensional dynamical systems as it requires computing the maximum of finite eigenvalues. Therefore, it does not work for any partial differential equation since it corresponds to an infinite-dimensional dynamical system with an unbounded maximum eigenvalue. It is indeed an interesting question to investigate the effects of perturbations in the initial blocking amplitude $B_0$ and the preexisting synoptic-scale eddies $F_0$ on the motion of blocking. Since both $B_0$ and $F_0$ are 1-dimensional functions, it raises curiosity about how variations in these parameters affect the evolution behavior of blocking. Additionally, it is worth exploring how changes in the westerly wind $U$ interact with these perturbations to influence the motion of blocking. Understanding these relationships can provide valuable insights into the dynamics of blocking and its predictability. In this paper, we employ the conditional nonlinear optimal perturbation (CNOP) method, which was initially proposed by~\citet{mu2003conditional}, to explore the most influential perturbations and their effects.

%

In this scenario, we define $B(t,x;\cdot,\cdot,\cdot)$ as the reference blocking amplitude with time evolution in the configuration space, where the three dots represent the three factors influencing the motion of blocking as mentioned previously. Given the initial blocking amplitude $B_0$, the synoptic-scale eddies $F_0$, and the background westerly wind $U$, we can express the blocking amplitude as $B(t,x; B_0, F_0, U)$. To quantify the magnitude of the blocking amplitude $B$, we utilize the standard mass norm, also known as the energy norm~\citep{farrell1996generalized}. This norm is defined as
\begin{equation}
\label{eqn: energy-norm-B}
\|B(t)\| = \left(\int_{-L_x}^{L_x} |B(t,x)|^2 dx\right)^{\frac{1}{2}} 
\end{equation}
where $L_x$ represents the limits of integration. Similarly, for the synoptic-scale eddies, we also employ the standard mass norm given by
\begin{equation}
\label{eqn: energy-norm-F}
\|F\| = \left(\int_{-L_x}^{L_x} |F(x)|^2 dx\right)^{\frac{1}{2}}. 
\end{equation}
To enhance convenience in notation, we simplify the representation of the blocking amplitude as $B(t; B_0, F_0, U)$ by ignoring the less commonly used variable $x$, which helps to avoid any confusion and improve the overall understanding and readability of the paper. 

%

\paragraph{CNOP of initial blocking amplitude} If we consider the initial blocking amplitude as $B_0 + b_0$, where $b_0$ represents a perturbation of the initial blocking amplitude $B_0$, then the reference blocking amplitude at time $T$ can be expressed by $B(T;B_0 + b_0, F_0, U)$. Therefore, the two reference blocking amplitudes are given by $B(T;B_0,F_0,U)$ and $B(T;B_0+b_0;F_0,U)$. Based on the conditions of the synoptic-scale eddies $F_0$ and the westerly wind $U$, we can formulate the objective function for the initial perturbation $b_0$ about the initial blocking amplitude $B_0$ as
\begin{equation}
\label{eqn: obj-init}
J(b_0;B_0,F_0,U) = \left\| B(T;B_0 + b_0,F_0,U) - B(T;B_0,F_0,U) \right\|^2.
\end{equation}
The CNOP can then be computed by maximizing $J(b_0;B_0,F_0,U)$ while ensuring that the constraint $\|b_0\|\leq \rho$ is satisfied, where $\rho$ is a predetermined value that sets the upper limit for the norm of $b_0$.  In other words, our goal is to find the optimal solution that maximizes $J(b0; B0, F0, U)$ while adhering to the constraint of $\|b_0\| \leq \rho$, expressed as
\begin{equation}
\label{eqn: max-obj-init}
\max_{\|b_0\|\leq \rho }J(b_0;B_0,F_0,U).
\end{equation}
By abbreviating $J(b_0;B_0,F_0,U)$ as $J(b_0)$, we can simplify and make it more convenient for subsequent discussions or calculations. This abbreviation allows us to refer to $J(b_0)$ more easily and conveniently without losing any generality.

\paragraph{CNOP of preexisting synoptic-scale eddies} In a similar manner, if we consider the synoptic-scale eddies as $F_0 + f_0$, where $f_0$ represents a perturbation of the preexisting synoptic-scale eddies $F_0$, then the reference blocking amplitude at time $T$ can be expressed by $B(T;B_0, F_0 + f_0, U)$. Consequently, we have two reference blocking amplitudes $B(T;B_0,F_0,U)$ and $B(T;B_0;F_0+f_0,U)$. Based on the conditions of the initial blocking amplitude $B_0$ and the westerly wind $U$, we can formulate the objective function for the perturbation $f_0$ about the background synoptic-scale eddies $F_0$ as
\begin{equation}
\label{eqn: obj-synp}
J(f_0;B_0, F_0, U) = \left\| B(T;B_0,F_0 + f_0,U) - B(T;B_0,F_0,U) \right\|^2.
\end{equation}
The CNOP can then be stated as maximizing $J(f_0;B_0,F_0,U)$ while ensuring that the constraint $\|f_0\|\leq \rho$ is satisfied, where $\rho$ is a predetermined value that sets the upper limit for the norm of $f_0$.  In other words, we aim to find the optimal solution that maximizes $J(f0; B0, F0, U)$while adhering to the constraint of $\|f_0\| \leq \rho$, expressed as
\begin{equation}
\label{eqn: max-synp}
\max_{\|f_0\|\leq \rho }J(f_0;B_0,F_0,U).
\end{equation}
It is also mentioned here that we shorten the notation $J(f_0;B_0,F_0,U)$ as $J(f_0)$ for convenience.

\paragraph{Variations in the westerly wind} By utilizing the CNOP approach, we can compute the optimal disturbance of the initial blocking amplitude, denoted as $b_0$ in~\cref{eqn: max-obj-init}, and the optimal disturbance of the preexisting synoptic-scale eddies, denoted as $f_0$ in~\cref{eqn: max-synp}. To explore the influence of changes in the westerly wind $U$ on these optimal disturbances, we can assign different values to the westerly wind $U$ and observe the resulting changes, which allows us to understand how variations in the westerly wind affect the optimal disturbances.

\paragraph{Numerical implementation} 
In the theoretical analysis, the optimization problems concerning the optimal perturbations $b_0$ and $f_0$, as described in~\cref{eqn: max-obj-init} and~\cref{eqn: max-synp}, are directly derived from the forced NLS equation~\eqref{eqn: ampli-nls}, which is considered as an infinite-dimensional model. However, when implementing it numerically on a computer, the optimization problems,~\eqref{eqn: max-obj-init} and~\eqref{eqn: max-synp}, are reduced to finite-dimensional ones.

In this paper, we adopt the same method and numerical settings as described in~\citep{luo2005barotropic, luo2014nonlinear, luo2019nonlinear} to study the time evolution of the blocking amplitude $B$. To numerically simulate the forced NLS equation with the periodic boundary condition~\eqref{eqn: ampli-nls}, we utilize the high-order split-step Fourier scheme developed in~\citep{muslu2005higher}, which is known for its excellent performance. We set the nondimensional grid parameters, $\Delta x = 0.2296$  as the spatial grid size ($d=101$) and $\Delta t = 0.0864$ as the time step. Additionally, the boundary parameter is set as $L_x=11.48$ and the initial blocking amplitude is set as $B_0=0.4$. To compute the CNOP, we conventionally employ the second spectral projected gradient method (SPG2) proposed in~\citep{birgin2000nonmonotone}. The standard numerical gradient is computed with the step size $\epsilon=10^{-8}$. The energy norm of the blocking amplitude is numerically set as
\[
\|b_0\| \approx \left(\sum_{i=1}^{d}b_{0,i}^2\right)^{\frac{1}{2}} \sqrt{\Delta x} \leq \rho = \gamma\sqrt{\Delta x}.
\]
It is worth noting that an important observation from the empirical study~\citep{breeden2020optimal} is that the intensification of a blocking event often reaches its maximum around $10$ days from the onset. Therefore, in this analysis, the prediction time is set at day $T=10$ (unit: day). In line with the approach used in the previous studies~\citep{luo2005barotropic, luo2014nonlinear, luo2019nonlinear}, we set the initial blocking amplitude as $B_0 = 0.4$ and the preexisting synoptic-scale eddies as $F_0 = a\exp\left[ - \mu \epsilon^2(x + x_T)^2 \right]$.




\section{Optimal disturbance of the initial blocking amplitude}
\label{sec: cnop-init}

In this section, the main objective is to investigate the optimal disturbance of the initial blocking amplitude. Our aim is to gain a better understanding of spatial patterns and nonlinear growth that are associated with this disturbance. Additionally, we also explore how the total blocking evolves as the optimal disturbance increases in size.  Furthermore, we analyze the time-delay effect of the optimal initial disturbance and its relation with predictability.

\subsection{Spatial pattern and nonlinear growth}
\label{subsec: spng-ib}

\begin{figure}[htb!]
\centering
\includegraphics[scale=0.26]{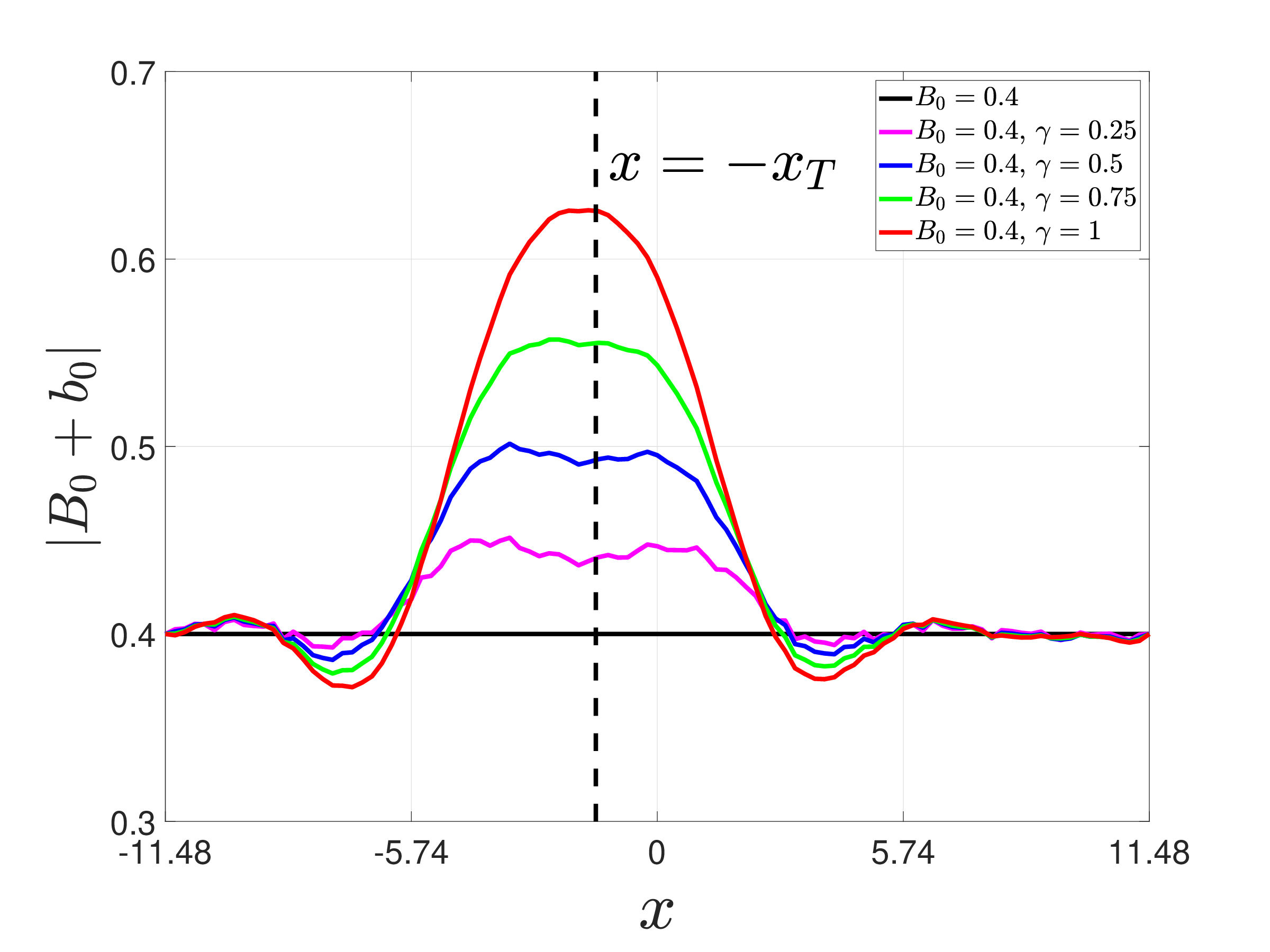}
\caption{Spatial patterns (nondimensionalization) of the optimal disturbance $b_0$ under the initial blocking amplitude $B_0=0.4$ varies with the incremental increase of the size parameter $\gamma$.}
\label{fig: nmi-cnop-init}
\end{figure}

In the given text, our goal is to numerically compute the optimal disturbance of the initial blocking amplitude, denoted as $b_0$,  which involves considering the governing forced NLS equation with the periodic boundary condition (2.9) and the initial blocking amplitude $B_0=0.4$. This computation can be achieved by maximizing the constrained objective function~\eqref{eqn: obj-init}. Increasing the size of the optimal disturbance allows for a more in-depth analysis of the numerical performance of spatial patterns. These spatial patterns are visualized in~\Cref{fig: nmi-cnop-init}, clearly representing how the spatial pattern varies with the incremental increase of the size parameter $\gamma$. It is truly fascinating to observe the spatial pattern of the optimal disturbance of the initial blocking amplitude in relation to the slow-varying preexisting synoptic-scale eddies $F_0 = a_0  \exp\left[ - \mu \epsilon^2(x + x_T)^2 \right]$.  In~\Cref{fig: nmi-cnop-init}, we can clearly see a bulge around the zonal location $x = -x_{T}$, accompanied by two small dents beside it, resembling a solitary wave, as noted in~\citep{Zabusky:2010}. As we increase the size parameter $\gamma$ incrementally from 0.25 to 1, the optimal disturbance exhibits a more pronounced solitary wave-like behavior. Specifically, the bulge becomes highly concentrated around the zonal center $x = -x_T$, with a sharply rising peak. This suggests that, in the context of blocking events in the real world, the largest deviation in the initial blocking amplitude $B_0$  occurs due to a positive incremental increase in the vicinity of the zonal location $x = -x_T$, which corresponds to the location of the synoptic-scale eddies acting as external forces. Additionally, the center of the optimal disturbance of the initial blocking amplitude is slightly offset to the left of the zonal center $x = -x_T$, as depicted in~\Cref{fig: nmi-cnop-init}.

\begin{figure}[htb!]
\centering
\begin{subfigure}[t]{0.48\linewidth}
\centering
\includegraphics[scale=0.18]{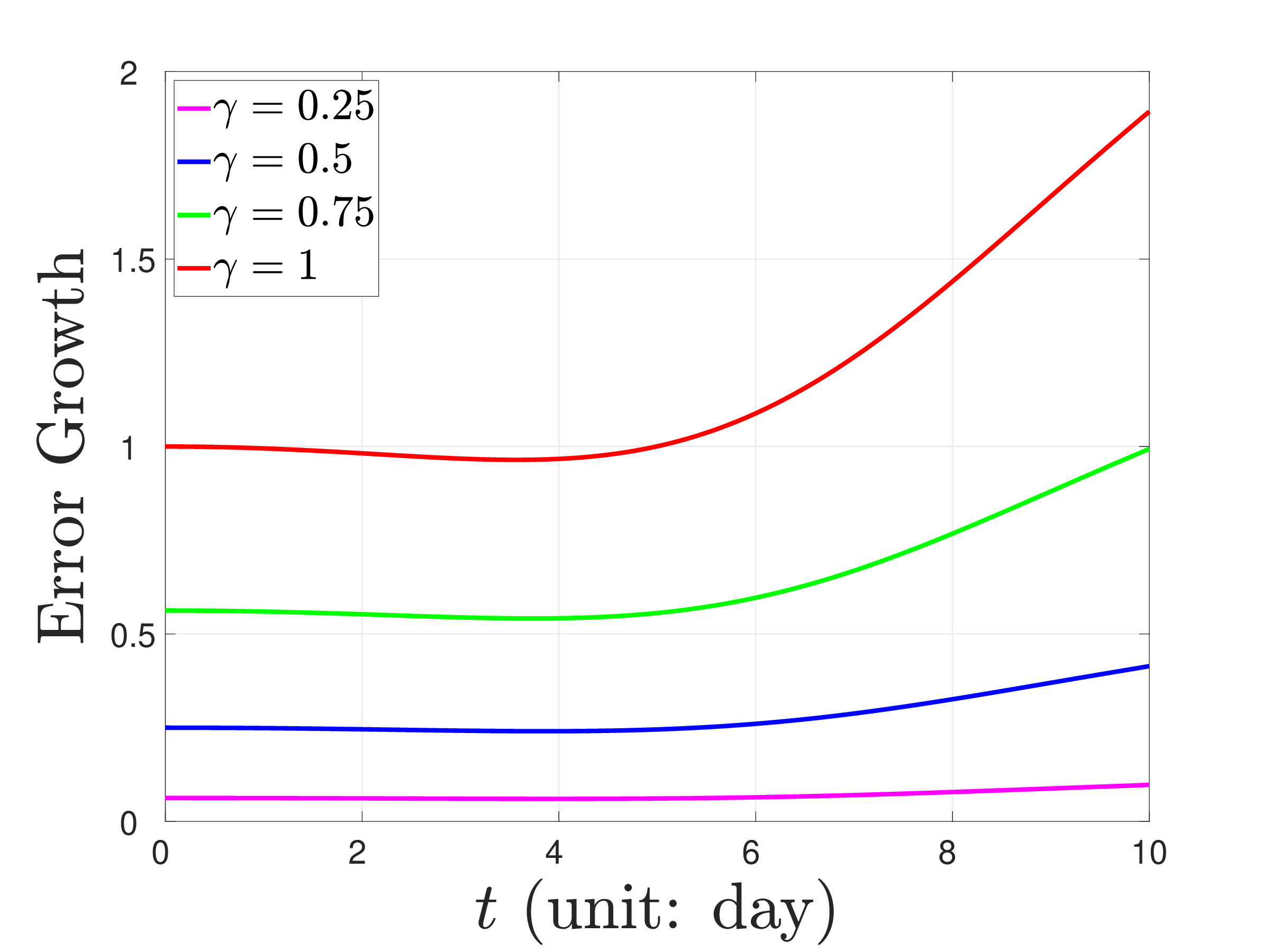}
\caption{(standard) Error Growth}
\label{fig: stan-init}
\end{subfigure}
\begin{subfigure}[t]{0.48\linewidth}
\centering
\includegraphics[scale=0.18]{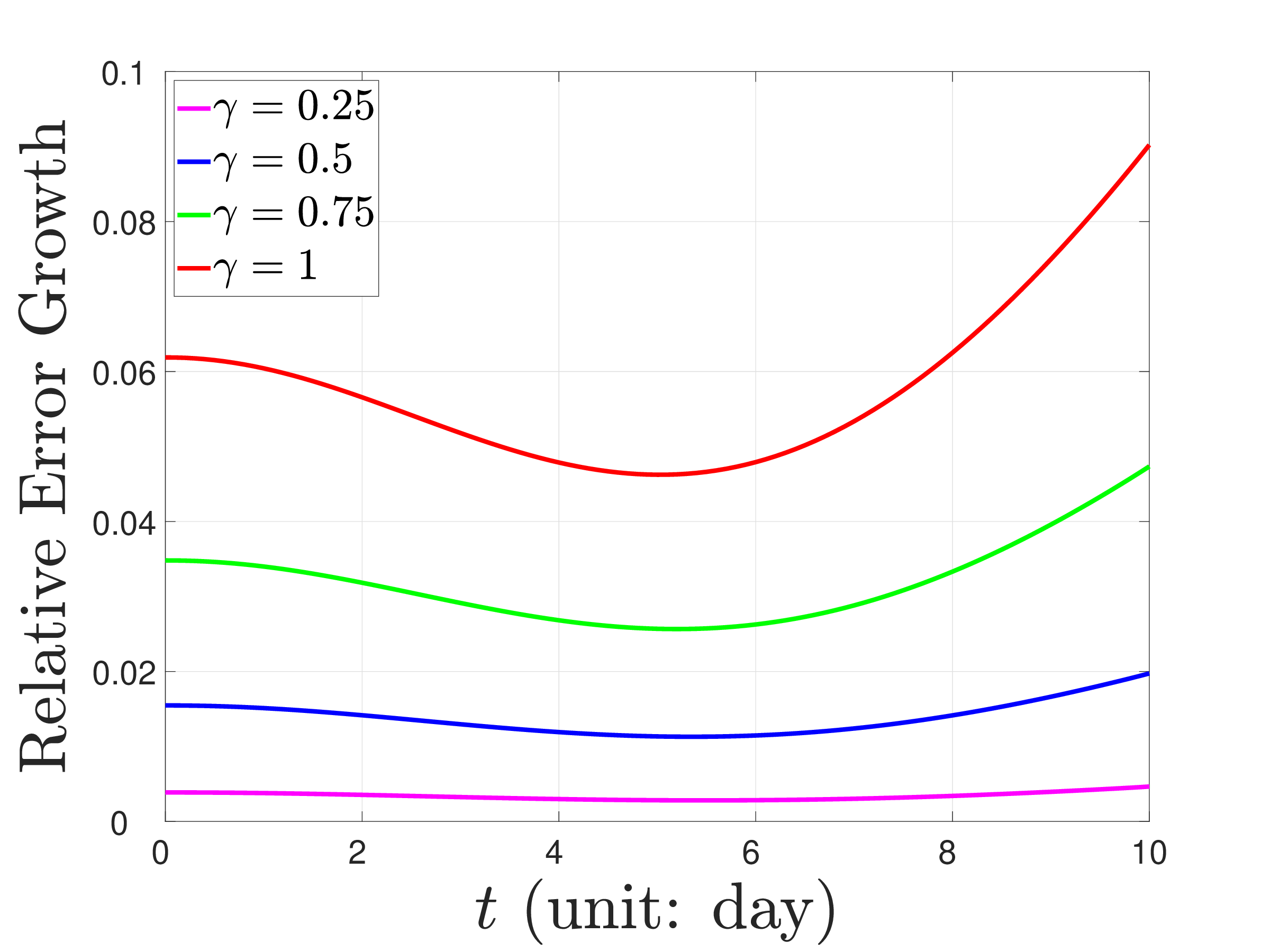}
\caption{Relative Error Growth}
\label{fig: rel-init}
\end{subfigure}
\caption{Nonlinear growth of the optimal disturbance given by~\eqref{eqn: error-init} and~\eqref{eqn: rel-error-init} varies with the incremental increase of the size parameter $\gamma$. }
\label{fig: error-growth-init}
\end{figure}

Then, we utilize the energy norm~\eqref{eqn: energy-norm-B} to characterize how the nonlinear growth of the optimal disturbance varies as the size increases, that is,
\begin{equation}
\label{eqn: error-init}
\frac{\|b(t)\|^2}{\Delta x} = \frac{\|B(t;B_0+b_0,F_0,U) - B(t;B_0,F_0,U)\|^2}{\Delta x},
\end{equation}
where the nonlinear evolution of the optimal disturbance $\frac{\|b(t)\|^2}{\Delta x}$ measures the difference between the blocking amplitudes $B$ at time $t$, taking into account both the initial blocking amplitude $B_0$ and the most perturbed initial blocking amplitude $B_0+b_0$, while keeping the synoptic-scale eddies $F_0$ and the westerly wind speed $U$ fixed. This allows for a comparison of the effects of the optimal disturbance on the blocking amplitude. The numerical performance of the nonlinear growth behavior is visualized in~\Cref{fig: stan-init}.
To further understand the nonlinear growth behavior,  it is indeed important to investigate the relative nonlinear growth of the optimal disturbance. This can be done by comparing the nonlinear growth of the optimal disturbance $\frac{\|b(t)\|^2}{\Delta x}$ with the blocking amplitude $\frac{\|B(t)\|^2}{\Delta x}$ using their ratio, that is,
\begin{equation}
\label{eqn: rel-error-init}
\frac{\|b(t)\|^2}{\|B(t)\|^2} = \frac{\|B(t;B_0+b_0,F_0,U) - B(t;B_0,F_0,U)\|^2}{\|B(t;B_0,F_0,U)\|^2}.
\end{equation}
By examining the relative nonlinear growth of the optimal disturbance~\eqref{eqn: rel-error-init}, it is observed that the nonlinear growth of the optimal disturbance is slower compared to the growth of the blocking amplitude during the period of blocking growth, while faster during other periods. The numerical performance is shown in~\Cref{fig: rel-init}. From both the subfigures in~\Cref{fig: error-growth-init}, it is evident that there is a fixed turning-time point in the nonlinear growth of the optimal disturbance. Prior to reaching this turning-time point, the nonlinear growth is relatively slow. However, once the turning-time point is reached, the nonlinear growth accelerates rapidly, and its growth pattern undergoes a significant change. It is also noteworthy that as the size parameter $\gamma$ increases incrementally, the turning-time point occurs earlier and the nonlinear growth of the optimal disturbance becomes more pronounced. This suggests that the size of the optimal disturbance has a positive impact on the timing and magnitude of its nonlinear growth. Meanwhile, we show the ratios of the nonlinear growth of the optimal disturbance in terms of norm squares in~\Cref{tab: error-growth-init}, where it is observed that the ratio increases as the size increases incrementally, further indicating a rising growth rate. 
\begin{table}[htb!]
\centering
\begin{tabular}{l|cccc}
   \toprule
                                       & $\gamma=0.25$                & $\gamma=0.5$                   & $\gamma=0.75$                 & $\gamma=1$  \\
   \midrule
     $\frac{\|b(10)\|^2}{\|b(0)\|^2}$  & $1.5616$                     & $1.6568$                       & $1.7664$                     & $1.8935$ \\
    \bottomrule
\end{tabular}
\caption{The ratio of the nonlinear growth of the optimal disturbance in terms of norm squares, $\|b(10)\|^2/\|b(0)\|^2$.}
\label{tab: error-growth-init}
\end{table}
The quantitative evidence in~\Cref{tab: error-growth-init} supports the idea that the size of the optimal disturbance has a positive effect on accelerating its nonlinear growth. Additionally, we compare the ratio of the relative nonlinear growth to further quantify the positive effect of the size of the optimal disturbance on its nonlinear growth, which is demonstrated in~\Cref{tab: relative-error-init}. 
\begin{table}[htb!]
\centering
\begin{tabular}{c|cccc}
   \toprule
                                                    & $\gamma=0.25$                & $\gamma=0.5$                 & $\gamma=0.75$                & $\gamma=1$  \\
   \midrule
   $\frac{\|b(0)\|^2}{\|B(0)\|^2}$                  & $0.39\%$                     & $1.55\%$                      & $3.48\%$                     & $6.19\%$ \\
   $\frac{\|b(10)\|^2}{\|B(10)\|^2}$                & $0.47\%$                     & $1.97\%$                      & $4.73\%$                     & $9.02\%$                     \\
   \midrule
   $\frac{\|b(10)\|^2}{\|B(10)\|^2} / \frac{\|b(0)\|^2}{\|B(0)\|^2}$  & $1.2051$   & $1.2710$                      & $1.3592$  & $1.4572$  \\
    \bottomrule
\end{tabular}
\caption{The relative nonlinear growth of the optimal disturbance in terms of norm squares, $\|b(0)\|^2/\|B(0)\|^2$ and $\|b(10)\|^2/\|B(10)\|^2$, and the ratio between them.}
\label{tab: relative-error-init}
\end{table}
It is worth noting that the nonlinear growth behavior observed in~\Cref{tab: relative-error-init} aligns with the findings in~\Cref{tab: error-growth-init}. This consistency provides further evidence to support the idea that the size of the optimal disturbance does have a positive effect on accelerating its nonlinear growth. 

\begin{figure}[htb!]
\centering
\begin{subfigure}[t]{0.48\linewidth}
\centering
\includegraphics[scale=0.18]{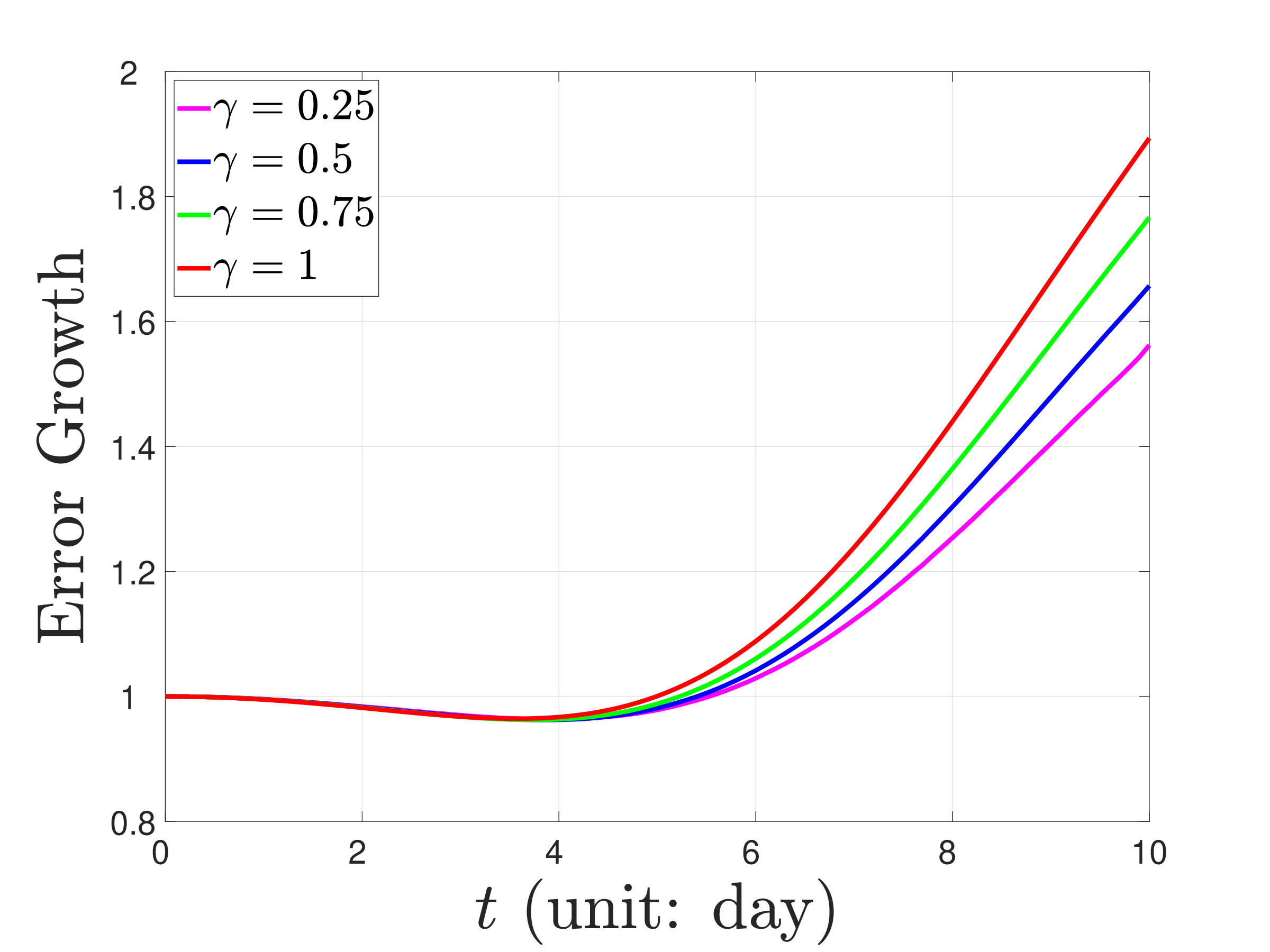}
\caption{(standard) Error Growth}
\label{fig: norm-stan-init}
\end{subfigure} 
\begin{subfigure}[t]{0.48\linewidth}
\centering
\includegraphics[scale=0.18]{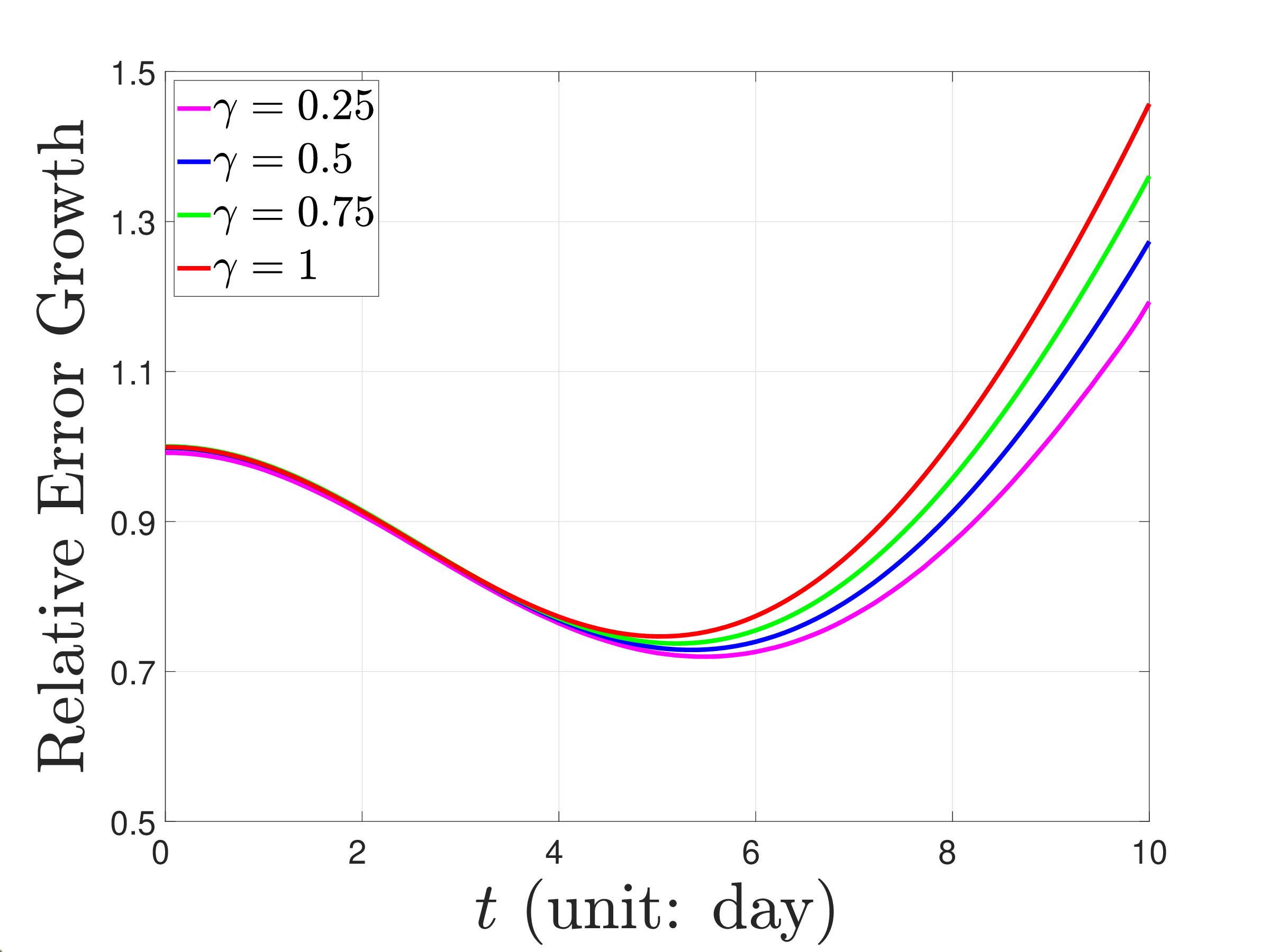}
\caption{Relative Error Growth}
\label{fig: norm-rel-init}
\end{subfigure}
\caption{Nonlinear growth of the optimal disturbance  given by~\eqref{eqn: error-init} and~\eqref{eqn: rel-error-init} under initial normalization varies with the increase of the size parameter $\gamma$. }
\label{fig: norm-error-growth-init}
\end{figure}

By normalizing the initial conditions for the growth curves in~\Cref{fig: error-growth-init}, we demonstrate that increasing the size of the optimal disturbance can indeed accelerate its nonlinear growth. The normalization for the nonlinear growth is given by $\frac{\|b(t)\|^2}{\|b(0)\|^2}$, while the relative nonlinear growth is given by $\frac{\|b(t)\|^2}{\|B(t)\|^2}/\frac{\|b(0)\|^2}{\|B(0)\|^2}$. This normalization indeed provides a clearer visualization of the nonlinear growth patterns of the optimal disturbances in terms of norm square, as shown in~\Cref{fig: norm-error-growth-init}. This normalization allows us to observe how fast the optimal disturbance grows over time as the size parameter $\gamma$ increases incrementally from $0.25$ to $1$. Furthermore,~\Cref{fig: norm-error-growth-init} shows that the nonlinear evolution of the optimal disturbances, transitioning from units to ratios, which aligns with the data shown in~\Cref{tab: error-growth-init} and~\Cref{tab: relative-error-init}. This comprehensive representation provides us with a better understanding of how the optimal disturbances change and grow throughout the process. By comparing~\Cref{fig: norm-stan-init} and~\Cref{fig: norm-rel-init}, it further supports the idea that the growth of the optimal disturbance is weaker than that of the blocking amplitude during the period of blocking growth, while faster during other periods.


\subsection{Temporal evolution of blocking under the optimal disturbance}
\label{subsec: block-simulation-init}

After analyzing the spatial patterns and the nonlinear growth of the optimal disturbance, it would indeed be valuable to explore how the motion of the blocking is influenced by the optimal disturbance. Understanding the dynamic relationship between the optimal disturbance and the evolution of the blocking can provide further insights into the impact of the optimal disturbance on the overall behavior of the blocking system. In particular, it would be interesting to investigate how the blocking evolves with time when the optimal disturbance is added to the initial blocking amplitude.

\begin{figure}[htb!]
\centering
\begin{subfigure}[b]{0.32\linewidth}
\centering
\includegraphics[scale=0.125]{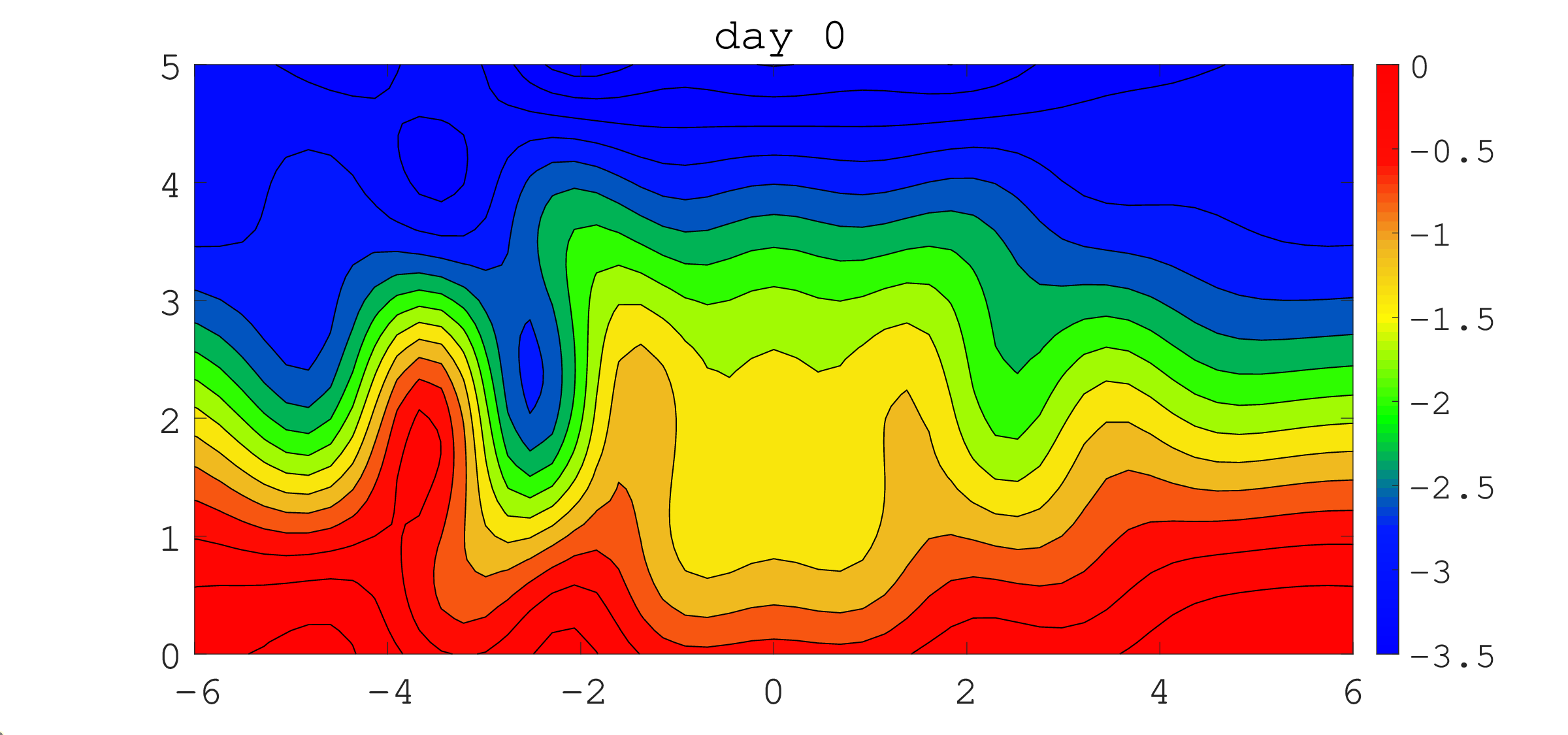}
\includegraphics[scale=0.125]{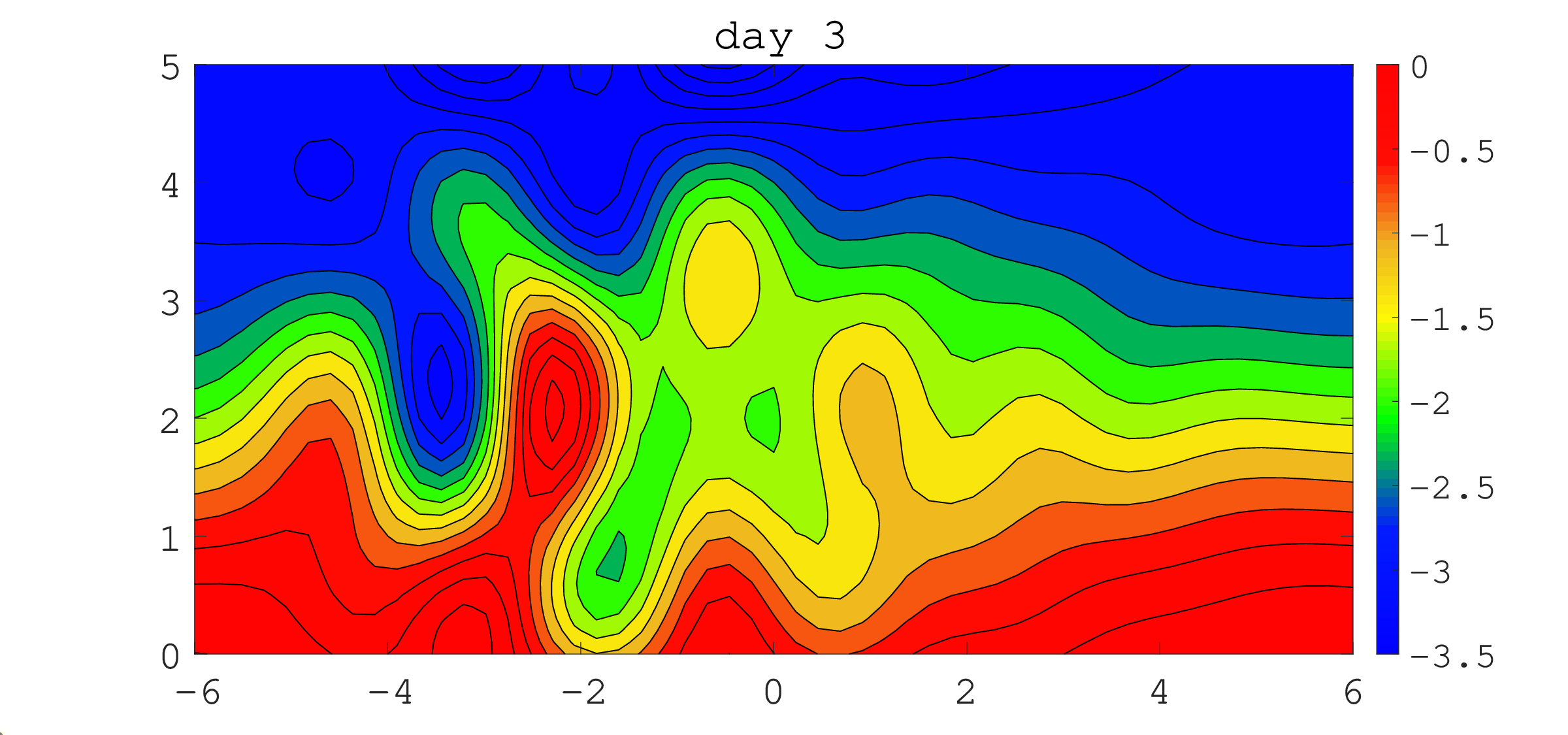}
\includegraphics[scale=0.125]{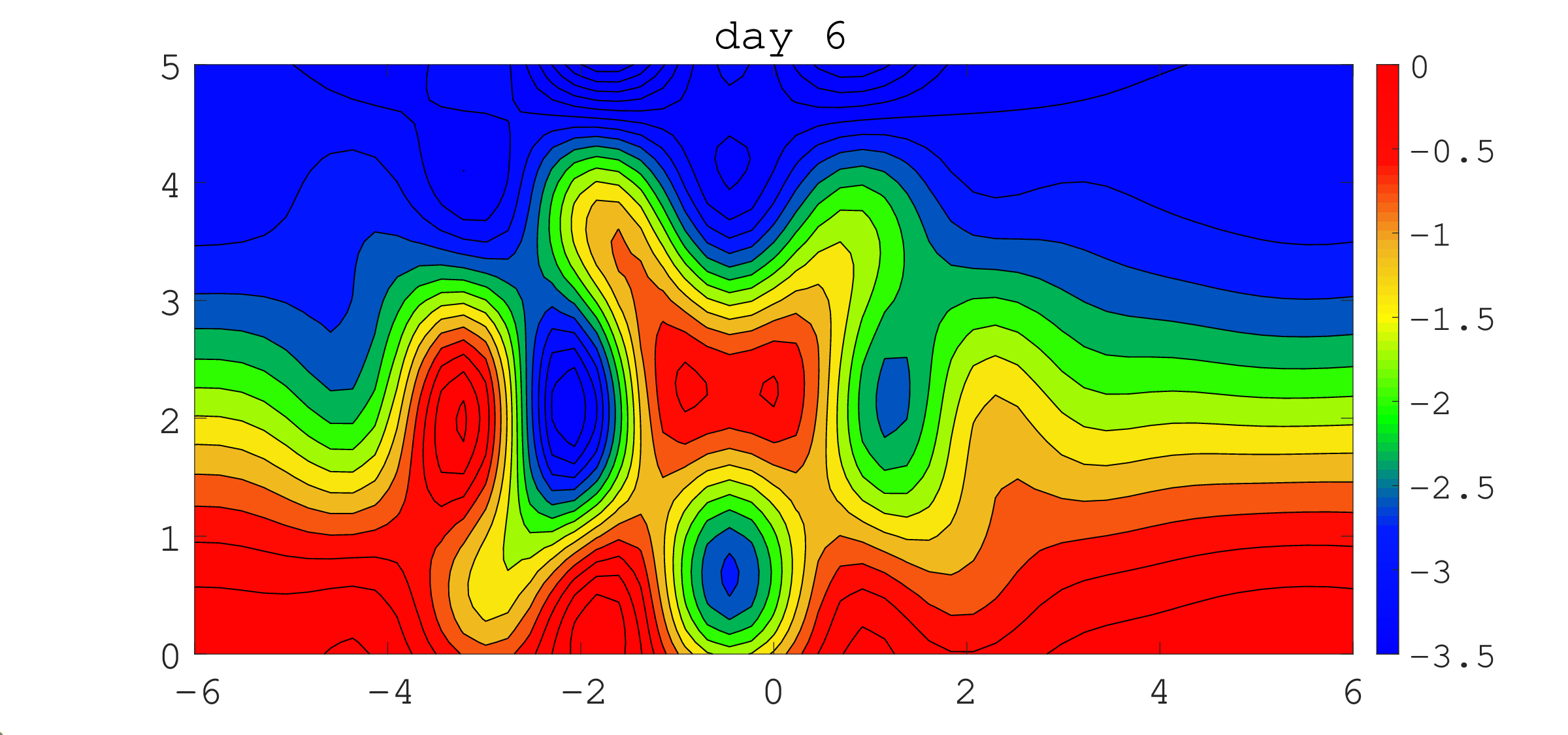}
\includegraphics[scale=0.125]{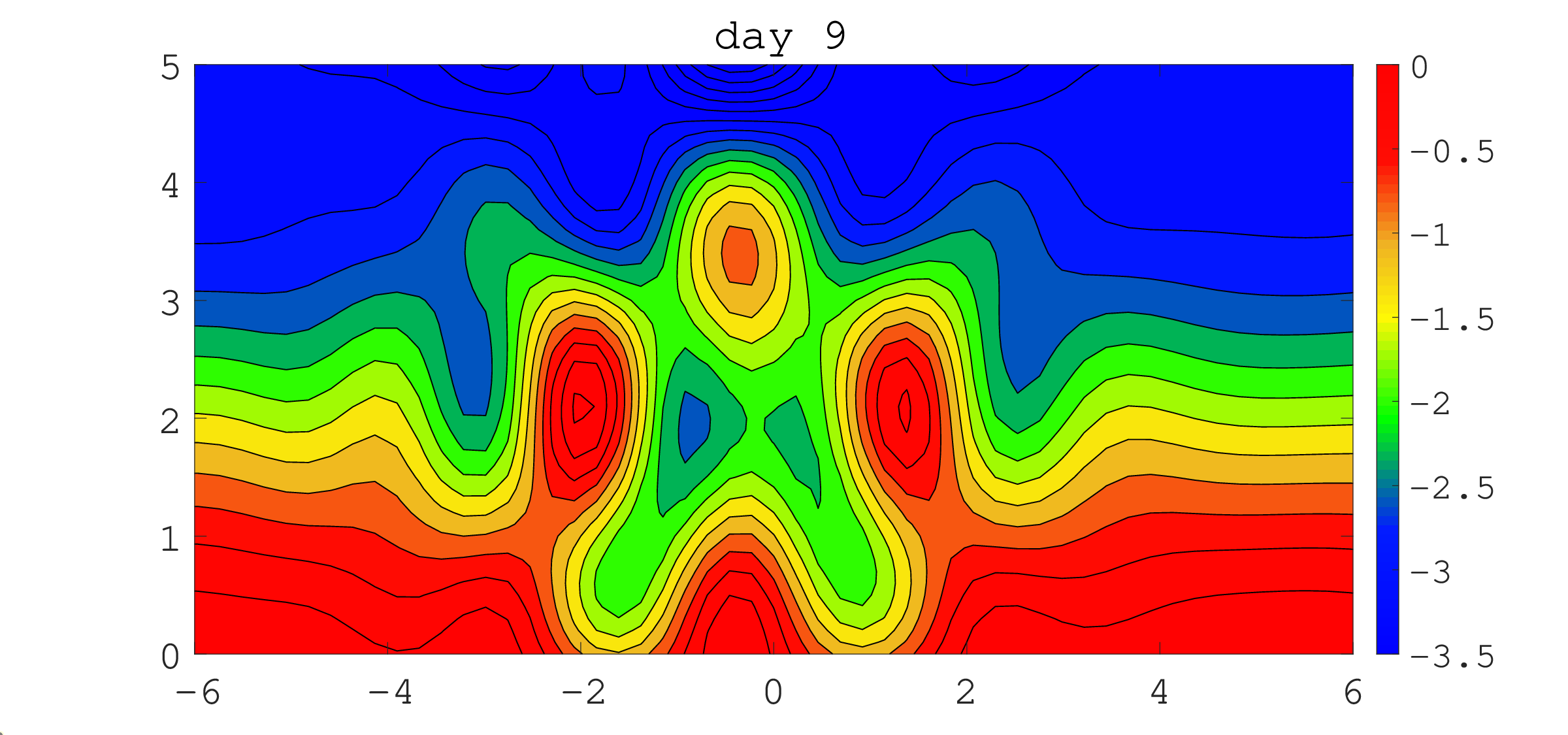}
\includegraphics[scale=0.125]{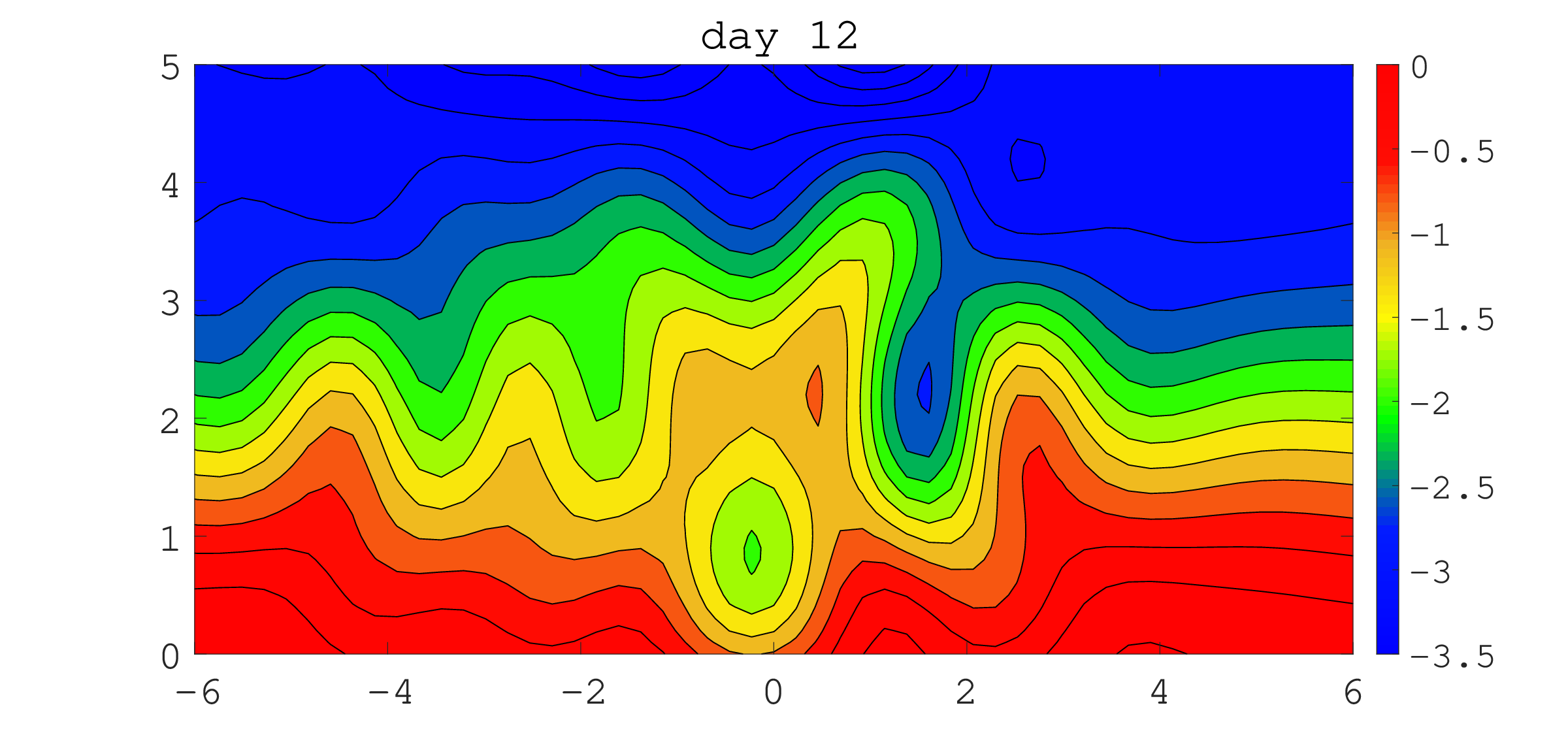}
\includegraphics[scale=0.125]{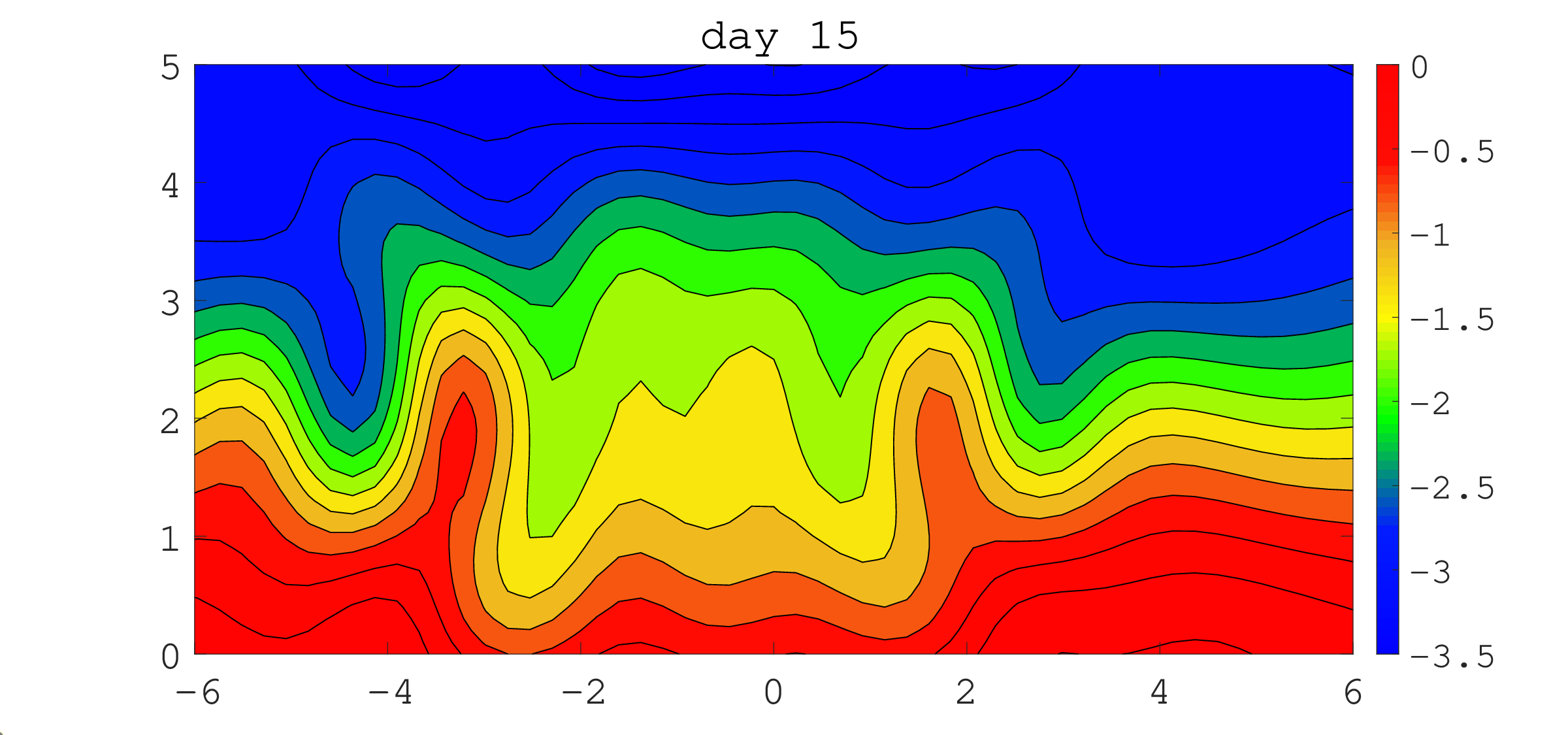}
\caption{$B_0=0.4$}
\label{fig:b0-init}
\end{subfigure}
\begin{subfigure}[b]{0.32\linewidth}
\centering
\includegraphics[scale=0.125]{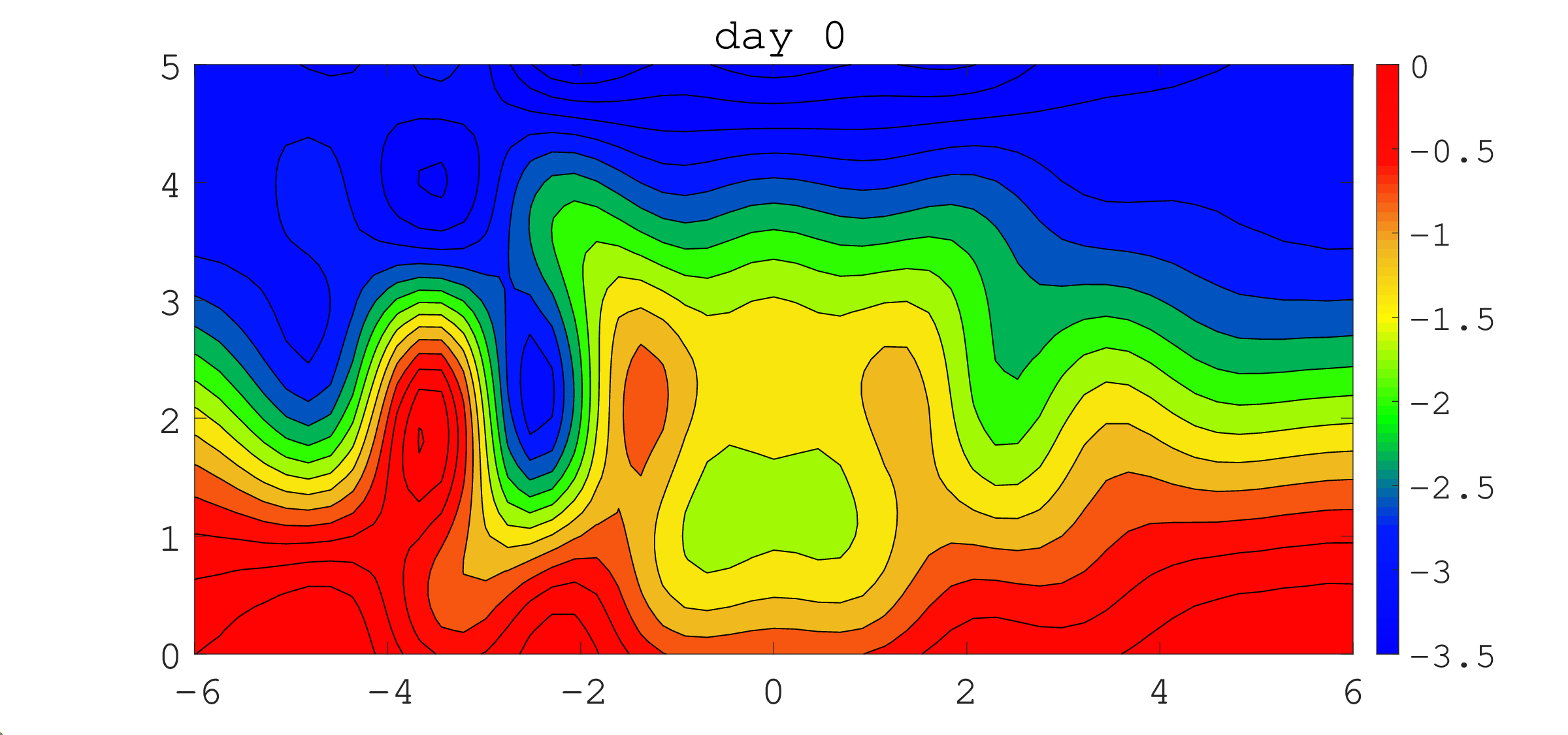}
\includegraphics[scale=0.125]{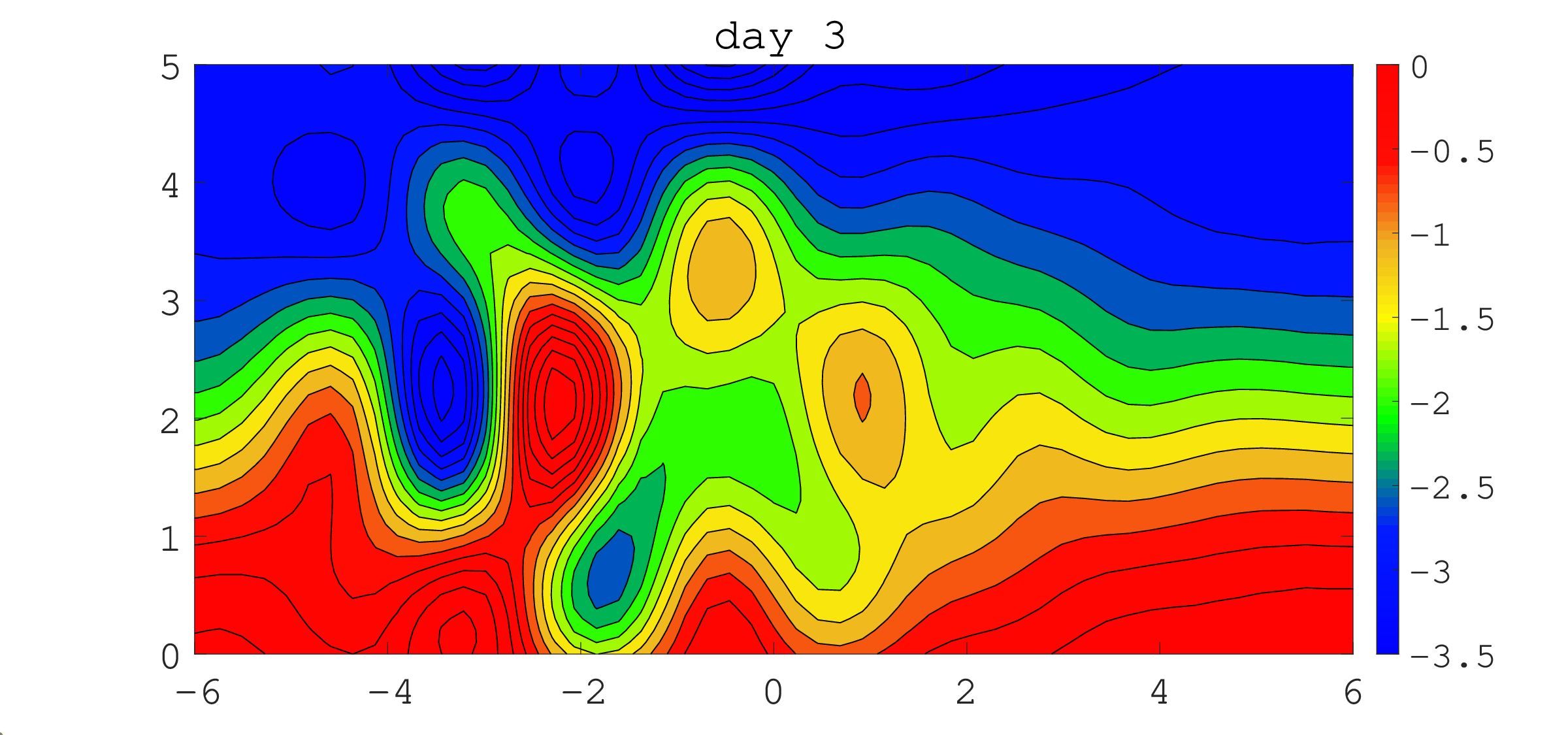}
\includegraphics[scale=0.125]{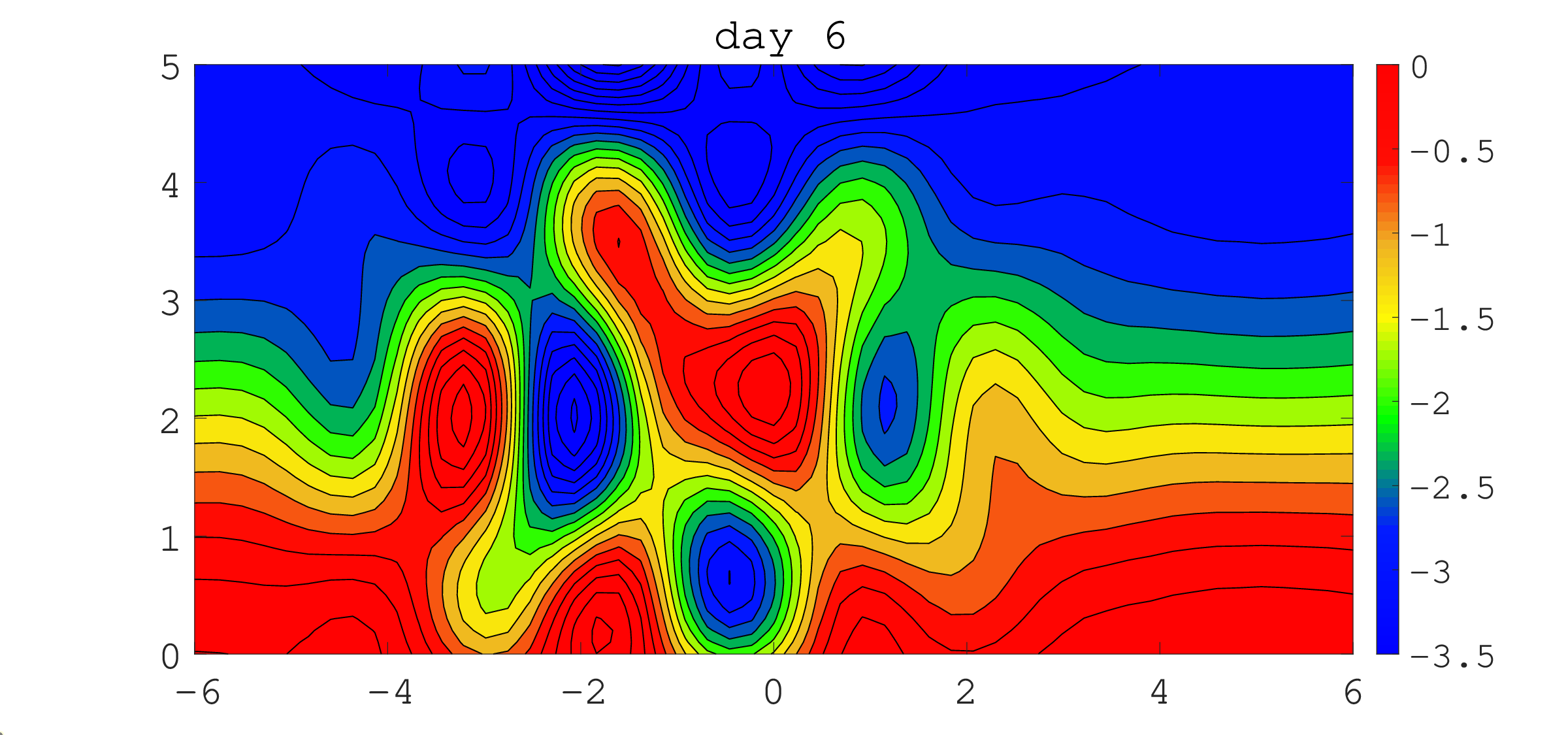}
\includegraphics[scale=0.125]{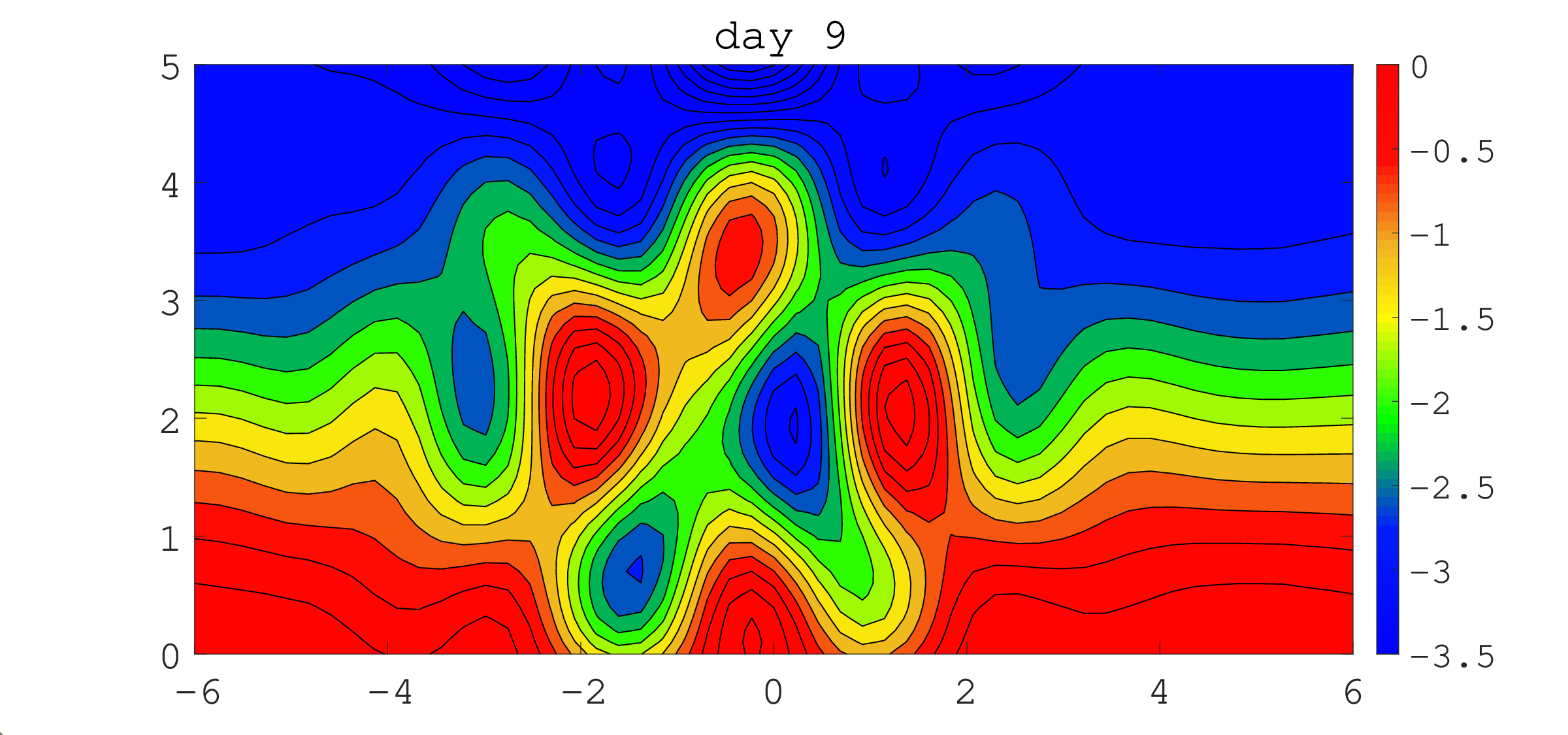}
\includegraphics[scale=0.125]{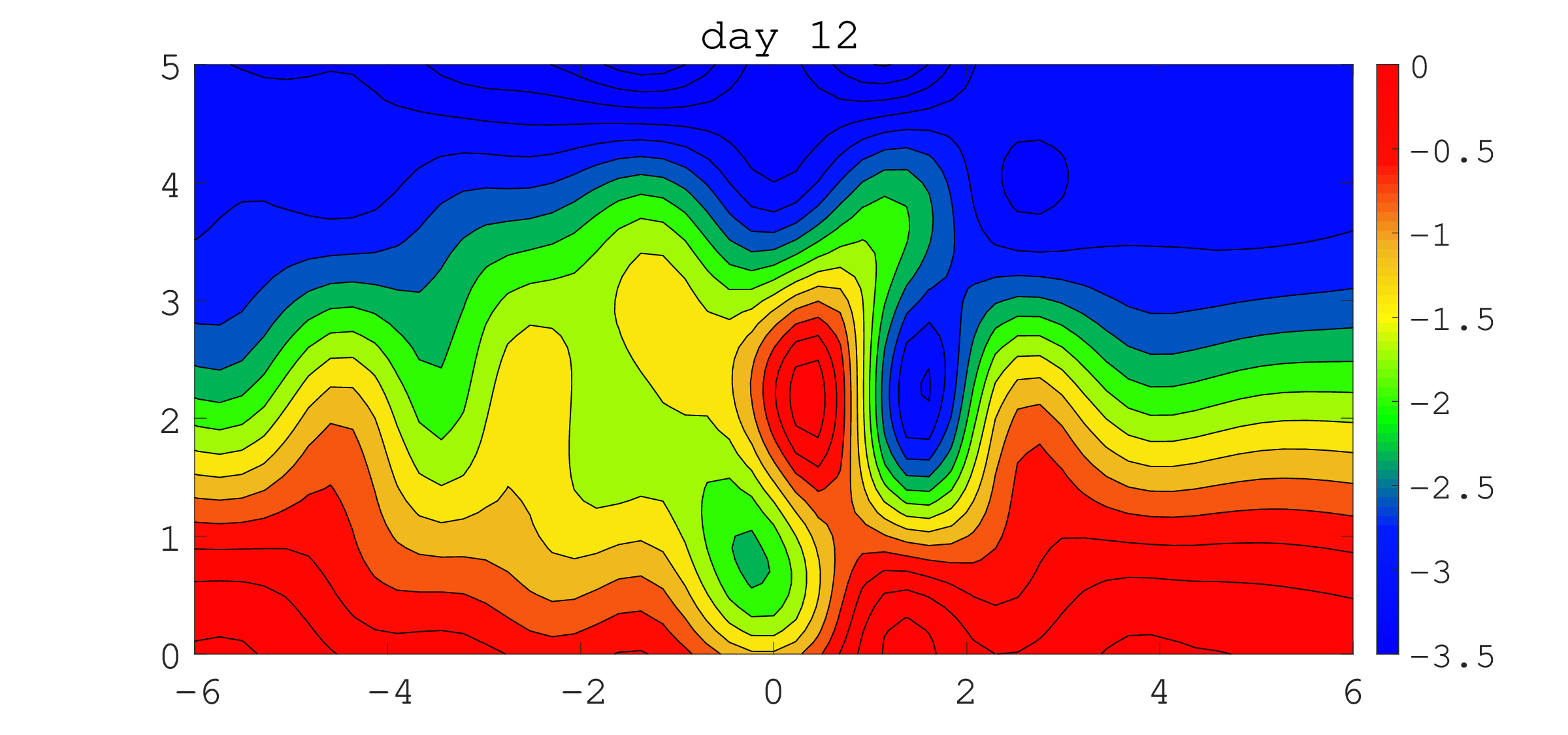}
\includegraphics[scale=0.125]{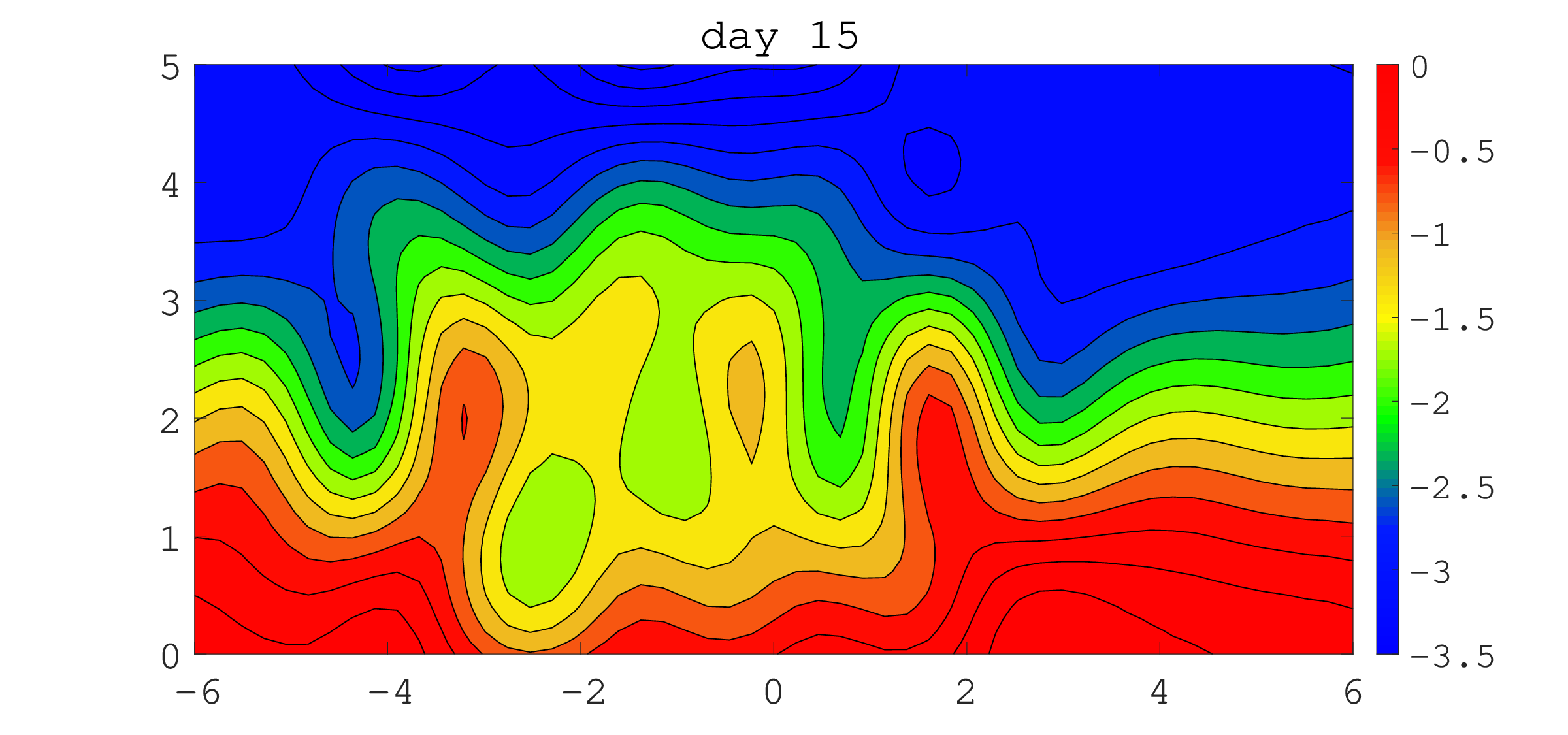}
\caption{$B_0=0.4,\;\gamma=0.5$}
\label{fig:b0-init-gamma}
\end{subfigure}
\begin{subfigure}[b]{0.32\linewidth}
\centering
\includegraphics[scale=0.125]{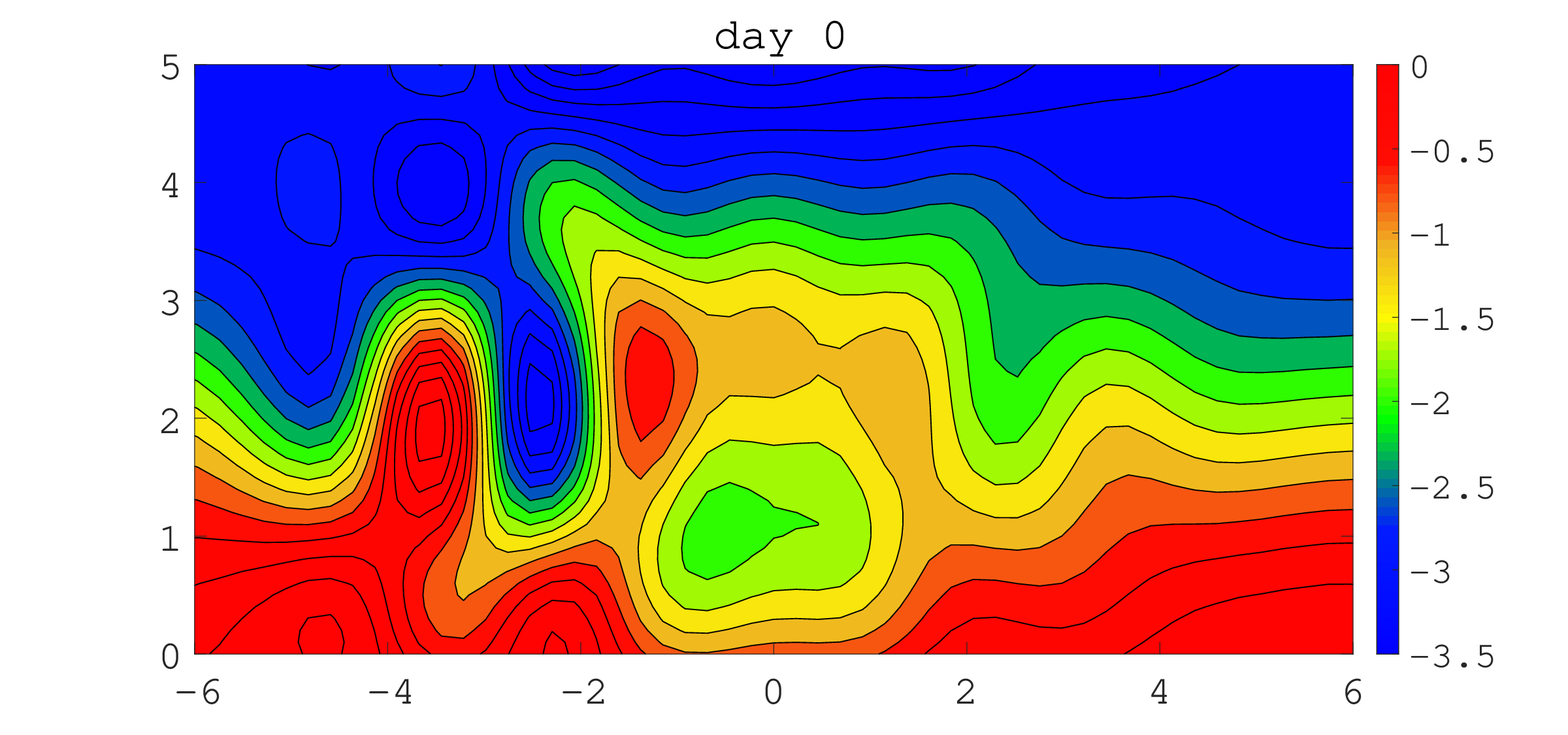}
\includegraphics[scale=0.125]{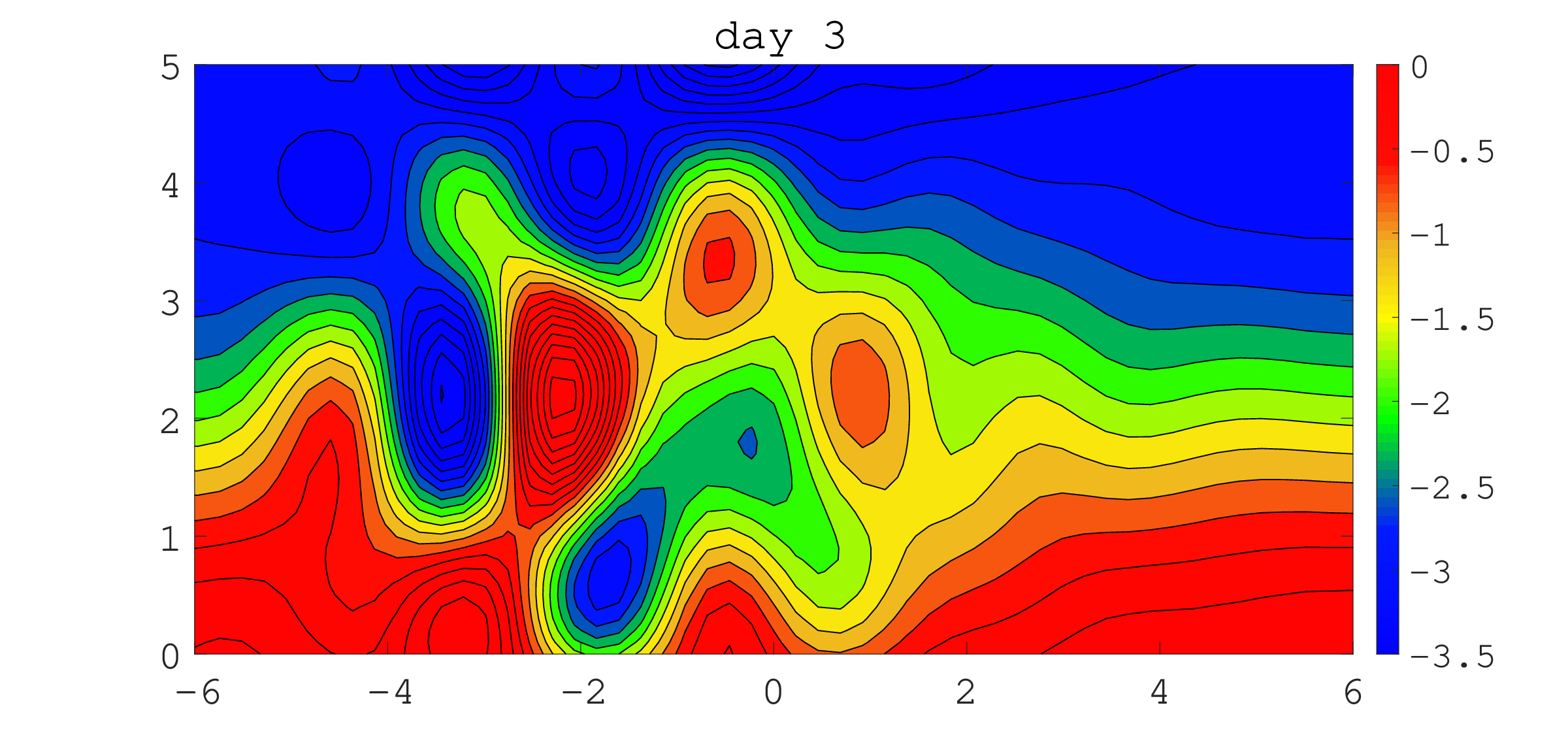}
\includegraphics[scale=0.125]{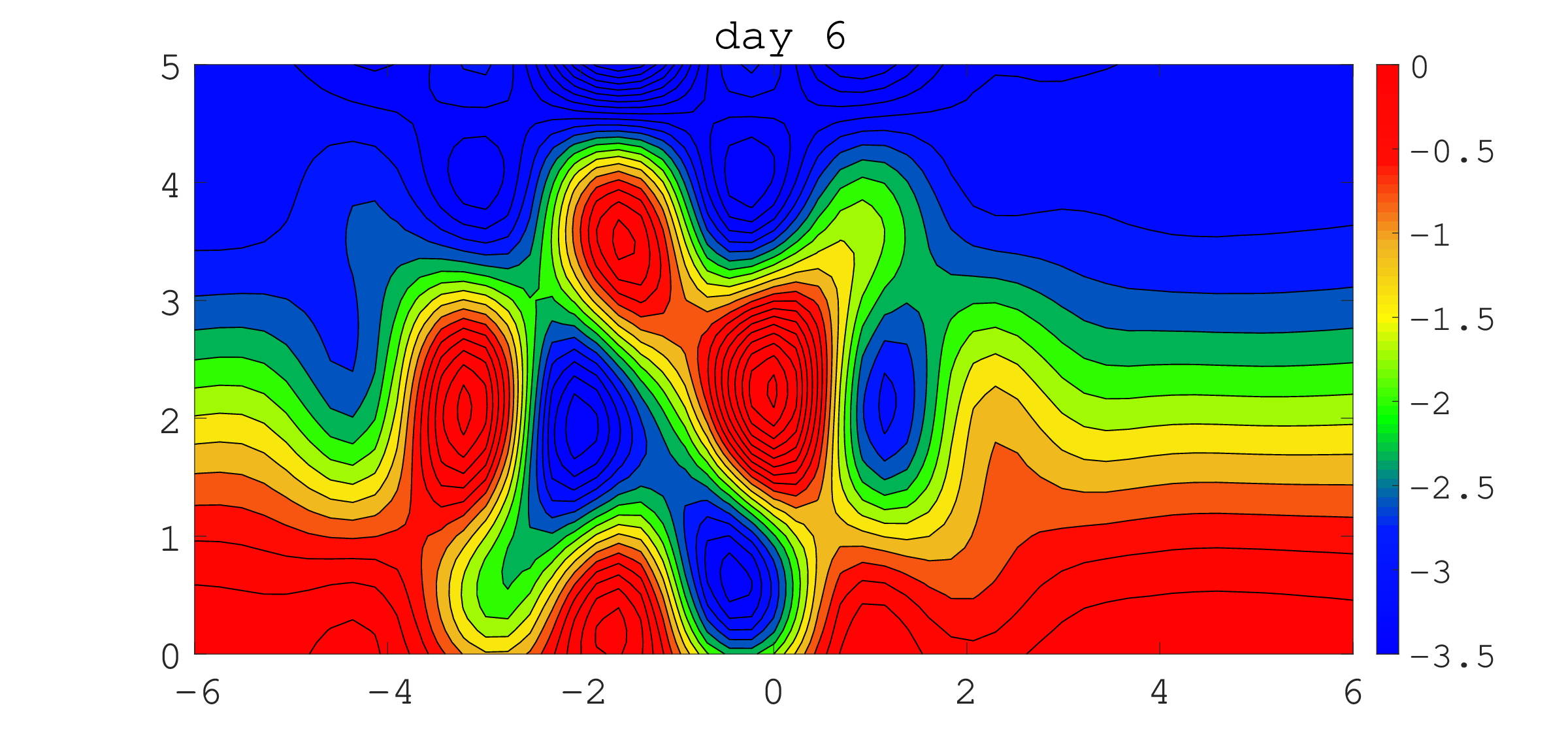}
\includegraphics[scale=0.125]{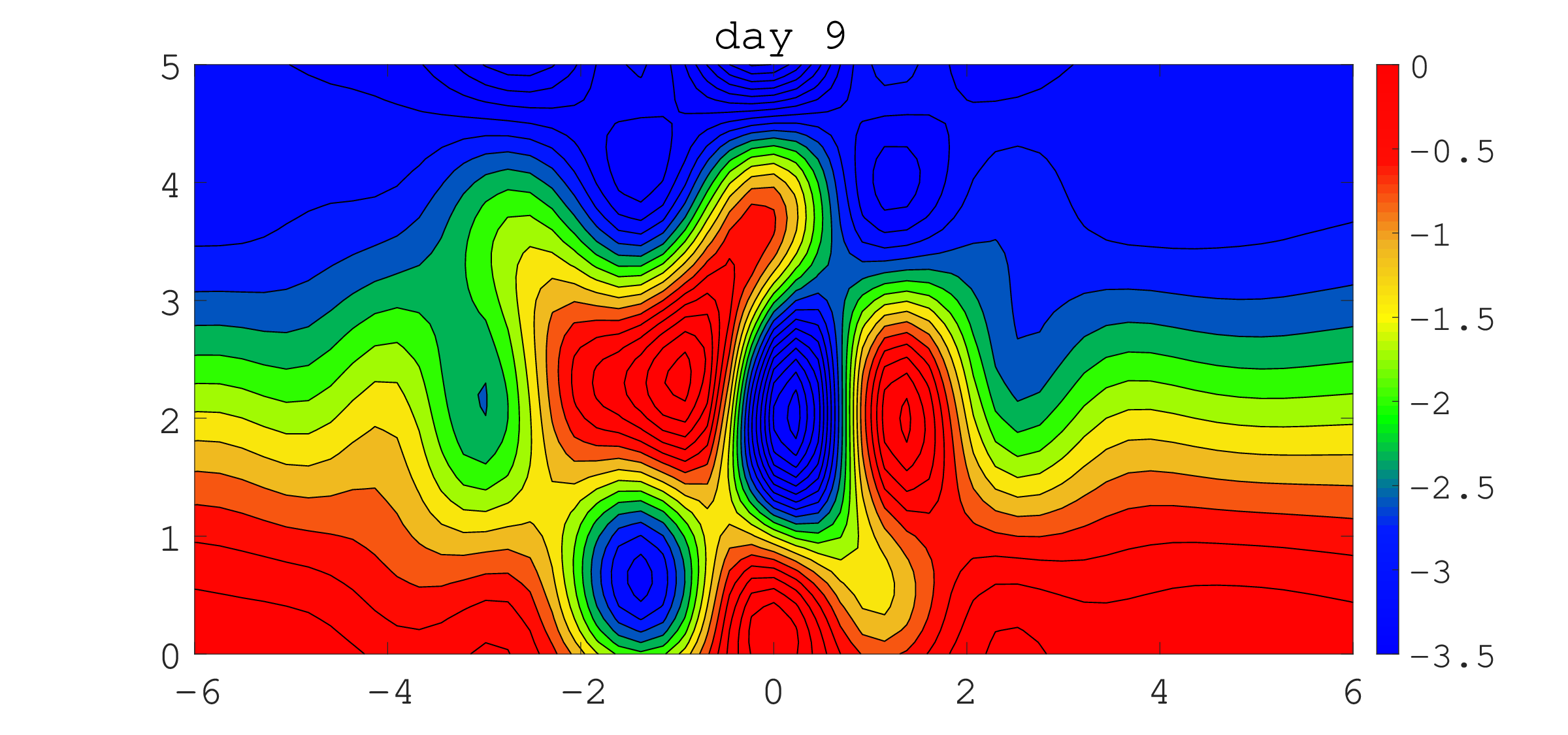}
\includegraphics[scale=0.125]{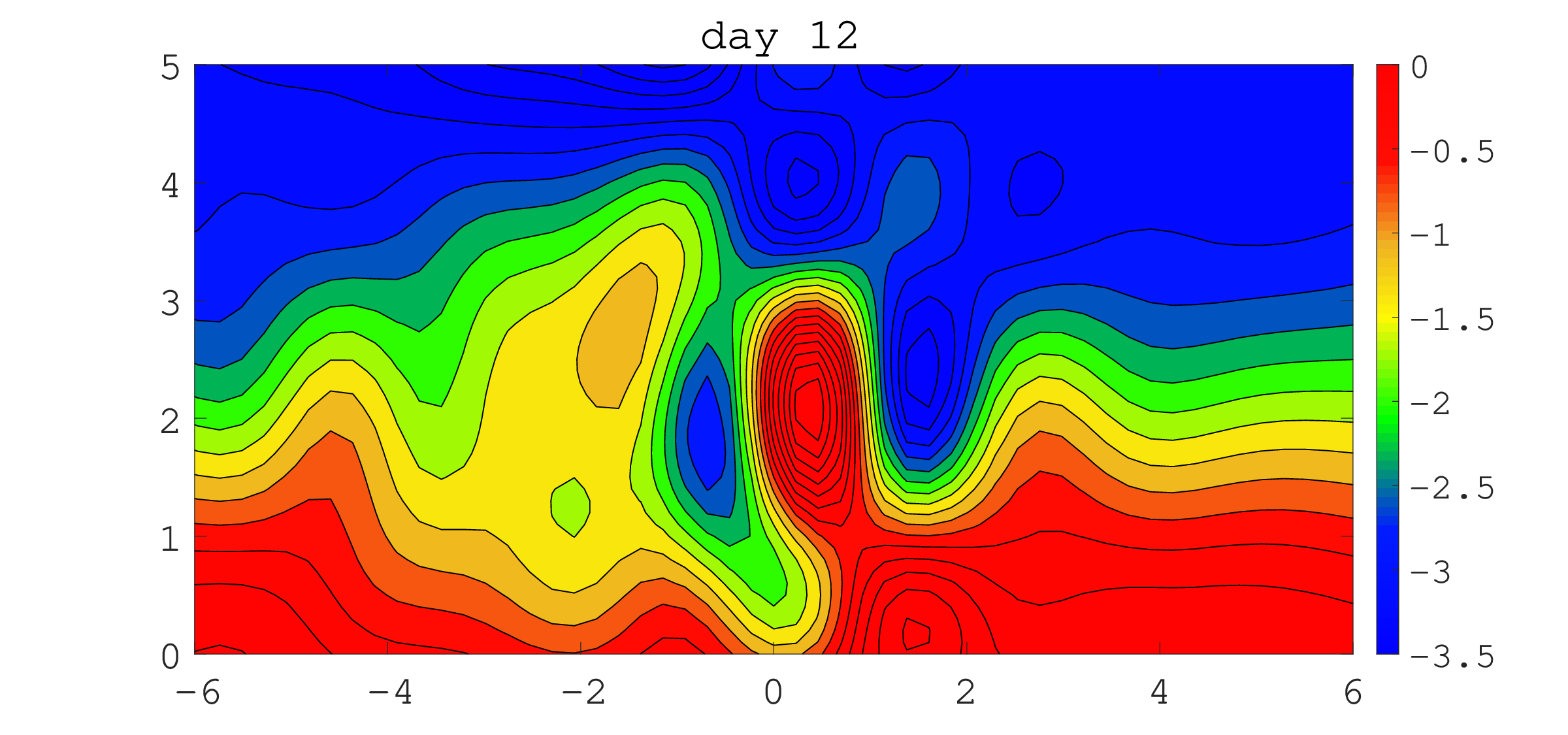}
\includegraphics[scale=0.125]{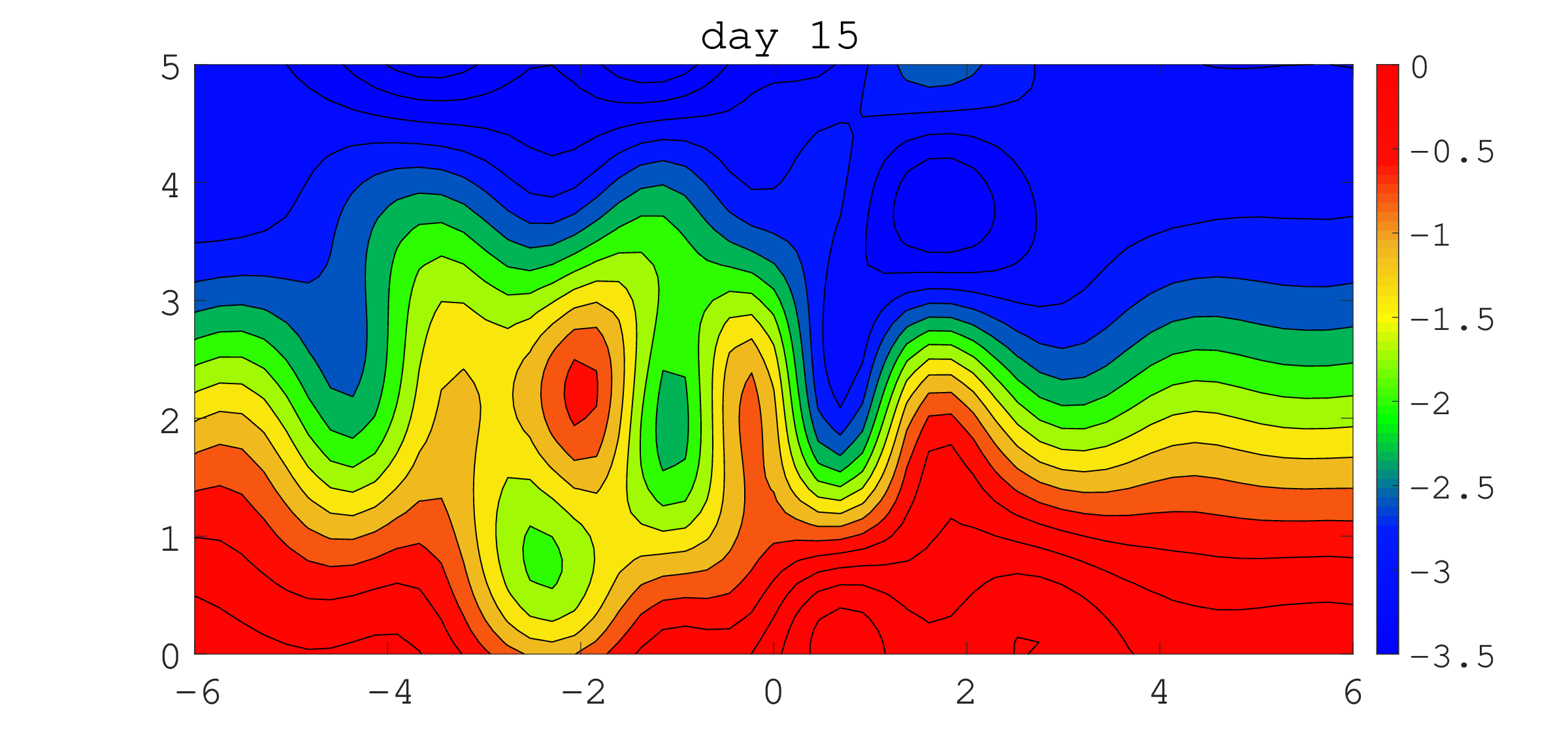}
\caption{$B_0=0.4,\;\gamma=1$}
\label{fig:b0-init-gamma1}
\end{subfigure}
\caption{Nonlinear evolution of the instantaneous total streamfunction field $\psi_T$ when the initial blocking amplitude is added by the optimal disturbance with the incremental increase of the size parameter $\gamma$.}
\label{fig:sf-init}
\end{figure}

Based on the provided expressions~\eqref{eqn: blocking-wave-1} and~\eqref{eqn: associate-2}, the blocking wavy anomaly is represented as $\psi_B = \epsilon \psi_1$ and the associated zonal-mean anomaly is represented as $\psi_m = \epsilon^2\psi_2$. Additionally, based on~\cref{eqn: synoptic-1}, the streamfunction of the preexisting synoptic-scale eddies is approximated by $\psi' \approx \epsilon^{\frac{3}{2}}\psi_1' + \epsilon^{\frac52}\psi_2'$.\footnote{The derivation of the deformed synoptic-scale eddy $\psi_2'$ is so tedious and circumstantial that we postpone it to the supplementary materials. } According to the expression~\eqref{eqn: total}, the total streamfunction can be approximated as $\psi_T  = \overline{\psi} + \psi + \psi' \approx  \overline{\psi}+\psi_B + \psi_m + \epsilon^{\frac{3}{2}}\psi_1' + \epsilon^{\frac52}\psi_2'$. In~\Cref{subsec: nmi-model}, it is derived that the blocking amplitude is governed by the forced NLS equation with the periodic boundary condition~\eqref{eqn: ampli-nls}. As mentioned in~\citep{luo2019nonlinear}, the evolution of the instantaneous total streamfunction $\psi_T$ is entirely dependent on the blocking amplitude throughout the whole life of the blocking. In~\Cref{fig:sf-init}, we show the evolution of the instantaneous total streamfunction $\psi_T$ when the initial blocking amplitude is added by the optimal disturbances. This visualization provides a clear representation of how the total streamfunction changes as the size parameter $\gamma$ increases incrementally.
\Cref{fig:b0-init} visualizes the evolution of the instantaneous total streamfunction without any perturbations, highlighting two predominant phenomena commonly associated with the blocking events: eddy straining and wave breaking. These phenomena, as described in~\citep{shutts1983propagation, pelly2003new}, are known to play a significant role in the occurrence of blocking events. By comparing~\Cref{fig:b0-init} to~\Cref{fig:b0-init-gamma} and~\Cref{fig:b0-init-gamma1} from left to right, it is observed that as the size parameter $\gamma$ increases, the phenomena of eddy straining and wave breaking become more prominent.  However, both the position and the period of the blocking remain almost unchanged. During the maintenance period of the blocking, the intensification of eddy straining and wave breaking becomes extremely dominant, but their positions and periods remain mostly invariant. From a physics perspective, it appears that the optimal disturbance of initial blocking amplitude tends to intensify the strength of the blocking without causing any other significant changes. This intensification becomes more pronounced as the size of the optimal disturbance increases.  This observation provides valuable insights into the motion of the blocking and the role of the size of the optimal disturbance in their intensification. It is possible that the nonlinear overgrowth caused by the optimal disturbance of the initial blocking amplitude, as mentioned in~\citep{bengtsson1981numerical, tibaldi1990operational, burroughs_1997}, could be a contributing factor to the frequent occurrence of extreme weather events and the subsequent decrease in predictability. This suggests that the nonlinear growth of the optimal disturbance, dominated by its size, may lead to the complex dynamics of the blocking.

\subsection{Less predictability on the medium-range}
\label{subsec: less-predictability-init}

Weather prediction systems have indeed made significant progress over the years, thanks to advancements in technology, data collection, and modeling techniques. These advancements have greatly improved the accuracy and reliability of weather forecasts. However, despite these improvements,  accurately predicting blocking on the medium-range weather timescale remains a challenge, as highlighted in~\citep{kautz2022atmospheric}. Additionally,~\citet{hamill2014skill} have found that forecast errors tend to be larger for European blocking compared to other regimes, particularly during the transition phases into or out of a blocking regime. These findings emphasize the complexity and difficulty in accurately forecasting blocking events.
 
\begin{figure}[htpb!]
\centering
\begin{subfigure}[t]{0.325\linewidth}
\centering
\includegraphics[scale=0.14]{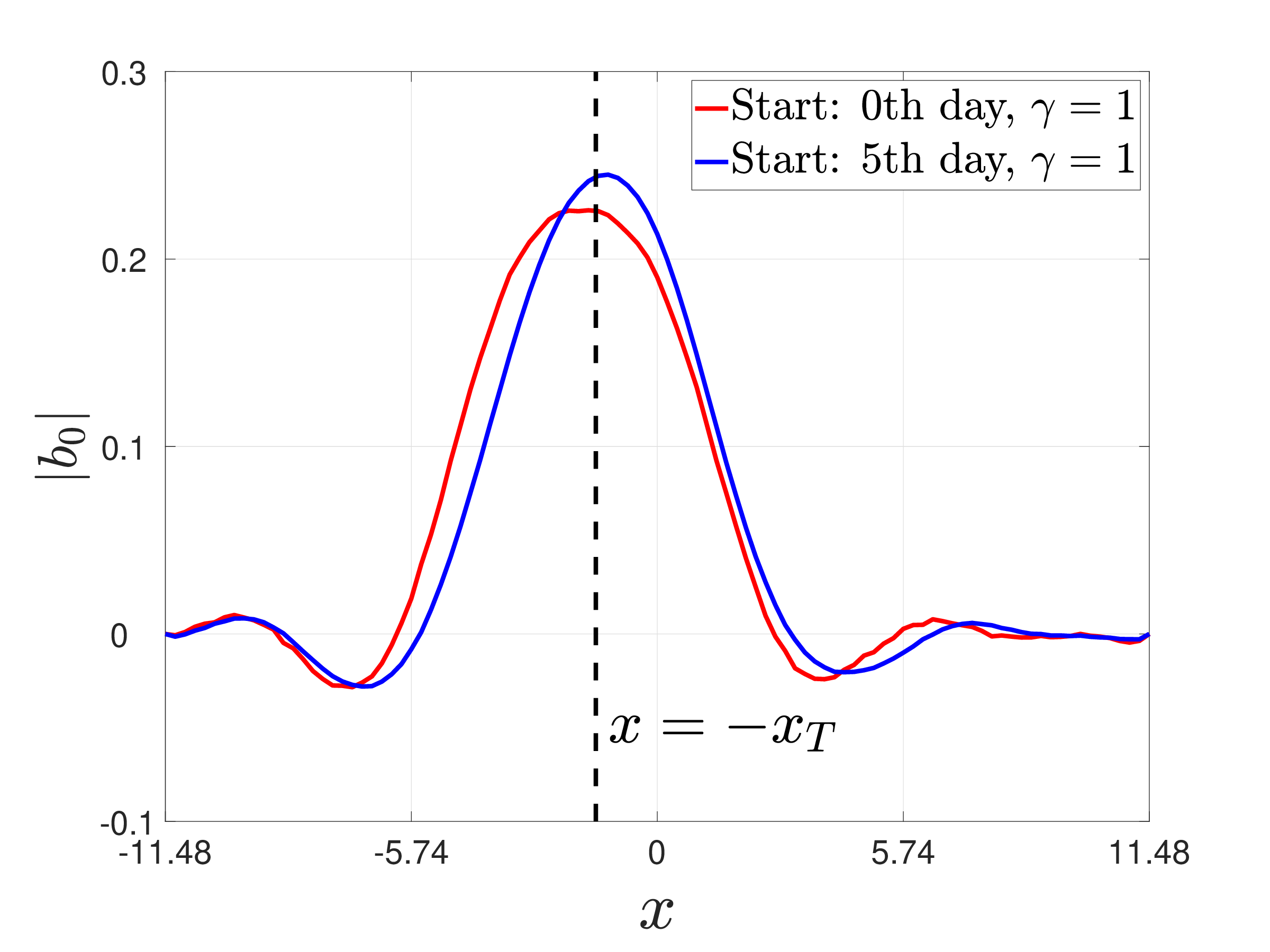}
\caption{Spatial Pattern}
\label{subfig: td-sp-cnop-init}
\end{subfigure}
\begin{subfigure}[t]{0.325\linewidth}
\centering
\includegraphics[scale=0.14]{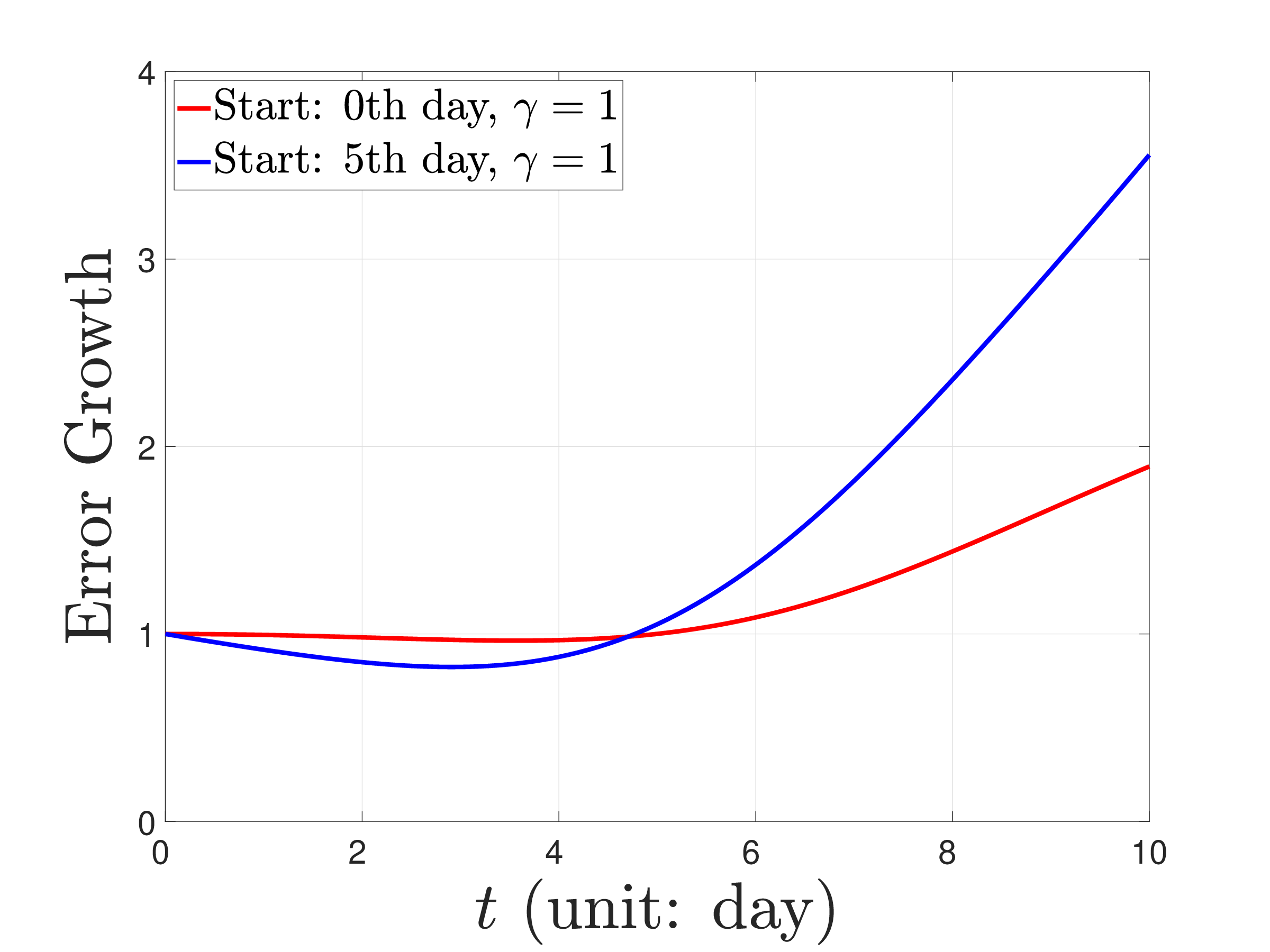}
\caption{(standard) Error Growth}
\label{subfig: tp-eg-cnop-init}
\end{subfigure}
\begin{subfigure}[t]{0.325\linewidth}
\centering
\includegraphics[scale=0.14]{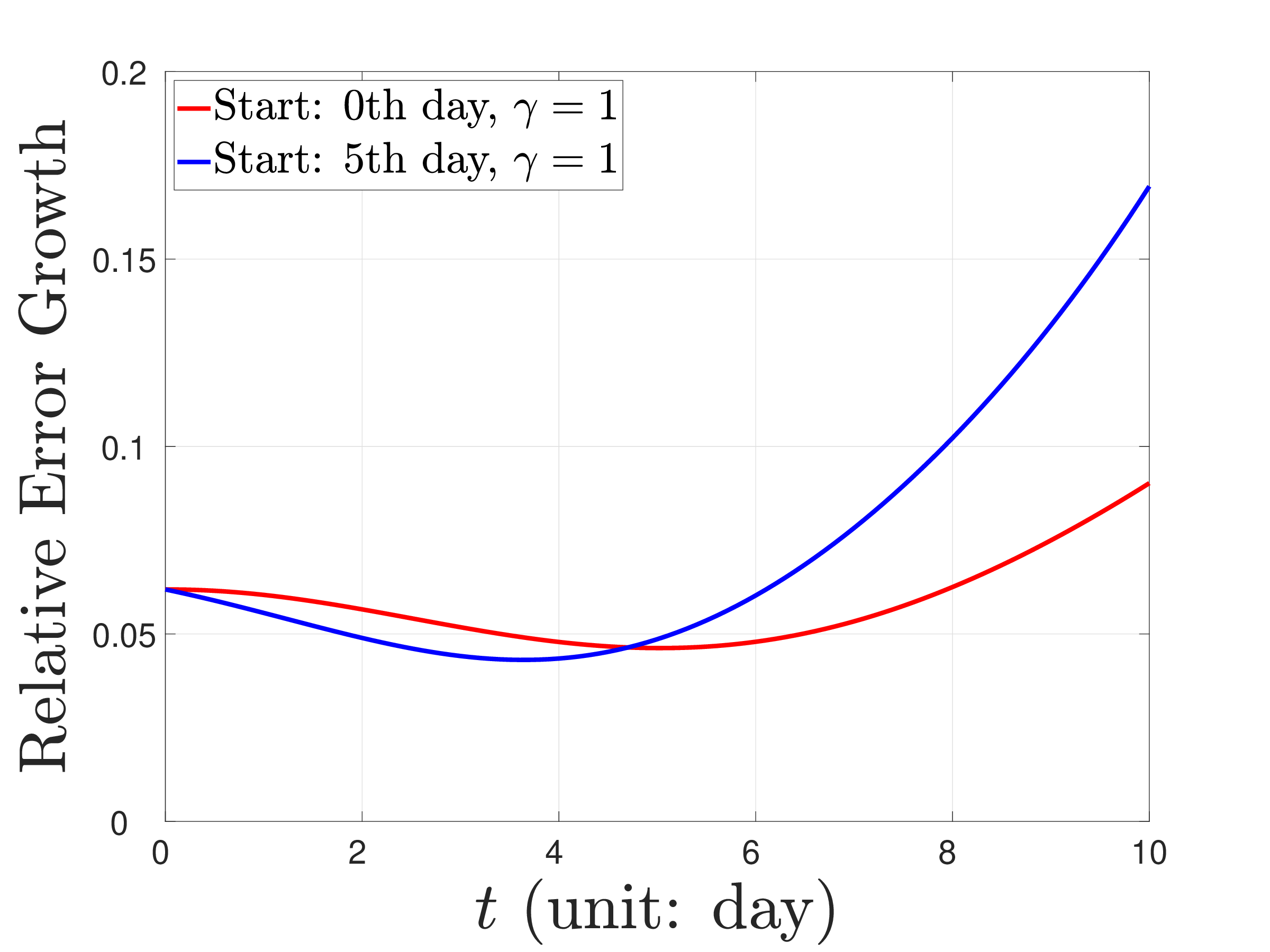}
\caption{Relative Error Growth}
\label{subfig: td-reg-cnop-init}
\end{subfigure}
\caption{Spatial pattern and nonlinear growth of the optimal disturbance at a later stage in comparison to the baseline at the initial stage. } 
\label{fig: td-cnop-init}
\end{figure}

It sounds like a reasonable approach for conducting a numerical experiment using the NMI model to explore whether there are larger forecast errors caused by the optimal disturbance. To start, we run the forced NLS equation~\eqref{eqn: ampli-nls} for 5 days without any perturbations. After that, using the blocking amplitude on the 5th day as the initial condition,  our goal is to obtain the optimal disturbance during the period from 5 to 15 days. The spatial pattern of the optimal disturbance at a later stage, compared with the initial stage, is shown in~\Cref{subfig: td-sp-cnop-init}, where it is observed that the peak of the solitary wave slightly offsets to the right and becomes sharper. The nonlinear growth and relative nonlinear growth of the optimal disturbances are depicted in~\Cref{subfig: tp-eg-cnop-init} and~\Cref{subfig: td-reg-cnop-init}, as described by~\cref{eqn: error-init} and~\cref{eqn: rel-error-init}, respectively. It is worth noting that when the time interval $T$ and the size parameter $\gamma=1$ are fixed, the optimal disturbance at a later stage leads to a larger error, which is demonstrated quantitatively by the ratios between the later stage and the initial stage are $1.8776$ for the nonlinear growth and $1.8781$ for the relative nonlinear growth, respectively. Additionally, the experiment reveals that the nonlinear evolution of the optimal disturbance is more pronounced during the decay of the blocking, while the error is smaller during the maintenance of the blocking. This finding aligns with the less predictability of blockings on the medium-range, as mentioned in~\citep{hamill2014skill, ferranti2015flow, zhang2019predictability}.  This suggests that the presence of the optimal disturbance at a later stage, as observed in the numerical experiment, can contribute to larger forecast errors in predicting blockings.

\section{Optimal disturbance of the preexisting synoptic-scale eddies}
\label{sec: cnop-synp}

In this section, our focus is on investigating the optimal disturbance of the preexisting synoptic-scale eddies. Our aim is to understand the spatial patterns and the nonlinear growth of the error associated with this disturbance. We also explore how the total blocking evolves as the optimal disturbance incrementally increases in size.  Furthermore, we analyze the time-delay effect of the optimal disturbance and its relation with predictability.

\subsection{Spatial pattern and nonlinear growth}
\label{subsec: spng-synp}

\begin{figure}[htb!]
\centering
\includegraphics[scale=0.26]{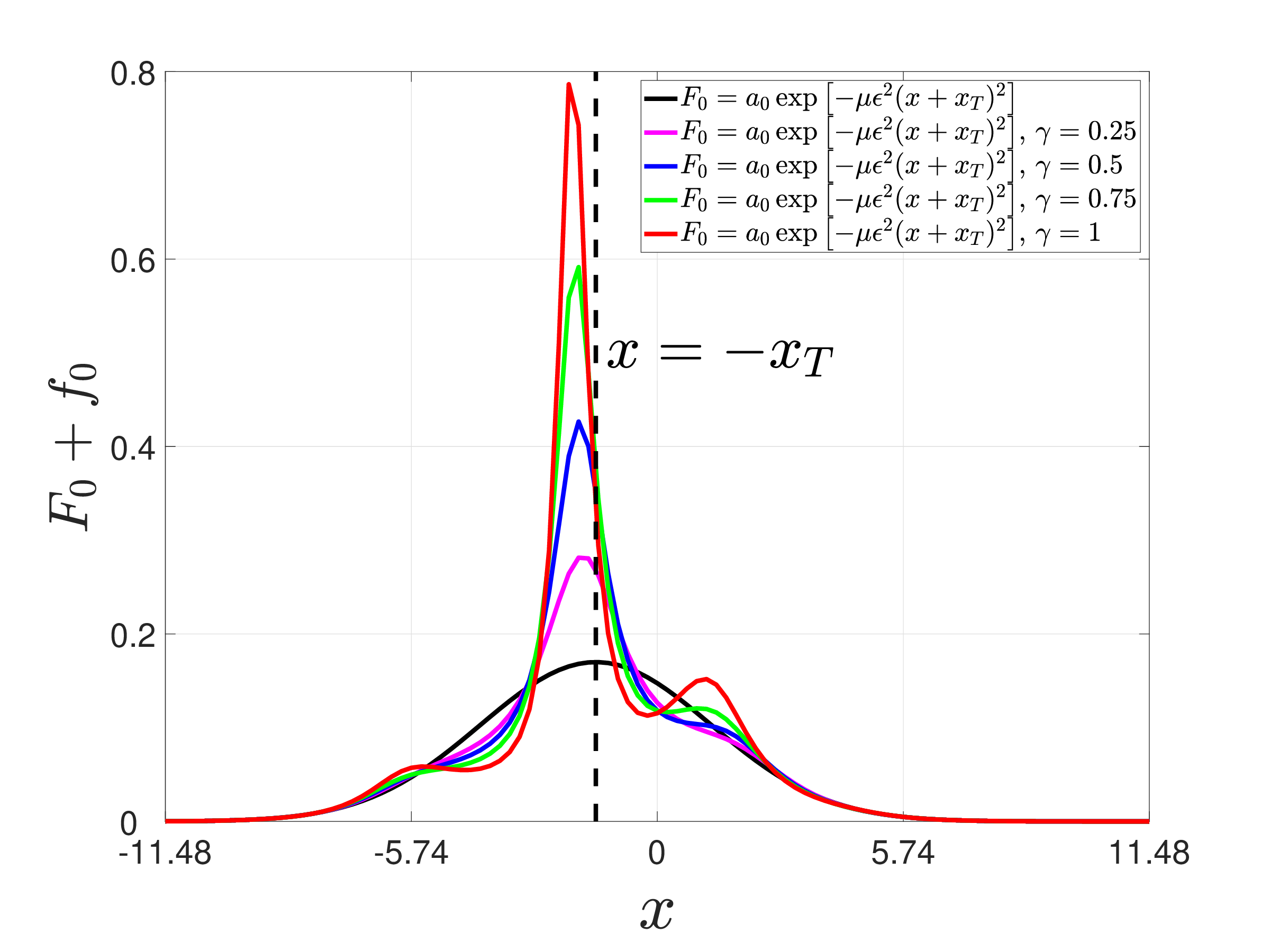}
\caption{Spatial patterns (nondimensionalization) of the optimal disturbance $f_0$ under the preexisting synoptic-scale eddies varies with the increase of the size parameter $\gamma$.}
\label{fig: nmi-cnop-synp}
\end{figure}

In this given context, our goal is to numerically compute the optimal disturbance of the preexisting synoptic-scale eddies, denoted as $f_0$. This computation is based on the forced NLS equation with the periodic boundary condition~\eqref{eqn: ampli-nls} and the preexisting synoptic-scale eddies $F_0 = a_0  \exp\left[ - \mu \epsilon^2(x + x_T)^2 \right]$. To achieve this, we can maximize the constrained objective function~\eqref{eqn: obj-synp}. Increasing the size of the optimal disturbance, denoted by the parameter $\gamma$, allows for a more in-depth analysis of the numerical performance of spatial patterns. In~\Cref{fig: nmi-cnop-synp}, we provide a clear visualization of the spatial patterns, showcasing how they vary as the size parameter $\gamma$ increases incrementally. The optimal disturbance is observed to concentrate sharply around a slight offset to the left of the zonal center $x = -x_T$, which appears as a sharp bulge with two small dents on either side and two small bulges beside them. As the size parameter $\gamma$ increases incrementally from $0.25$ to $1$, the optimal disturbance becomes even more pronouncedly sharp. Specifically, the bulge becomes highly concentrated, resembling a sharply rising peak, which appears to be more predominant than the optimal disturbance of the initial blocking amplitude. This suggests that, in the context of blocking events in the real world, the largest deviation in the preexisting synoptic-scale eddies concentrates sharply around a slight offset to the left of the zonal center $x = -x_T$.

\begin{figure}[htpb!]
\centering
\begin{subfigure}[t]{0.48\linewidth}
\centering
\includegraphics[scale=0.18]{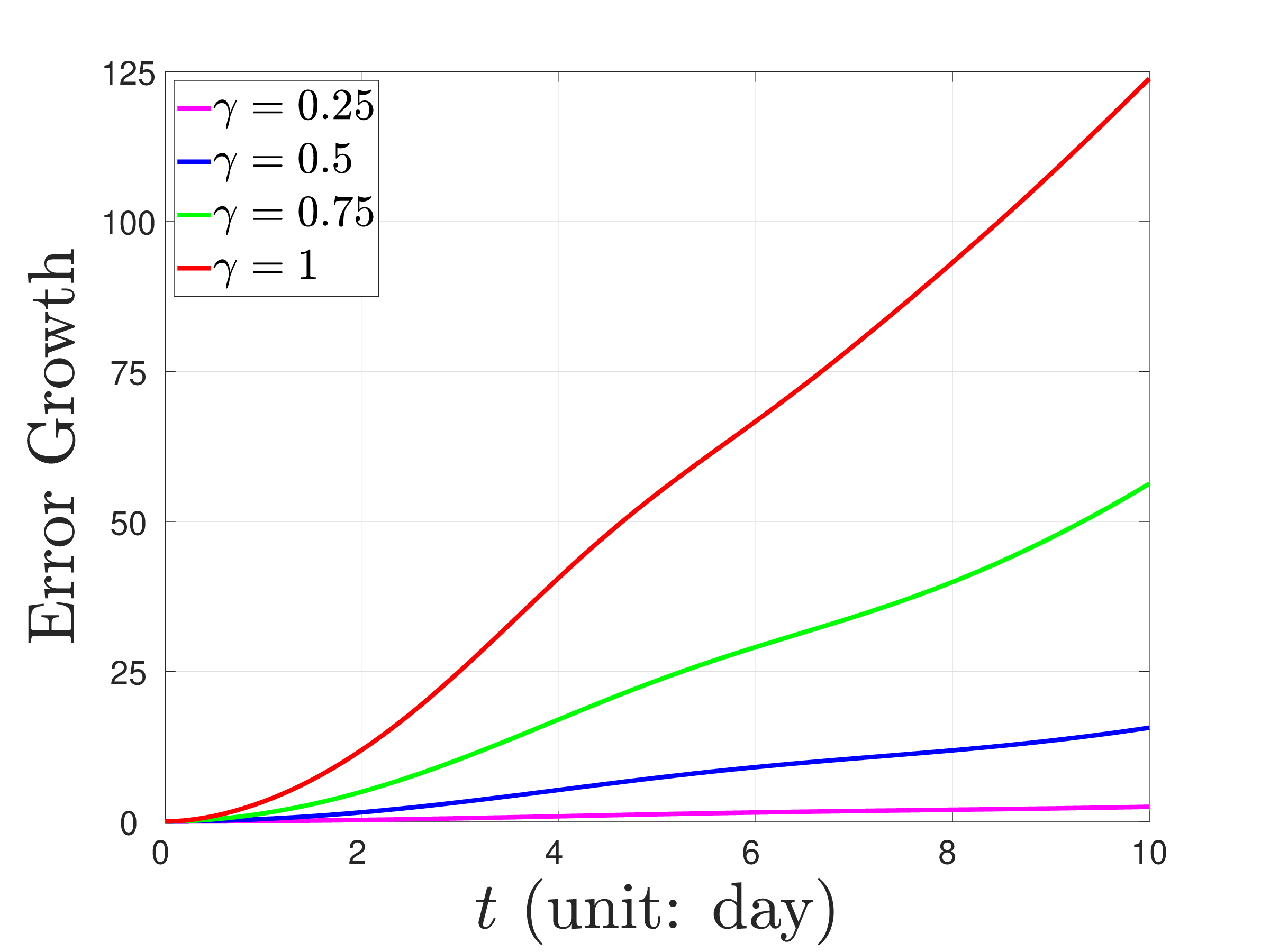}
\caption{(standard) Error Growth}
\label{fig: norm-stan-synp}
\end{subfigure} 
\begin{subfigure}[t]{0.48\linewidth}
\centering
\includegraphics[scale=0.18]{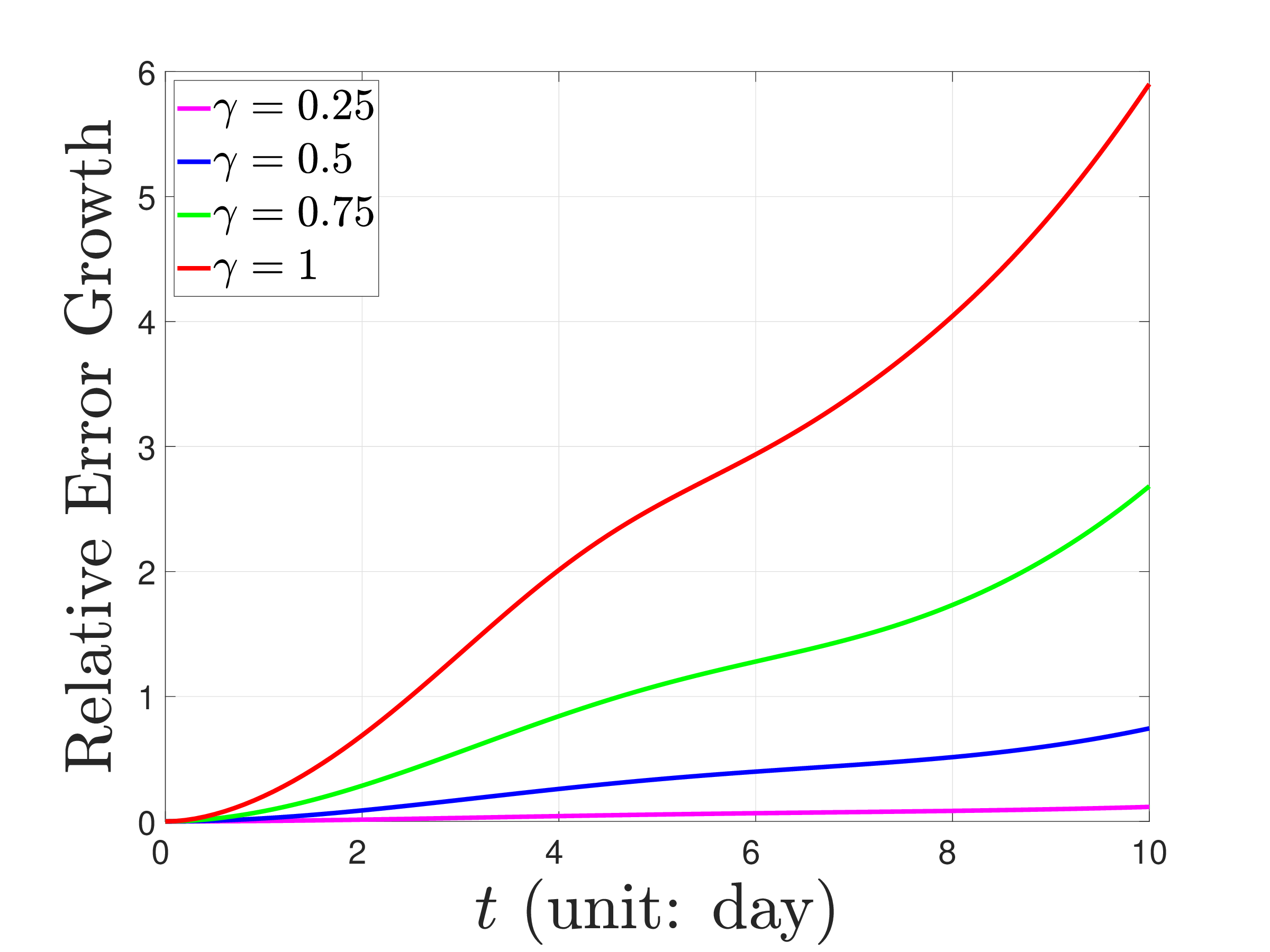}
\caption{Relative Error Growth}
\label{fig: norm-rel-synp}
\end{subfigure}
\caption{Nonlinear growth of the error generated by the optimal disturbance of the preexisting synoptic-scale eddies given by~\eqref{eqn: error-synp} and~\eqref{eqn: rel-error-synp} varies with the increase of the size parameter $\gamma$.}
\label{fig: error-growth-synp}
\end{figure}

Next, we explore the nonlinear growth of the error caused by the optimal disturbance of the preexisting synoptic-scale eddies. The energy norm~\eqref{eqn: energy-norm-B} is utilized to quantify this growth as 
\begin{equation}
\label{eqn: error-synp}
\frac{\|b(t)\|^2}{\Delta x} = \frac{\|B(t;B_0,F_0+f_0,U) - B(t;B_0,F_0,U)\|^2}{\Delta x},
\end{equation}
which measures the difference between the blocking amplitudes $B$ at time $t$, considering both the preexisting synoptic-scale eddies $F_0$ and the most perturbed preexisting synoptic-scale eddies $F_0+f_0$, while keeping the initial blocking amplitude $B_0$ and the westerly wind speed $U$ fixed. This allows for a comparison of the effects of the optimal disturbance on the blocking amplitude. The numerical performance of the nonlinear growth behavior is visualized in~\Cref{fig: norm-stan-synp}.
It is intriguing to see that the nonlinear growth of the error caused by the optimal disturbance of the preexisting synoptic-scale eddies exhibits a striking rise as the size parameter $\gamma$ increases incrementally. This behavior seems to be fully distinct from the nonlinear growth of the optimal disturbance of the initial blocking amplitude, as shown in~\Cref{fig: stan-init} and~\Cref{fig: norm-stan-init}. The phenomenon of the error growing several times aligns with weather predictions in the real world, as mentioned by~\citet{zhang2019predictability}. Similarly, we also explore the relative nonlinear growth of the error generated by the optimal disturbance, which can be calculated by taking the ratio between the nonlinear growth of the error $\frac{\|b(t)\|^2}{\Delta x}$  and the blocking amplitude $\frac{\|B(t)\|^2}{\Delta x}$ as
\begin{equation}
\label{eqn: rel-error-synp}
\frac{\|b(t)\|^2}{\|B(t)\|^2} = \frac{\|B(t;B_0,F_0+f_0,U) - B(t;B_0,F_0,U)\|^2}{\|B(t;B_0,F_0,U)\|^2}.
\end{equation}
It is interesting to note that in~\Cref{fig: norm-rel-synp}, the relative nonlinear evolution of the error caused by the optimal disturbance also exhibits a significant growth, which is consistent with the nonlinear growth observed in~\Cref{fig: norm-stan-synp}. Comparing these subfigures in~\Cref{fig: error-growth-synp} with the nonlinear growth of the optimal disturbance of the initial blocking amplitude shown in~\Cref{fig: error-growth-init} and~\Cref{fig: norm-error-growth-init} can help identify their differences. It appears that the optimal disturbance of the preexisting synoptic-scale eddies results in a considerably large error growth. This finding provides further insight into the role played by the fast-moving short-lived (high-frequency) synoptic-scale eddies in the blocking system, which has been previously studied in~\citep{berggren1949aerological, shutts1983propagation, hoskins1985use, luo2014nonlinear, luo2019nonlinear}. This highlights the potential impact of such disturbance on blockings, which could be a probable cause of weather extremes and reduce predictability. 
To further provide a more comprehensive characterization of the striking growth of the error caused by the optimal disturbance of the preexisting synoptic-scale eddies, it is indeed important to include the quantitative measurements of the nonlinear growth and the relative nonlinear growth. These quantitative measurements are calculated and presented in~\Cref{tab: relative-error-synp}, which further provides the evidence that the nonlinear growth and the relative nonlinear growth appear more pronouncedly sharp as the size parameter $\gamma$ increases incrementally from $0.25$ to $1$ .
\begin{table}[htb!]
\centering
\begin{tabular}{c|cccc}
   \toprule
                                                & $\gamma=0.25$                & $\gamma=0.5$                 & $\gamma=0.75$                & $\gamma=1$  \\
   \midrule
   $\frac{\|b(10)\|^2}{\Delta x}$               & $2.4371$                     & $15.6206$                    & $56.2985$                          & $123.8228$ \\
    \midrule
   $\frac{\|b(10)\|^2}{\|B(10)\|^2}$           & $0.1161$                      & $0.7443$                     & $2.6826$                              & $5.9000$   \\
    \bottomrule
\end{tabular}
\caption{The nonlinear growth and relative nonlinear growth of the error caused by the optimal disturbance of the preexisting synoptic-scale eddies in terms of norm squares.}
\label{tab: relative-error-synp}
\end{table}

\subsection{Temporal evolution of blocking under the optimal disturbance}
\label{subsec: block-simulation-synp}

It is indeed a valuable step to explore how the motion of the blocking is influenced by the optimal disturbance.  Understanding the dynamic relationship between the optimal disturbance and the evolution of the blocking can provide further insights into the impact of the optimal disturbance on the overall behavior of the blocking system. In particular, it would be interesting to investigate how the blocking evolves with time when the optimal disturbance is added to the preexisting synoptic-scale eddies.

\begin{figure}[htb!]
\centering
\begin{subfigure}[b]{0.32\linewidth}
\centering
\includegraphics[scale=0.125]{figure/sf_0/day00.eps}
\includegraphics[scale=0.125]{figure/sf_0/day03.eps}
\includegraphics[scale=0.125]{figure/sf_0/day06.eps}
\includegraphics[scale=0.125]{figure/sf_0/day09.eps}
\includegraphics[scale=0.125]{figure/sf_0/day12.eps}
\includegraphics[scale=0.125]{figure/sf_0/day15.eps}
\caption{$\scriptstyle{F_0 = a_0  e^{ - \mu \epsilon^2(x + x_T)^2 }}$}
\label{fig:b0-synp}
\end{subfigure}
\begin{subfigure}[b]{0.32\linewidth}
\centering
\includegraphics[scale=0.125]{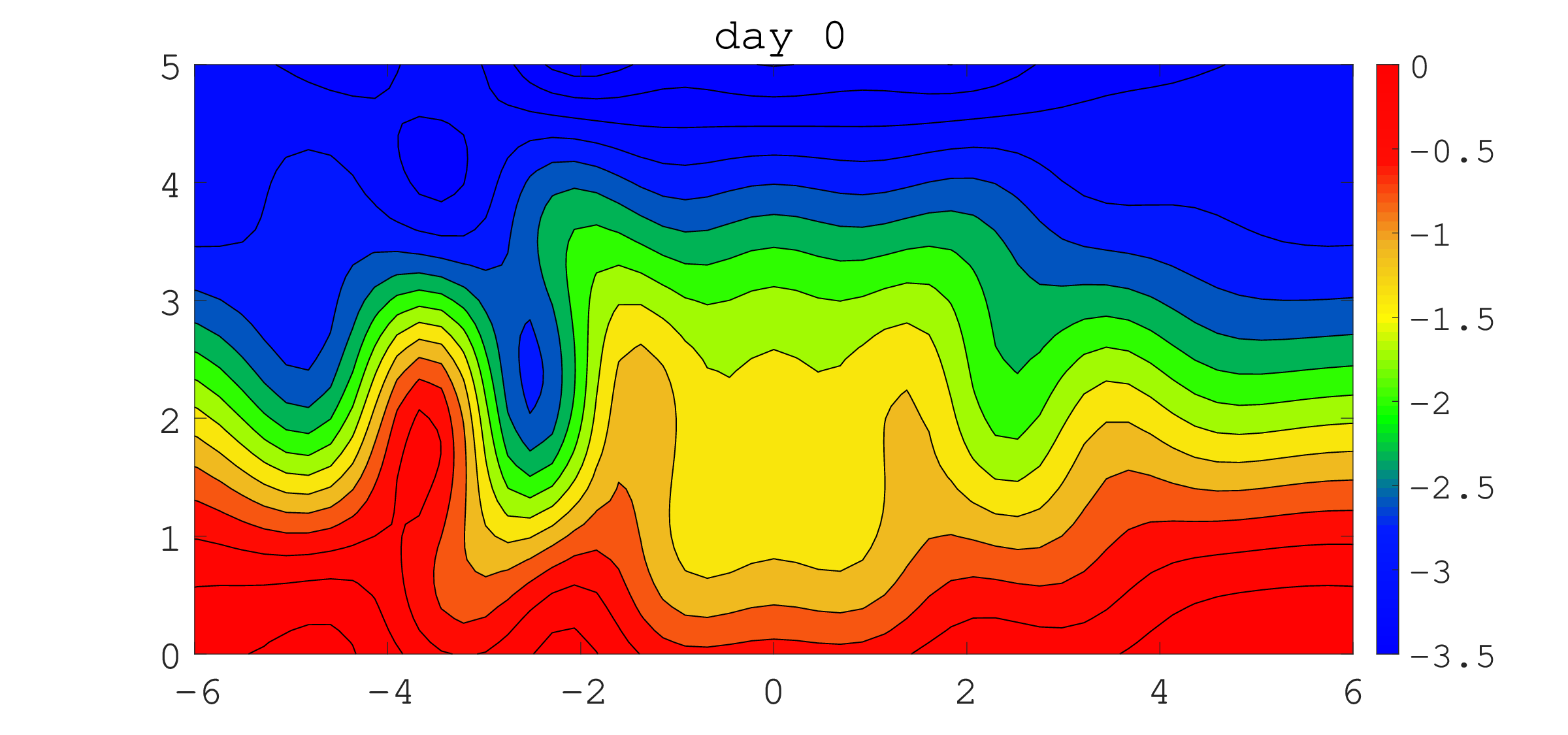}
\includegraphics[scale=0.125]{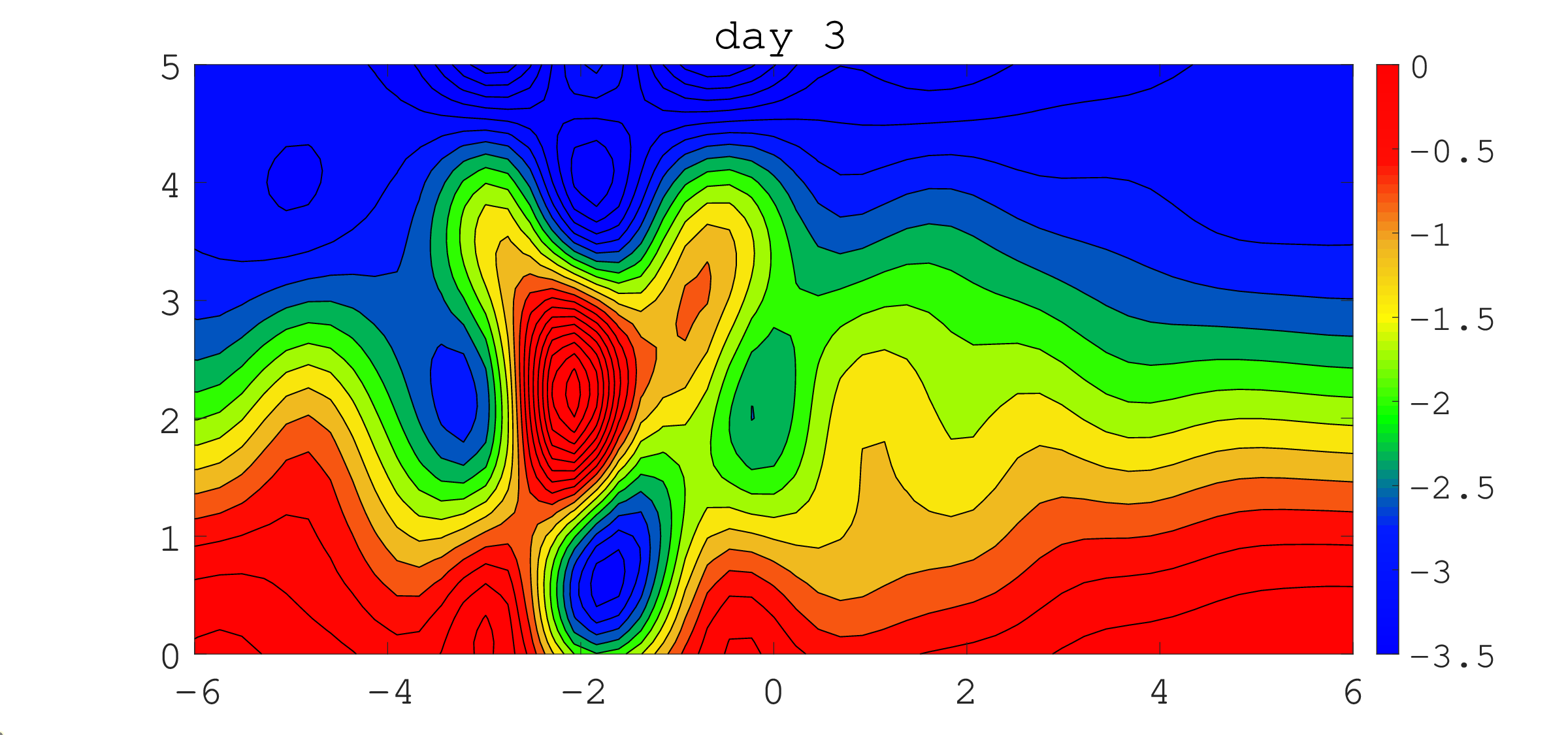}
\includegraphics[scale=0.125]{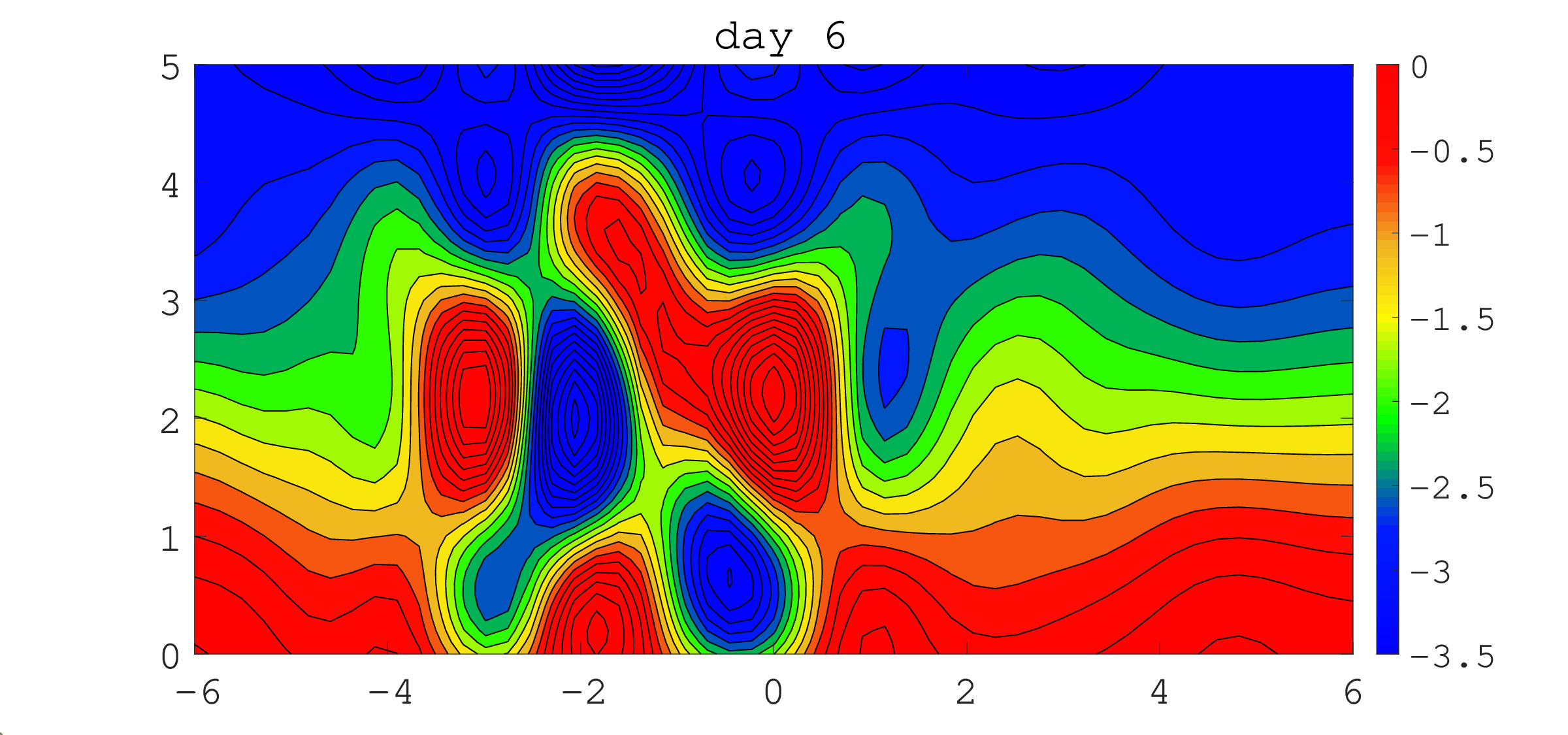}
\includegraphics[scale=0.125]{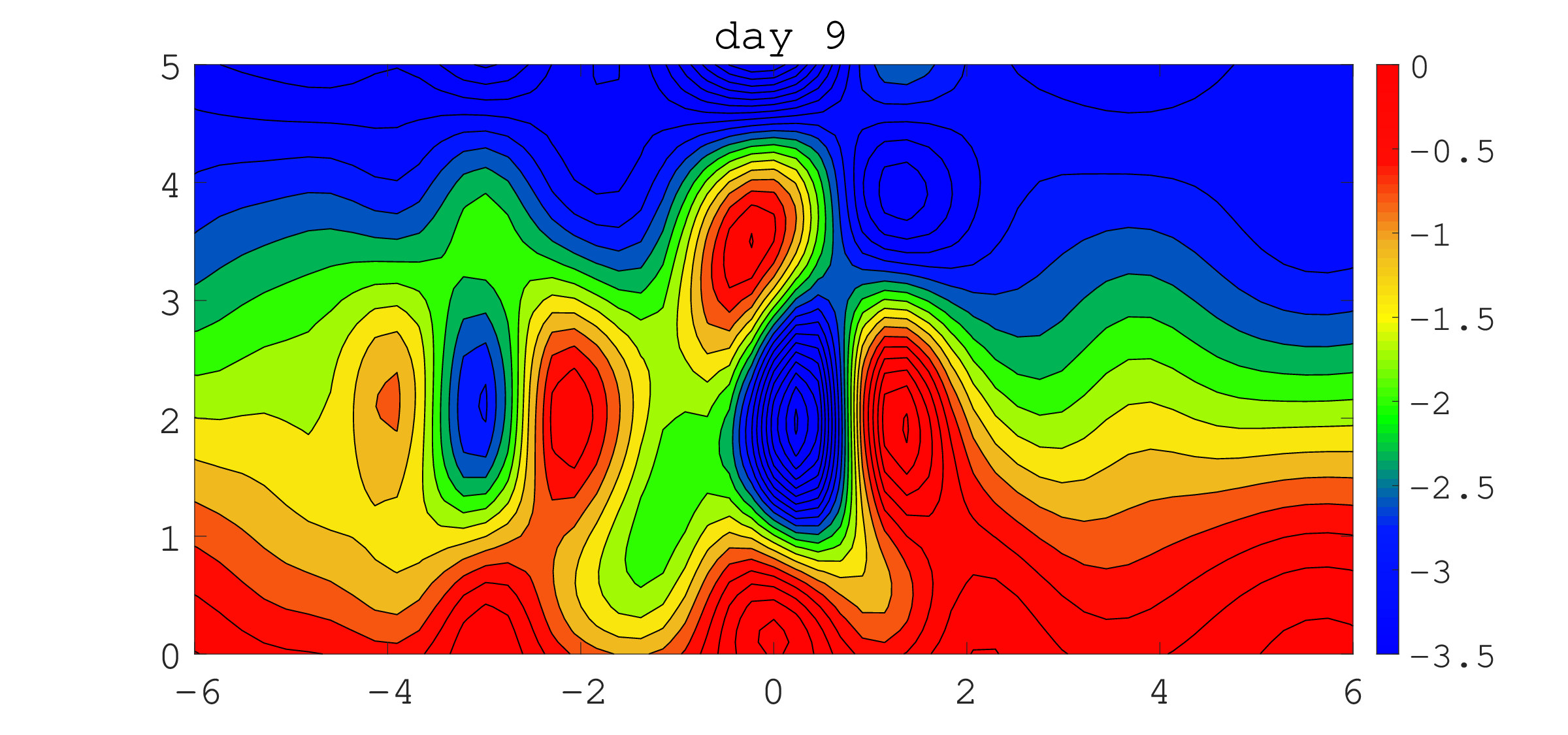}
\includegraphics[scale=0.125]{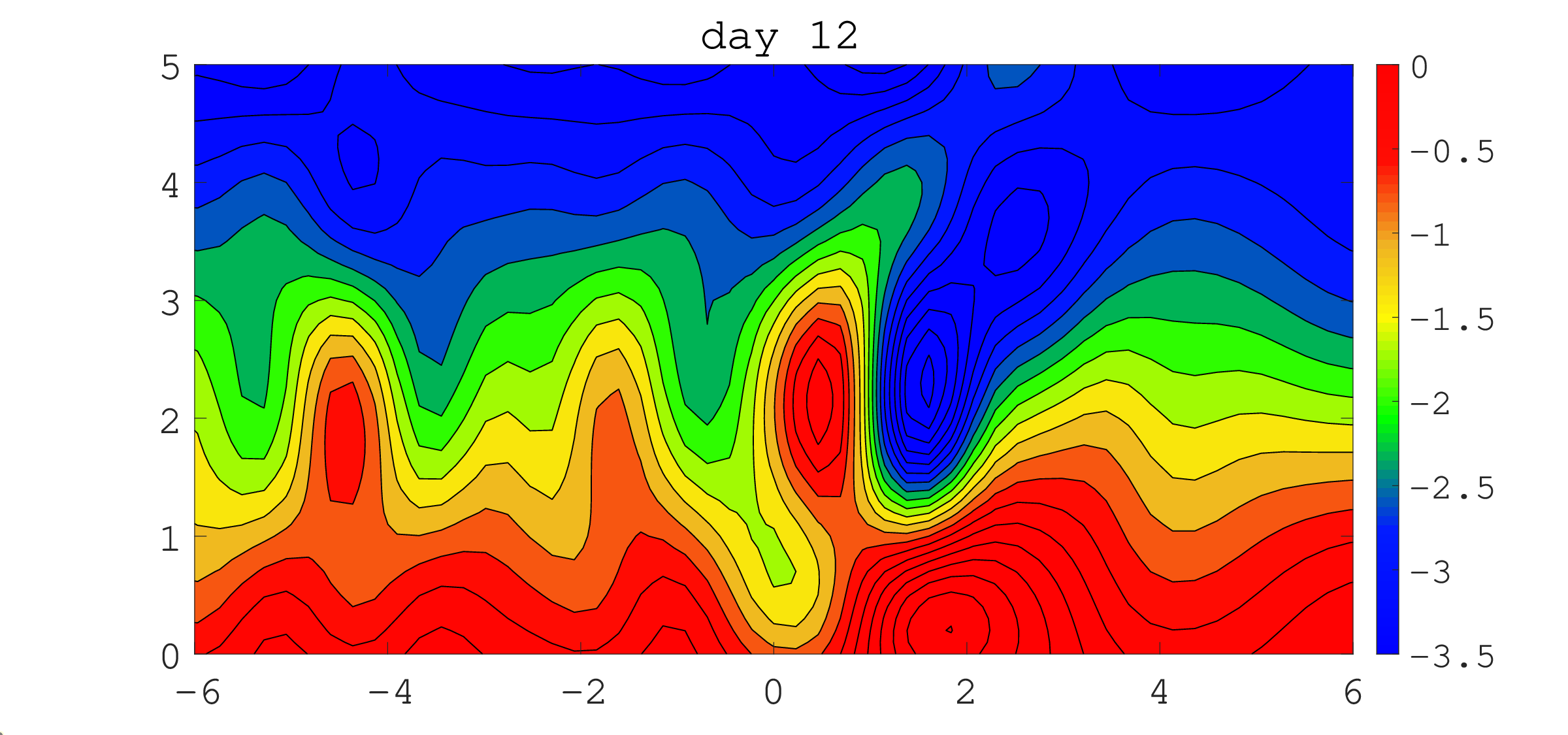}
\includegraphics[scale=0.125]{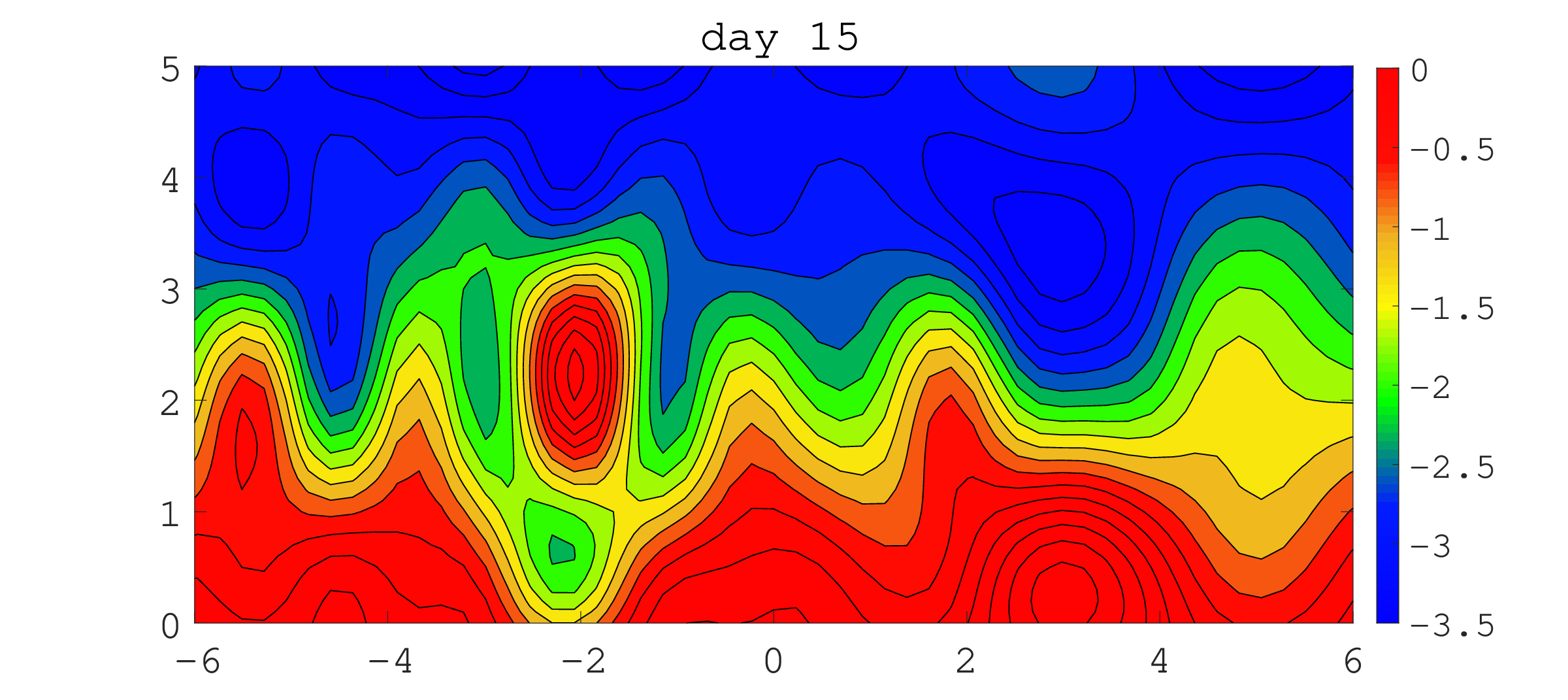}
\caption{$\scriptstyle{F_0 = a_0  e^{ - \mu \epsilon^2(x + x_T)^2 },\;\gamma=0.5}$}
\label{fig:b0-synp-gamma}
\end{subfigure}
\begin{subfigure}[b]{0.32\linewidth}
\centering
\includegraphics[scale=0.125]{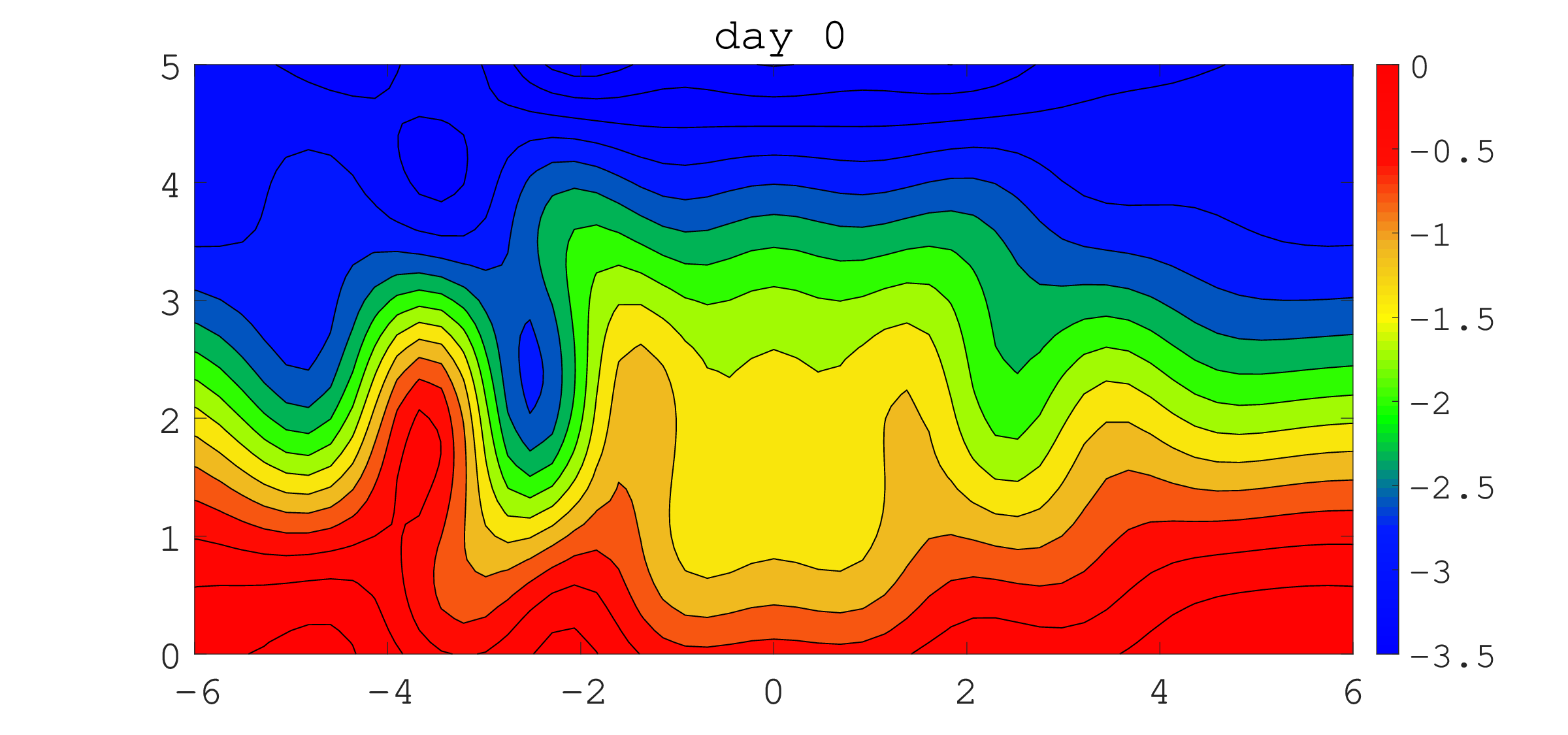}
\includegraphics[scale=0.125]{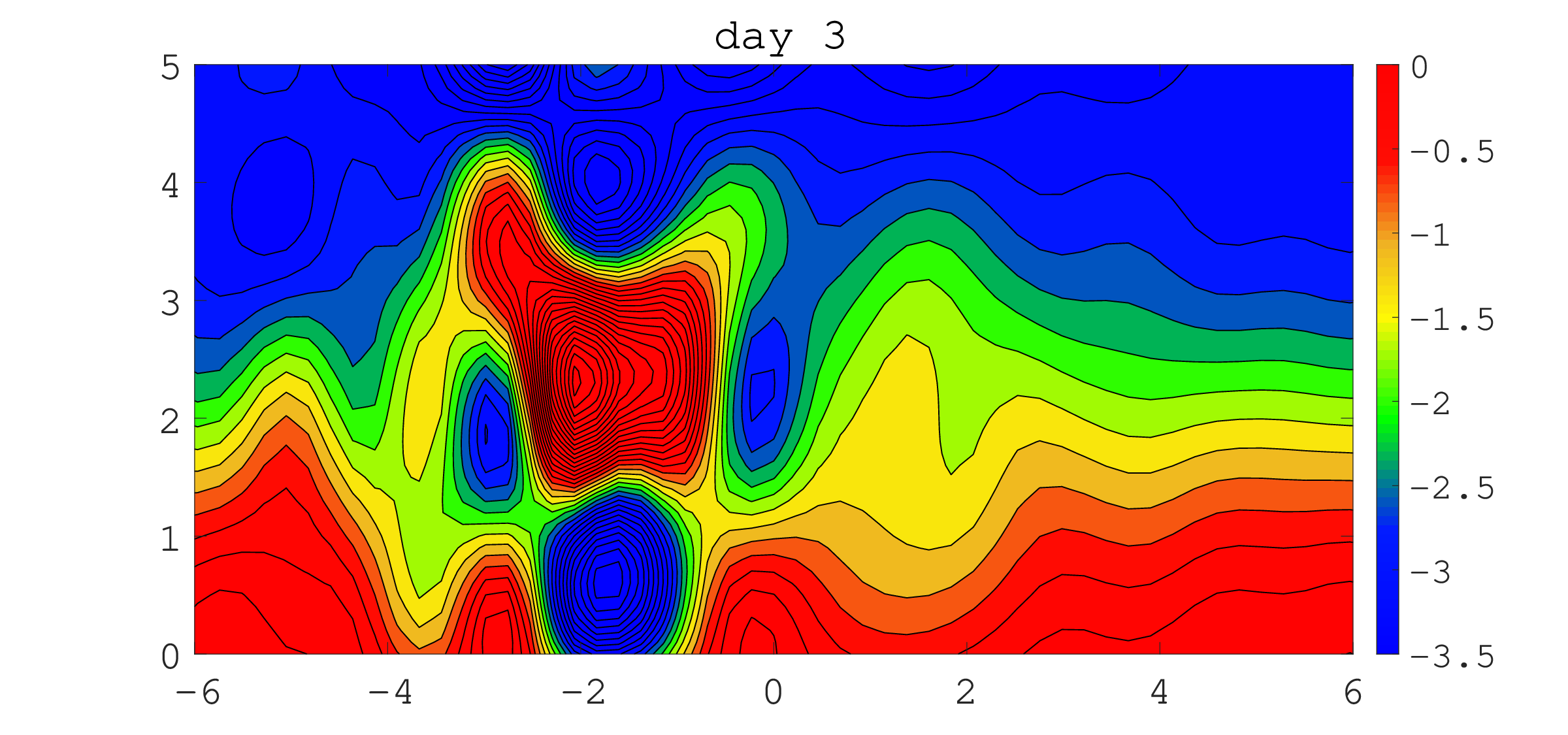}
\includegraphics[scale=0.125]{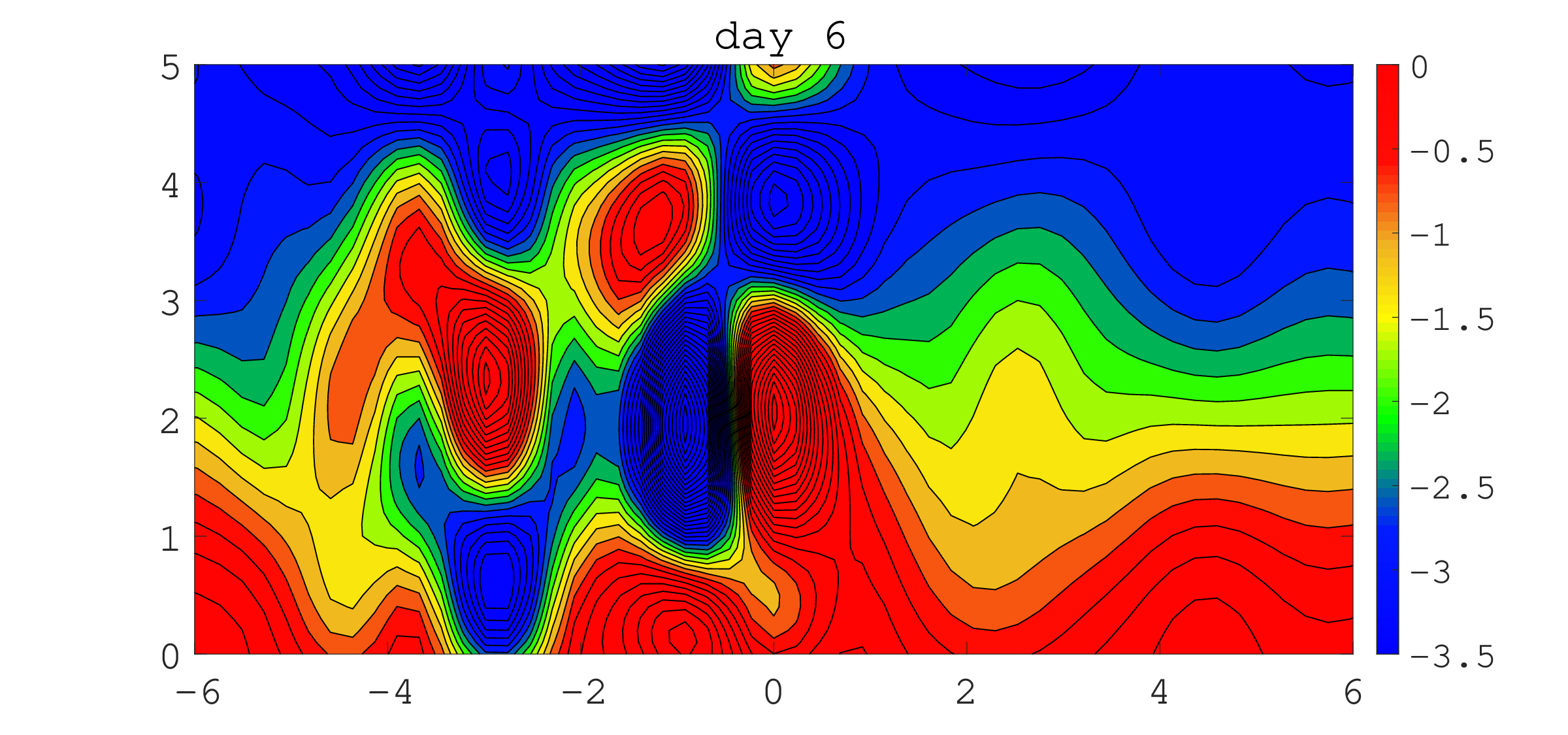}
\includegraphics[scale=0.125]{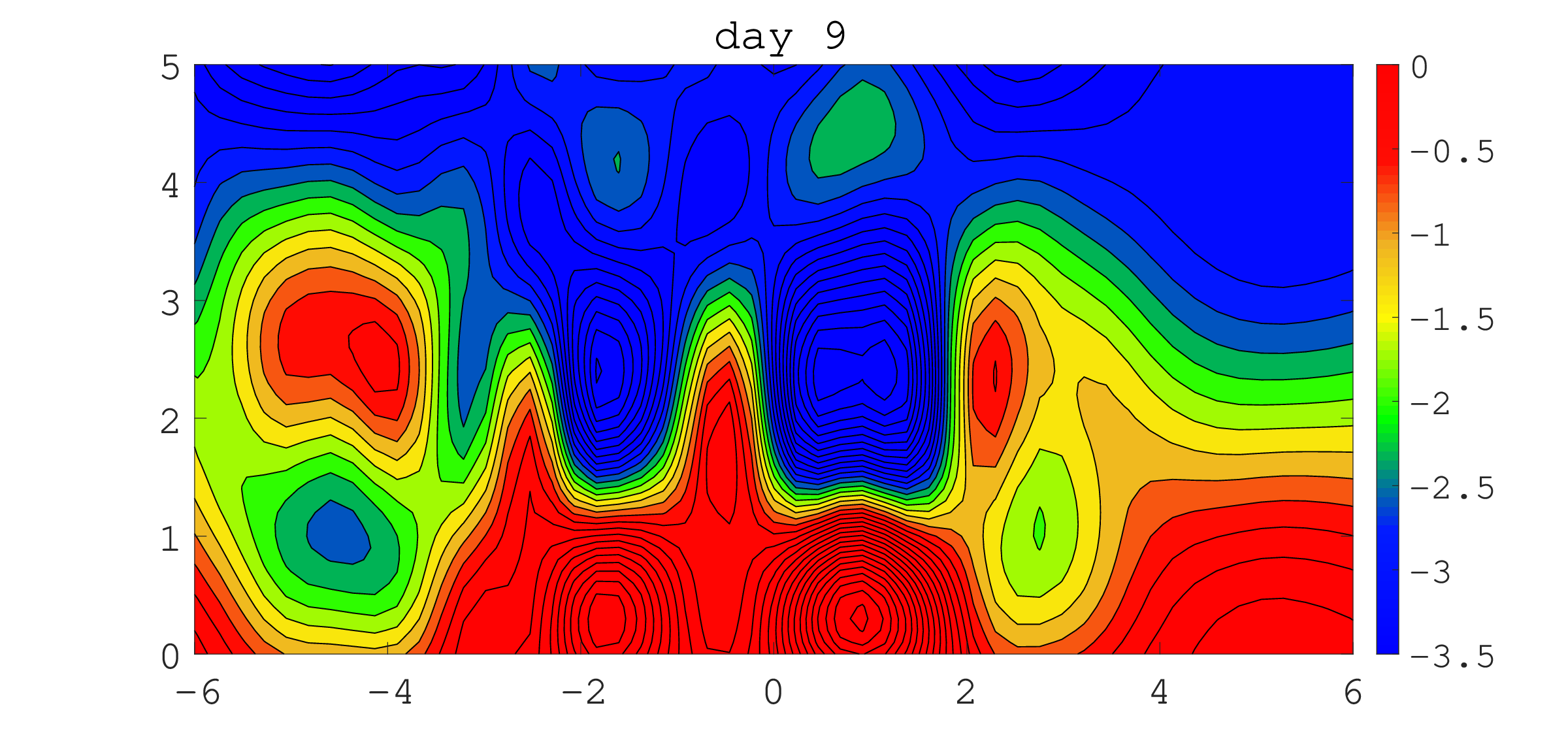}
\includegraphics[scale=0.125]{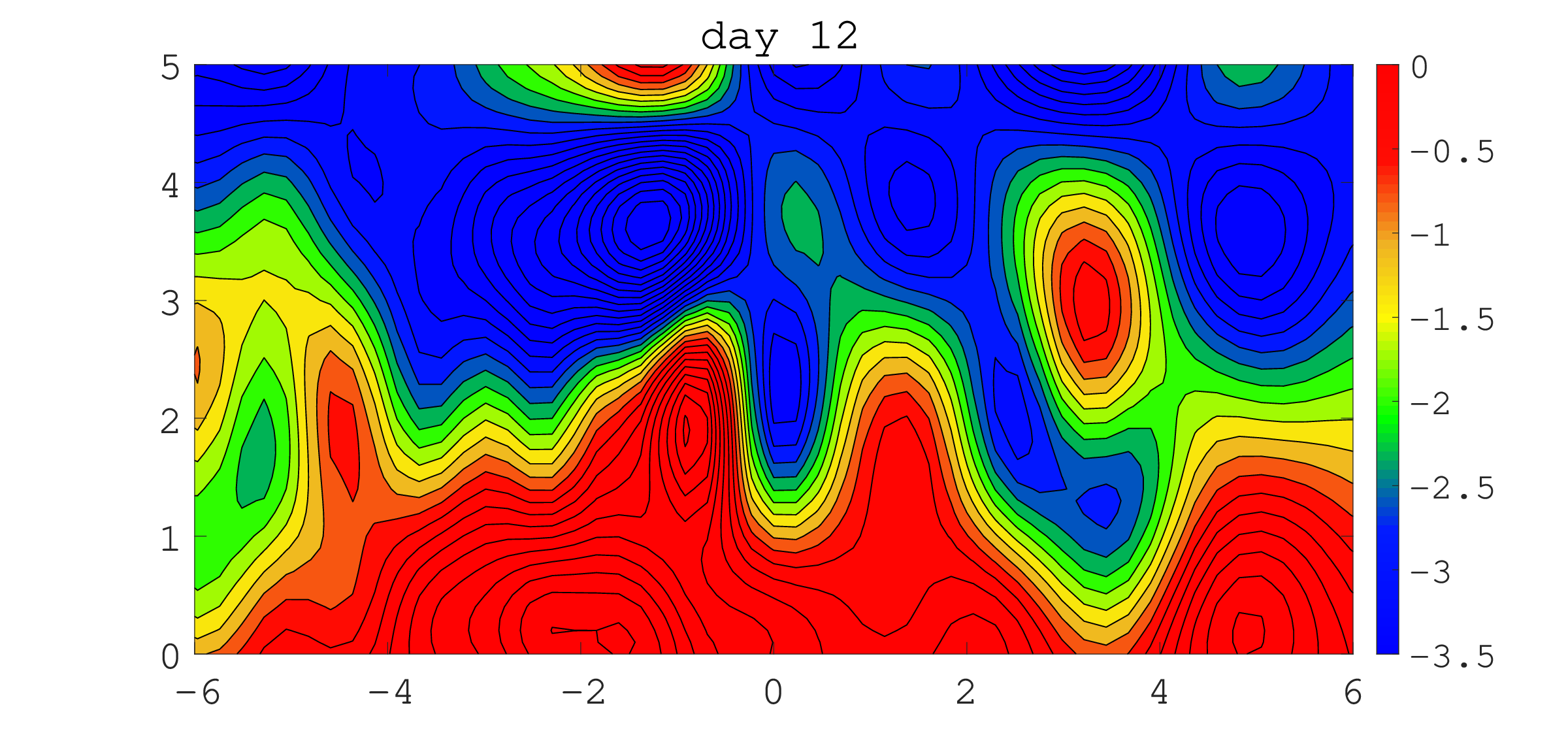}
\includegraphics[scale=0.125]{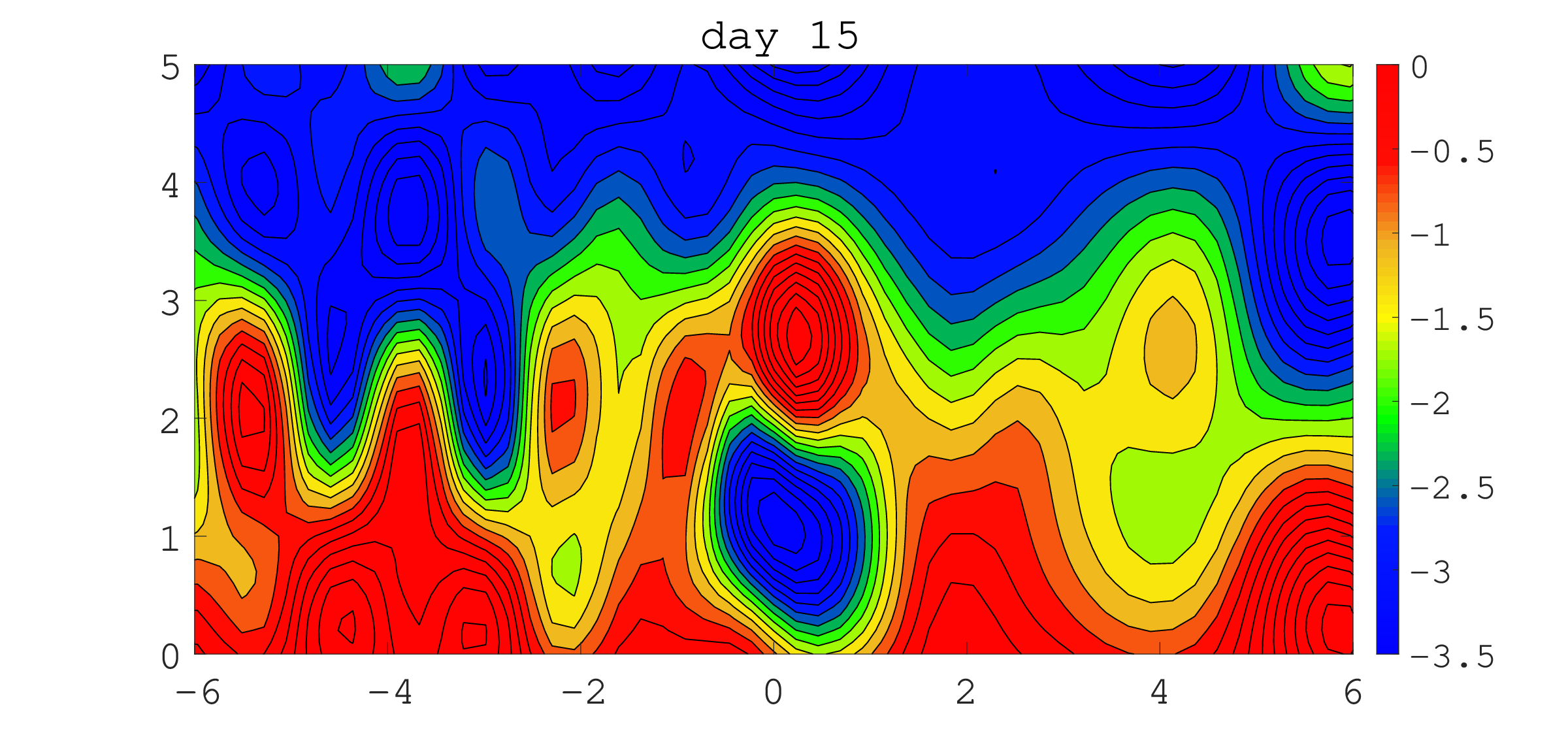}
\caption{$\scriptstyle{F_0 = a_0  e^{ - \mu \epsilon^2(x + x_T)^2 },\;\gamma=1}$}
\label{fig:b0-synp-gamma1}
\end{subfigure}
\caption{Nonlinear evolution of the instantaneous total streamfunction field $\psi_T$ when the preexisting synoptic-scale eddies are added by the optimal disturbance with the incremental increase of the size parameter $\gamma$.}
\label{fig:sf-synp}
\end{figure}

In~\Cref{fig:sf-synp}, we depict the evolution of the instantaneous total streamfunction $\psi_T$ when the preexisting synoptic-scale eddies is added by the optimal disturbances. This visualization offers a clear representation of how the total streamfunction changes as the size parameter $\gamma$ increases incrementally.~\Cref{fig:b0-synp} is the same as~\Cref{fig:b0-init}, which illustrates that the evolution of the instantaneous total streamfunction without any perturbations.
By comparing~\Cref{fig:sf-synp} with~\Cref{fig:sf-init}, it becomes evident that as the size parameter $\gamma$ increases incrementally, the phenomena related to blocking, eddy straining and wave breaking,  become more prominent in comparison to the motion of the blocking added by the optimal disturbance of the initial blocking amplitude.  Additionally, both the position and the period of the blocking exhibit significant changes and become chaotic. This observation indicates that the behavior of the blocking indeed becomes more unpredictable and less stable when perturbations occur in the preexisting synoptic-scale eddies. In other words, this highlights the sensitivity of the blocking to perturbations of the preexisting synoptic-scale eddies, which can potentially lead to weather extremes and pose challenges in accurately predicting them, as mentioned in~\citep{bengtsson1981numerical, tibaldi1990operational, burroughs_1997}. This further suggests that the error caused by the optimal disturbance of the preexisting synoptic-scale eddies can indeed contribute to the complex dynamics of the blocking. As the size of the disturbance increases, the complexity of the blocking can also increase. 
\subsection{Less predictability on the medium-range}
\label{subsec:  less-predictability-synp}

\begin{figure}[htb!]
\centering
\begin{subfigure}[t]{0.325\linewidth}
\centering
\includegraphics[scale=0.14]{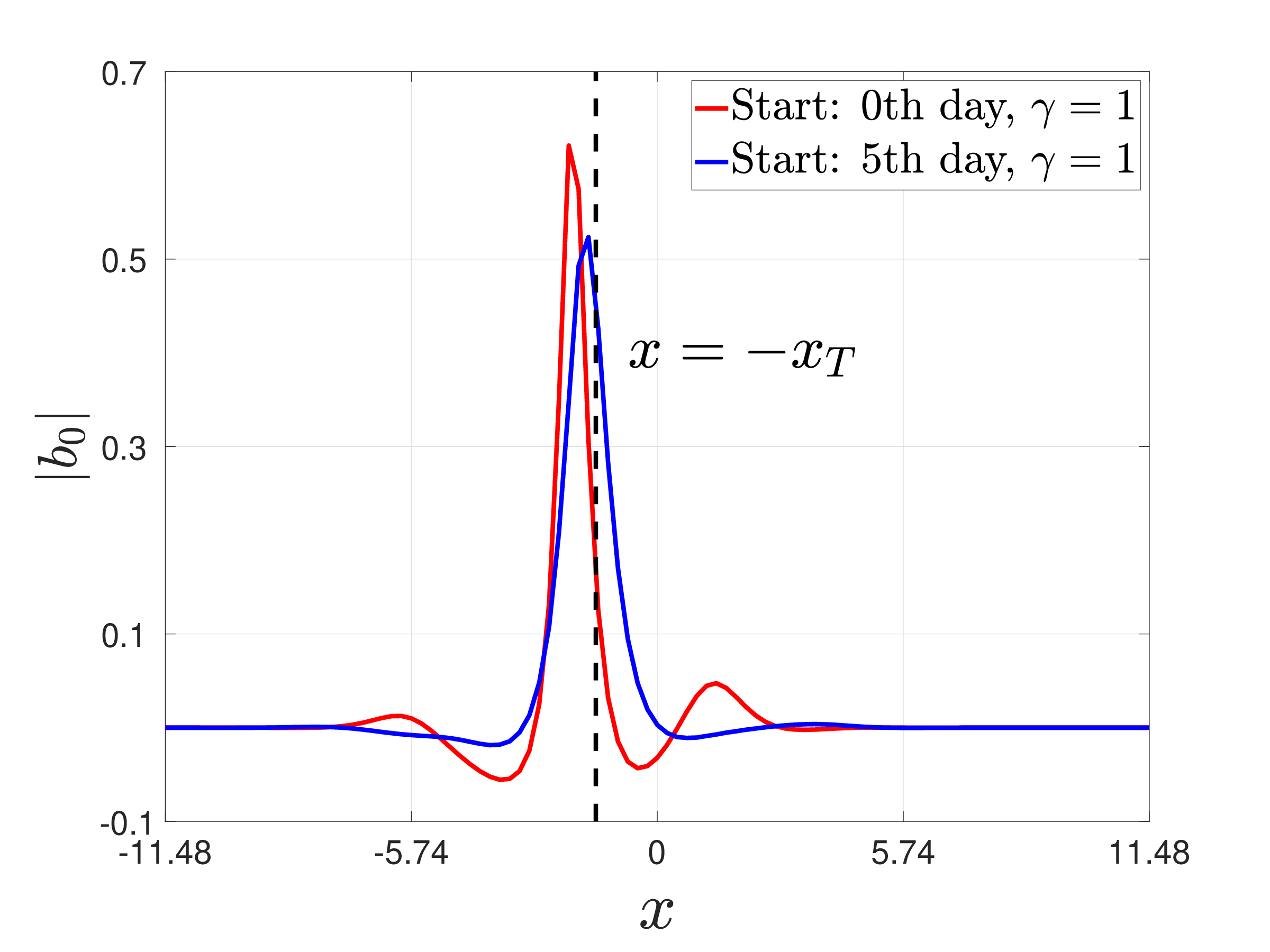}
\caption{Spatial Patterns}
\label{subfig: td-sp-cnop-synp}
\end{subfigure}
\begin{subfigure}[t]{0.325\linewidth}
\centering
\includegraphics[scale=0.14]{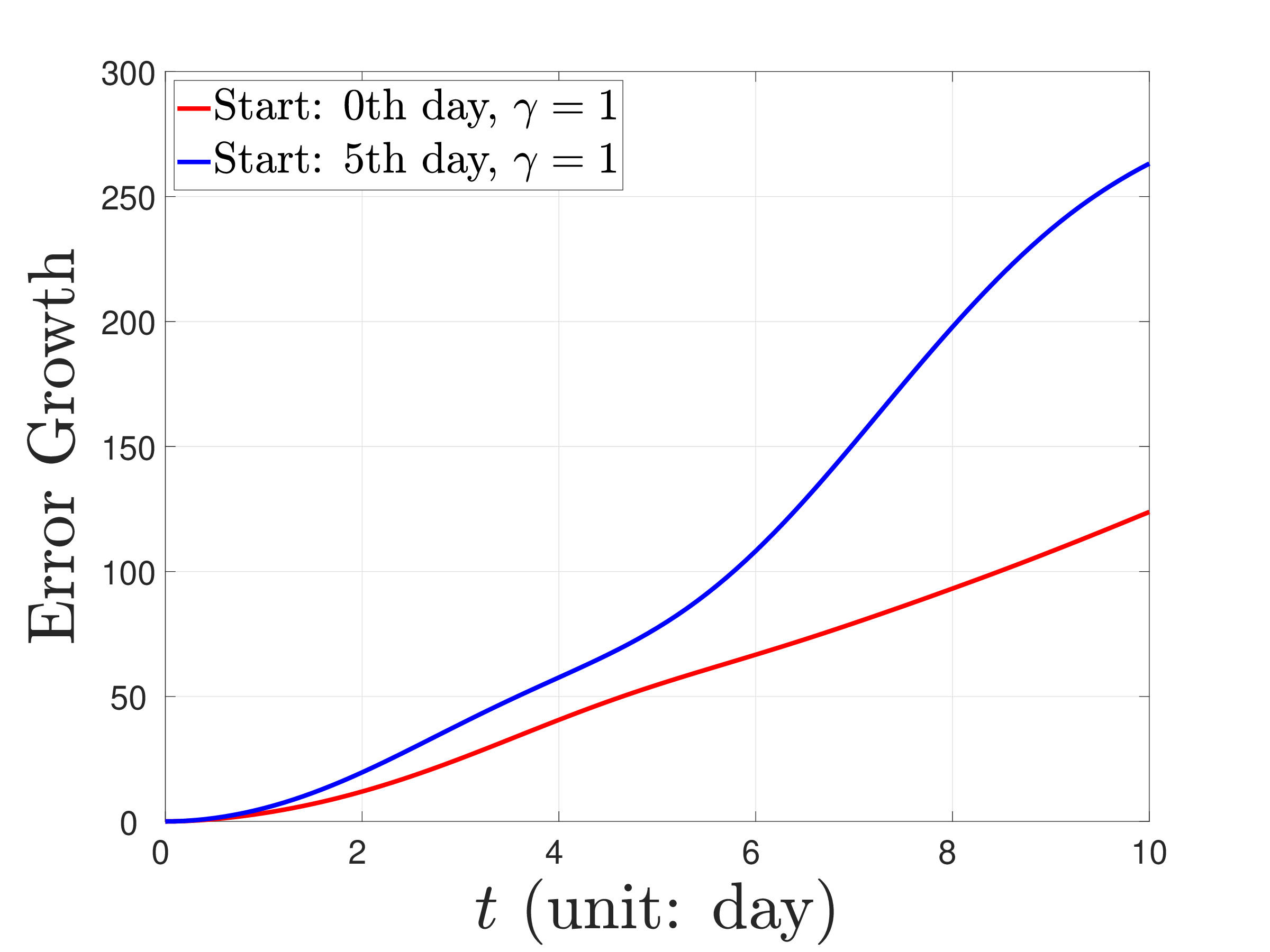}
\caption{(standard) Error Growth}
\label{subfig: tp-eg-cnop-synp}
\end{subfigure}
\begin{subfigure}[t]{0.325\linewidth}
\centering
\includegraphics[scale=0.14]{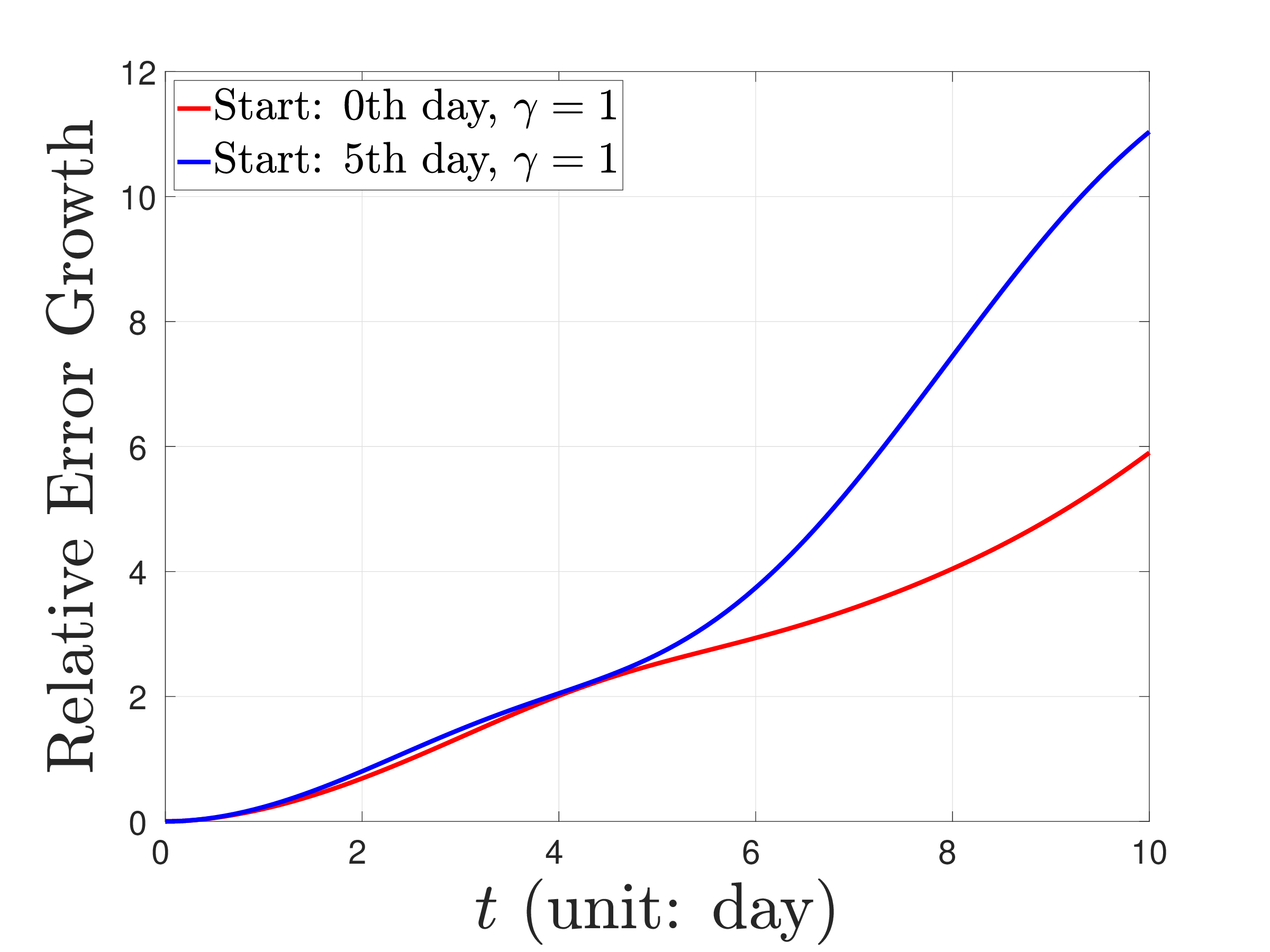}
\caption{Relative Error Growth}
\label{subfig: td-reg-cnop-synp}
\end{subfigure}
\caption{Spatial patterns and nonlinear growth of the error caused by the optimal disturbance at a later stage in comparison to the baseline at the initial stage. } 
\label{fig: td-cnop-synp}
\end{figure}

Similarly, it is also a valuable endeavor to conduct the numerical experiment to explore the potential impact of the optimal disturbance of preexisting synoptic-scale eddies on forecast errors. Specifically, we investigate whether there are larger forecast errors at a later stage by taking a comparison of optimal disturbances between different time ranges, such as an early stage from 0 to 10 days and a later stage from 5 to 15 days.  The spatial pattern of the optimal disturbance at a later stage, compared with an early stage, is shown in~\Cref{subfig: td-sp-cnop-synp}, where it is observed that the concentration distribution as a sharp peak slightly offsets to the right and flattens.
Additionally, the nonlinear growth and relative nonlinear growth of the error caused by the optimal disturbances are depicted in~\Cref{subfig: tp-eg-cnop-synp} and~\Cref{subfig: td-reg-cnop-synp}, as described by~\cref{eqn: error-synp} and~\cref{eqn: rel-error-synp}, respectively. Compared with the growth patterns of the optimal disturbance of the initial blocking amplitude shown in~\Cref{subfig: tp-eg-cnop-init} and~\Cref{subfig: td-reg-cnop-init}, it appears that the error caused by the optimal disturbance of the preexisting synoptic-scale eddies grows more predominantly during the decay period of the blocking. It is worth noting that when the time interval $T$ and the size parameter $\gamma=1$ are fixed, the optimal disturbance at a later stage leads to a larger error. This difference in error is further quantitatively highlighted by the ratios between the later stage and the initial stage, $2.1256$ for the nonlinear growth and $1.8706$ for the relative nonlinear growth. Additionally, the experiment reveals that the nonlinear evolution of the error caused by the optimal disturbance of the preexisting synoptic-scale eddies exhibits a sharp growth during the decay of the blocking. This finding about the optimal disturbance of the synoptic-scale eddies aligns with that of the initial blocking amplitude shown in~\Cref{subsec: less-predictability-init}, which leads to the less predictability of blockings on the medium-range, as mentioned in~\citep{hamill2014skill, ferranti2015flow, zhang2019predictability}.  This also support the idea that the presence of the optimal disturbance at a later stage, as observed in the numerical experiment, can contribute to larger forecast errors in predicting blockings.

\section{The impact of the background westerly wind}
\label{sec: role-westerly}

In this section, we take several numerical experiments to explore how the background westerly wind affects the optimal disturbances of both the initial blocking amplitude and the preexisting synoptic-scale eddies. Specifically, we investigate their spatial patterns and nonlinear growth of the error caused by the optimal disturbances, which provide insights into the influence of the background westerly wind on these weather phenomena.

\begin{figure}[htb!]
\centering
\begin{subfigure}[t]{0.325\linewidth}
\centering
\includegraphics[scale=0.14]{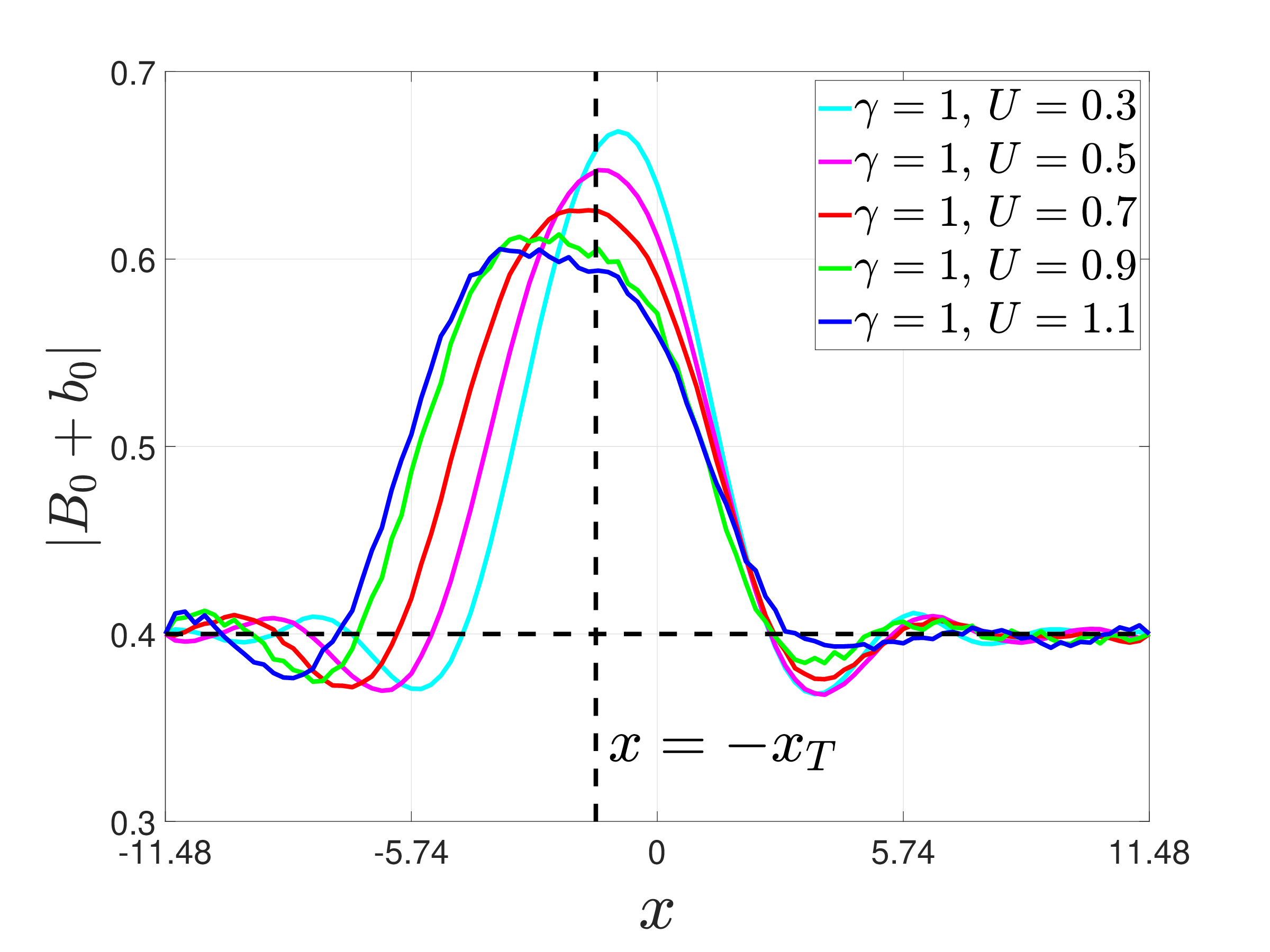}
\caption{Spatial Patterns}
\label{subfig: sp-ww-init}
\end{subfigure}
\begin{subfigure}[t]{0.325\linewidth}
\centering
\includegraphics[scale=0.14]{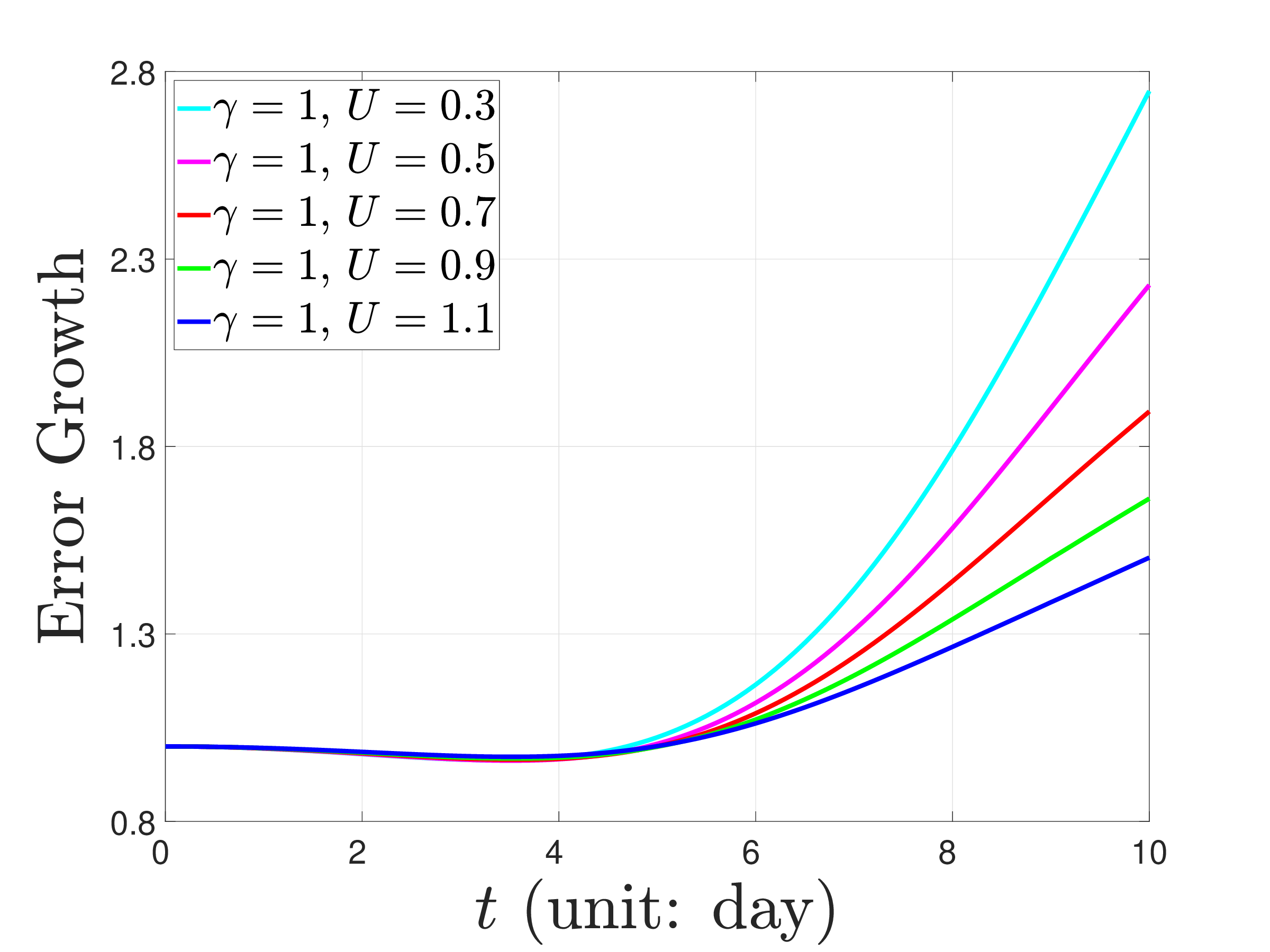}
\caption{(standard) Error Growth}
\label{subfig: eg-ww-init}
\end{subfigure}
\begin{subfigure}[t]{0.325\linewidth}
\centering
\includegraphics[scale=0.14]{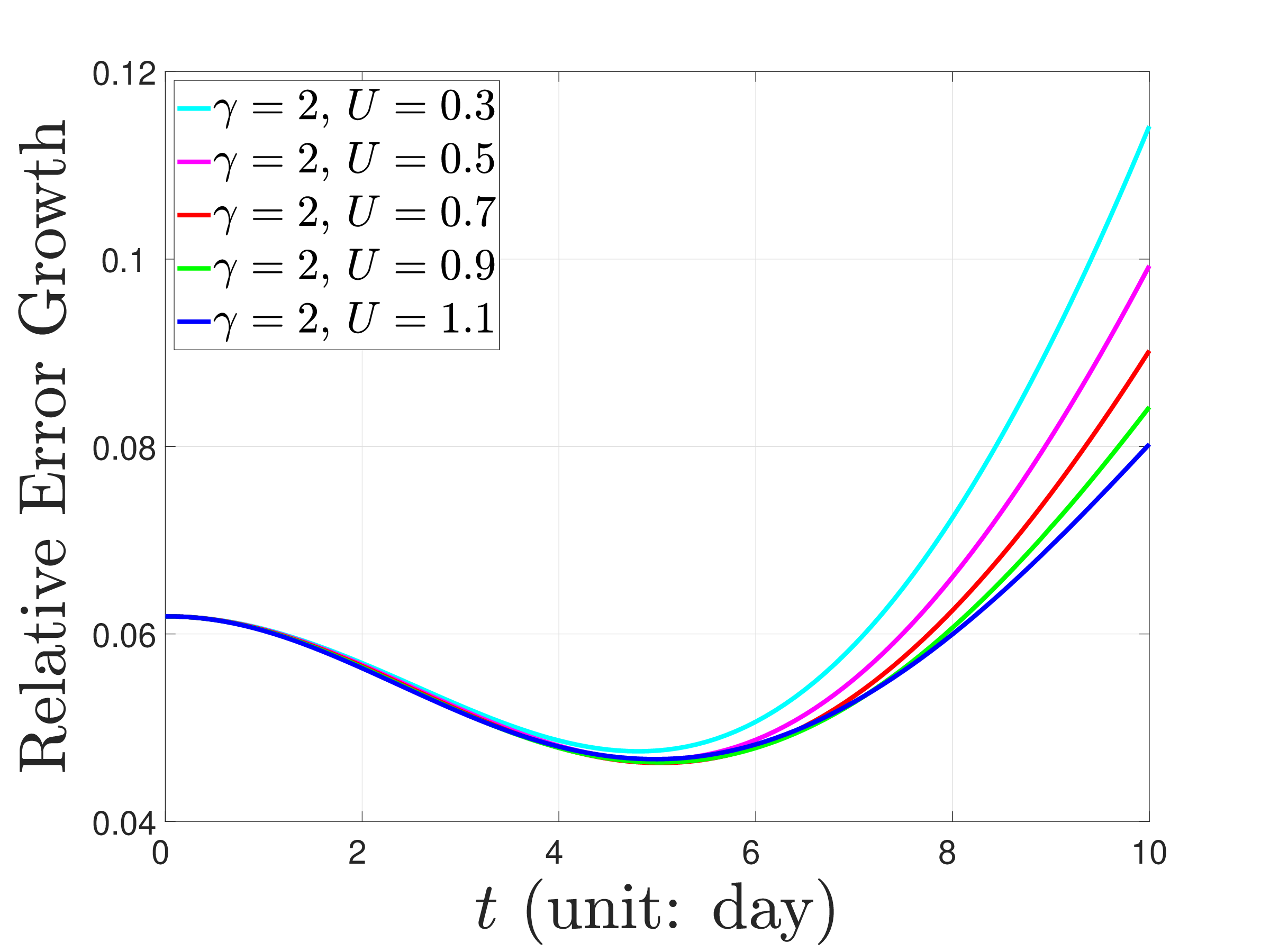}
\caption{Relative Error Growth}
\label{subfig: reg-ww-init}
\end{subfigure}
\caption{Spatial patterns and nonlinear growth of the optimal disturbance of the initial blocking amplitude varies with the changes of the background westerly winds.} 
\label{fig: westly-wind-init}
\end{figure}

\paragraph{The optimal disturbance of the initial blocking amplitude} In our numerical experiment, we fix the size parameter of the optimal disturbance as $\gamma=1$. Decreasing the westerly wind speed $U$ from $1.1$ to $0.3$ with a decrement of $0.2$ allows us to explore how the variation of wind speed affects the optimal disturbance of the initial blocking amplitude. The spatial patterns, as shown in~\Cref{subfig: sp-ww-init}, indicate that the solitary wave-like pattern becomes more concentrated, sharper, and gradually shifts to the right as the wind speed gradually dwindles. Additionally, the growth patterns depicted in~\Cref{subfig: eg-ww-init} and~\Cref{subfig: reg-ww-init} reveal that the nonlinear growth and relative nonlinear growth become gradually larger as the wind speed gradually dwindles.
These observations suggest that the westerly wind speed $U$ plays a role shaping the spatial and growth patterns. Furthermore, we quantitatively show the nonlinear growth~\eqref{eqn: error-init} and relative nonlinear growth~\eqref{eqn: rel-error-init} at the prediction time $T=10$ in~\Cref{tab: wester-wind-init-tab}. Indeed, it verifies that the nonlinear growth and relative nonlinear growth become gradually larger as the wind speed gradually dwindles.
\begin{table}[htb!]
\centering
\begin{tabular}{c|ccccc}
   \toprule
                                            & $U=0.3$               & $U=0.5$                  & $U=0.7$                    & $U=0.9$                  & $U=1.1$  \\
   \midrule
     $\frac{\|b(10)\|^2}{\Delta x}$         & $2.7481$              & $2.2307$                 & $1.8935$                   & $1.6610$                 & $1.5036$ \\
   \midrule
     $\frac{\|b(10)\|^2}{\|B(10)\|^2}$      & $0.1142$              & $0.0993$                 & $0.0902$                   & $0.0842$                 & $0.0802$ \\    
    \bottomrule
\end{tabular}
\caption{The growth patterns of the optimal disturbance of the initial blocking amplitude $B_0=0.4$.}
\label{tab: wester-wind-init-tab}
\end{table}

The influence of the westerly wind on the nonlinear growth behavior, as depicted in~\Cref{fig: westly-wind-init} and~\Cref{tab: wester-wind-init-tab}, aligns with the $PV_y$ theory for the NMI model proposed in~\citep{luo2019nonlinear}. The relationship between the potential vorticity and the westerly wind, as described in~\citep{pedlosky1987geophysical}, can be expressed as $PV=f_0 + \beta y - U_y - F\overline{\psi}$. When considering a uniform westerly wind, the meridional gradient of potential vorticity has a linear relation with the westerly wind, given by $PV_y = \beta + FU$. Based on the forced NLS equation~\eqref{eqn: ampli-nls} and the conditions of coefficients~\eqref{eqn: pvy-lambda-delta}, we can deduce that the coefficient of the dispersive term is proportional to $PV_y$, i.e., $\lambda \propto PV_y = \beta + FU$, and the coefficient of the nonlinear term is inversely proportional to $PV_y$, i.e., $\delta \propto 1/PV_y = 1/(\beta + FU)$. Therefore, when the westerly wind weakens, the meridional gradient of potential vorticity decreases, resulting in the suppression of the dispersive effect and the intensification of the nonlinear effect. This results in an increase in the nonlinear growth. Conversely, when the westerly wind strengthens, the meridional gradient of potential vorticity increases, resulting in the intensification of the dispersive effect and the suppression of the nonlinear effect, leading to a decrease in the nonlinear growth. Furthermore, as the coefficient of the nonlinear term is inversely proportional to $PV_y$,  when the westerly wind speed gradually dwindles, the rate of increase in $PV_y$ becomes fast, causing the nonlinear growth rate to become large.

\paragraph{The optimal disturbance of the preexisting synoptic-scale eddies}
Similarly, for the numerical experiment related to the preexisting synoptic-scale eddies, the size parameter of the optimal disturbance is also set as $\gamma=1$. Then, we explore how the variation of the wind speed affects the optimal disturbance of the preexisting synoptic-scale eddies by decreasing the westerly wind speed $U$ from $1.1$ to $0.3$ with a decrement of $0.2$. The spatial patterns, as shown in~\Cref{subfig: sp-ww-init}, demonstrate how the spatial patterns change as the wind speed varies.  
\begin{figure}[htb!]
\begin{subfigure}[t]{0.325\linewidth}
\centering
\includegraphics[scale=0.14]{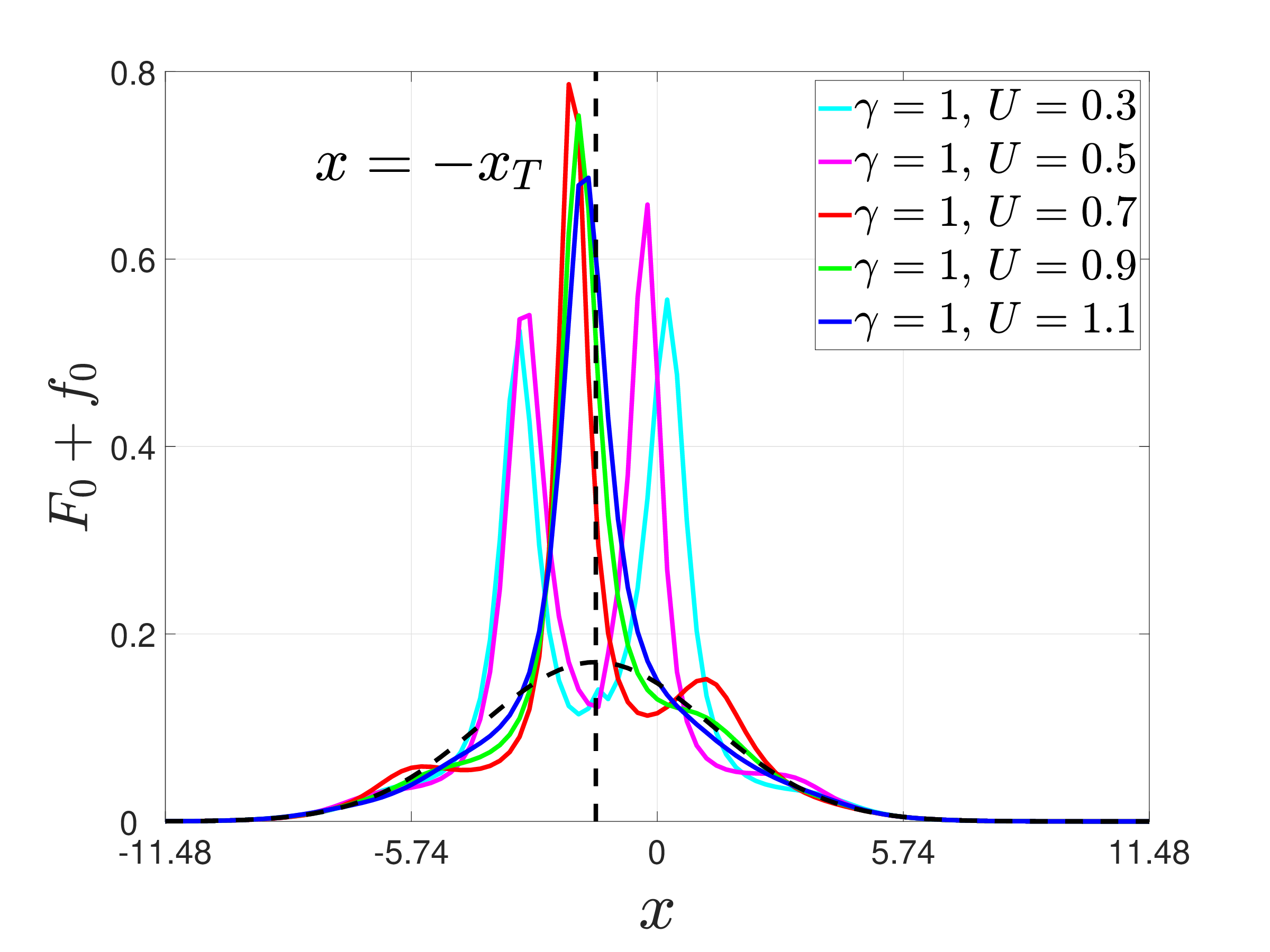}
\caption{Spatial Patterns}
\label{subfig: sp-ww-synp}
\end{subfigure}
\begin{subfigure}[t]{0.325\linewidth}
\centering
\includegraphics[scale=0.14]{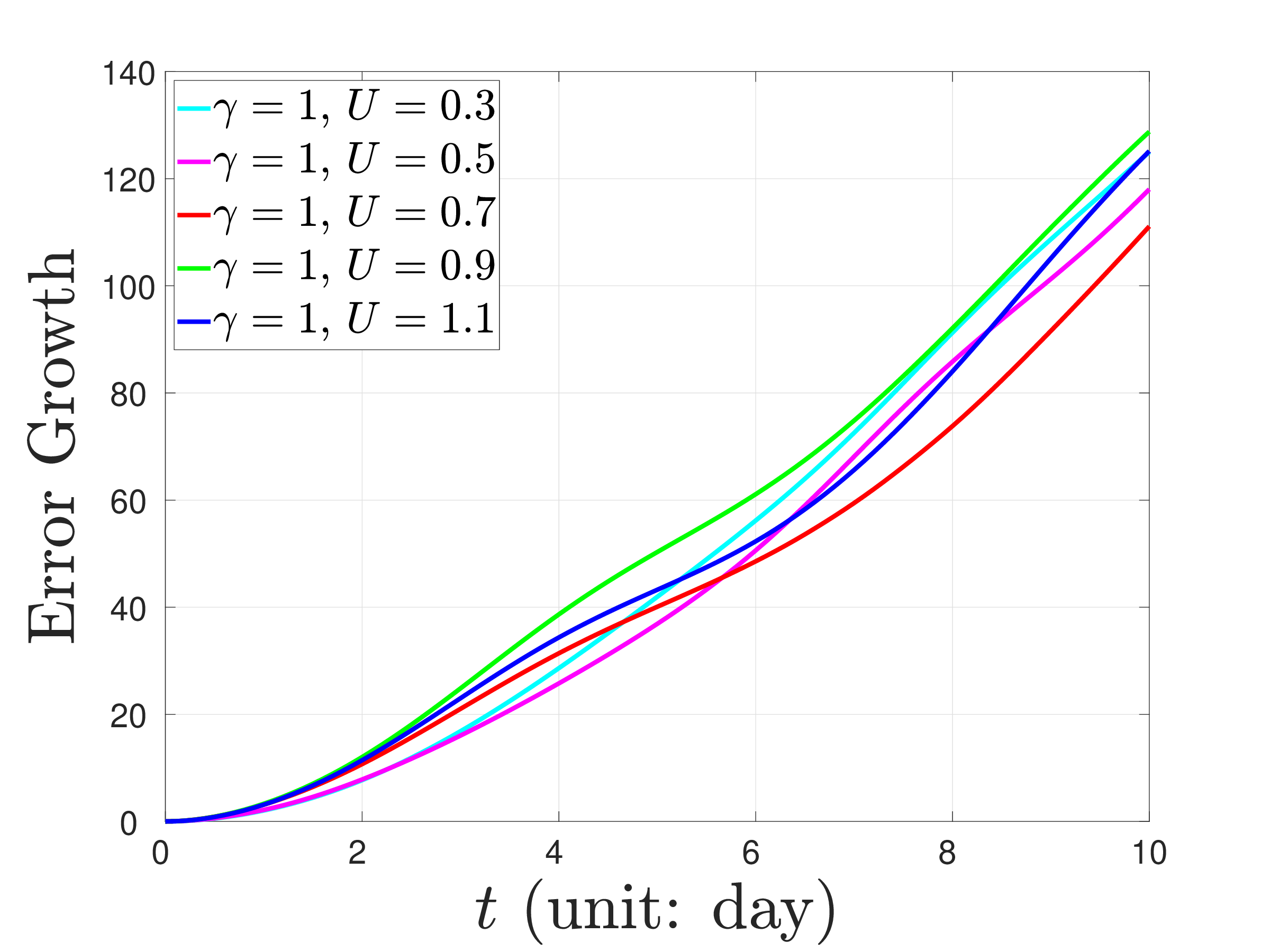}
\caption{(standard) Error Growth}
\label{subfig: eg-ww-synp}
\end{subfigure}
\begin{subfigure}[t]{0.325\linewidth}
\centering
\includegraphics[scale=0.14]{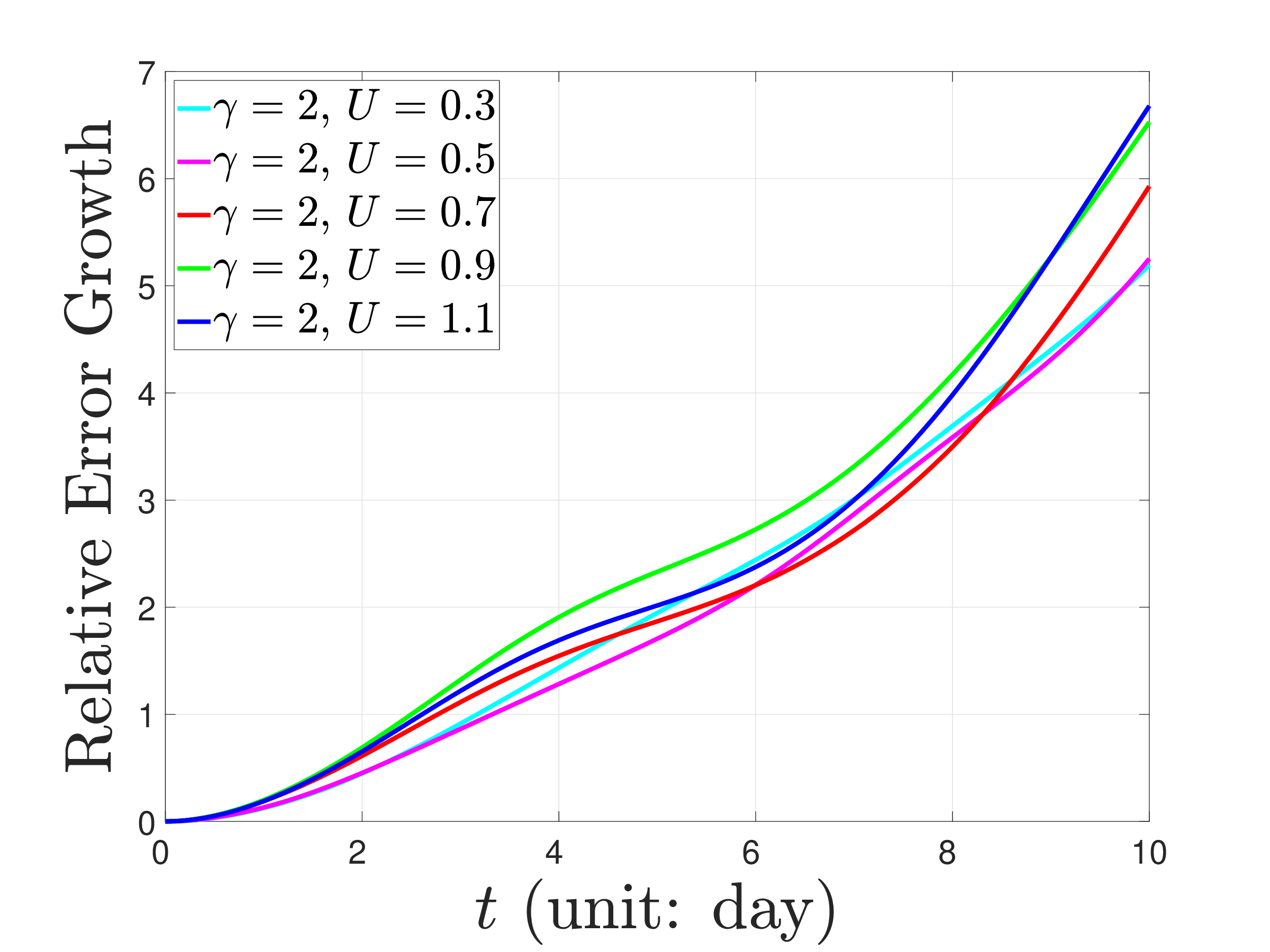}
\caption{Relative Error Growth}
\label{subfig: reg-ww-synp}
\end{subfigure}
\caption{Spatial patterns and nonlinear growth of the error caused by the optimal disturbance of the preexisting synoptic-scale eddies vary with the changes of the background westerly winds.} 
\label{fig: westerly-wind-synp}
\end{figure}
It is intriguing to observe the different behaviors of the peak-like pattern based on wind speed variations. When the wind speed is smaller than the standard speed $U=0.7$, the sharp peak-like pattern separates into two lower peaks on both sides. As the wind speed decreases further, the two peak-like pattern moves outside and becomes lower.  On the other hand, when the wind speed is larger than the standard speed $U=0.7$, the sharp peak-like pattern descends and shifts slightly to the right. As the wind speed increases further, the sharp peak-like pattern continues to descend and shift slightly to the right. However, it is worth noting that the growth behaviors of the error caused by the optimal disturbance of the preexisting synoptic-scale eddies, regardless of the nonlinear growth or the relative nonlinear growth, are rarely influenced by the variation of the wind speed, as shown in~\Cref{subfig: eg-ww-synp} and~\Cref{subfig: reg-ww-synp}.

\section{Summary and Discussion}
\label{sec: discussion}

Taking the barotropic NMI model developed in~\citep{luo2000planetary, luo2005barotropic, luo2014nonlinear, luo2019nonlinear} as a basis, we have specifically focused on exploring optimal disturbances of blocking by utilizing the CNOP approach in this paper. In the NMI model, the motion of blocking is governed by the forced NLS equation~\eqref{eqn: ampli-nls}, which provides a framework for studying the optimal disturbances of the initial blocking amplitude and preexisting synoptic-scale eddies. Our analysis of the optimal disturbances includes examining their spatial patterns, nonlinear growth patterns of the error caused by them,  their influence on the motion of the total blocking, and their time-delay effect. It is observed that the optimal disturbance of the initial blocking amplitude has a well-behaved impact solely on the blocking amplitude without any other influence.  Increasing the size of the optimal disturbance indeed accelerates the nonlinear growth of the error.   However, a striking phenomenon is observed in the optimal disturbance of the preexisting synoptic-scale eddies, which leads to a significant increase in error growth. As the size of the disturbance increases, the phenomena related to blocking, eddy straining and wave breaking, become highly noticeable. Specifically, this results in significant changes in both the position and period of the blocking, leading to chaotic behavior. This finding manifests that the blocking is extremely sensitive to perturbations of the fast-moving short-lived (high-frequency) synoptic-scale eddies in the blocking system, as mentioned in~\citep{bengtsson1981numerical, tibaldi1990operational, burroughs_1997}. Furthermore, it highlights the role played by the fast-moving short-lived (high-frequency) synoptic-scale eddies in the blocking system, as mentioned in~\citep{berggren1949aerological, shutts1983propagation, hoskins1985use, luo2014nonlinear, luo2019nonlinear}.  The perturbations of these eddies may be a probable cause of weather extremes and can reduce predictability. Additionally, both the optimal disturbances occurring at a later stage, particularly during the decay period of blocking, also contribute to accelerating the nonlinear growth of the error. This has implications for the predictability of blockings on the medium-range,  which aligns with the practical weather prediction, as mentioned in~\citep{hamill2014skill, ferranti2015flow, zhang2019predictability}.  Finally, we have analyzed the influences of the westerly wind on the optimal disturbances. Regarding the initial blocking amplitude, the nonlinear evolution behavior indicates that the influence of the westerly wind on the optimal disturbance aligns with the $PV_y$ theory proposed in~\citep{luo2019nonlinear}. However, when considering the preexisting synoptic-scale eddies, it is observed that the westerly wind has no impact on them. 

In this paper, our main focus is on studying the optimal disturbances of the initial blocking amplitude and preexisting synoptic-scale eddies, as well as the influence of the westerly wind. This study utilizes the 1-dimensional forced NLS equation that specifically considers the zonal direction. However, it is acknowledged that the meridional shear of the westerly wind, represented as $PV_y = \beta + FU - U_{yy}$, also plays a significant role in the meridional gradient of potential vorticity. Previous studies, such as~\citep{thorncroft1993two}, have observed that the meridional shear of the background westerly wind can break up synoptic-scale anticyclones or cyclones. The $PV_y$ theory proposed in~\citep{luo2019nonlinear}, it further suggests that it affects the dispersive and nonlinear effects.  When there is positive shear $U_{yy}>0$, the dispersive effect is suppressed and the nonlinear effect is intensified. On the other hand, there is the negative shear $U_{yy}<0$, the dispersive effect is intensified, and the nonlinear effect is suppressed. Therefore, it would be valuable to conduct further research to explore the optimal disturbances in the 2-dimensional NMI model. Additionally, the 3-dimensional baroclinic NMI model, developed in~\citep{luo2020nonlinear1, luo2020nonlinear2, luo2021nonlinear}, considers the non-homogeneous vertical structure. Hence, it would indeed be intriguing to further explore how the optimal disturbances change and their influence on the growth of errors by taking into account the effect of the horizontal temperature gradients.

There are indeed other theories and models related to blocking, such as the local wave activity proposed in~\citep{huang2016local}, the traffic jam theory in~\citep{nakamura2018atmospheric}, and the amplified Rossby wave theory in~\citep{kornhuber2020amplified}. It would be intriguing to investigate how different types of perturbations in these models influence the growth of errors. Exploring the impact of these perturbations can provide valuable insights into the behavior and dynamics of blocking systems. Furthermore,~\citet{mu2008method} started utilizing the CNOP approach in the T21L3 quasi-geostrophic model~\citep{vannitsem1997lyapunov} to investigate the initial perturbations that trigger blocking onset. It would be worth exploring the influence of the perturbations by employing the CNOP approach in the real numerical weather prediction model, as suggested in~\citep{zhang2019predictability}. Additionally, the planetary solitary waves are also large-scale important phenomena occurring in the atmosphere and ocean with diameters from a hundred kilometers to scale larger than the earth, such as vortices embedded in a shear zone, Rossby solitons, and equatorial Kelvin solitary waves among others~\citep{rizzoli1982planetary, boyd2007planetary}. It would also be valuable to investigate the influence of the perturbations on their motion by employing the CNOP approach. 

In the conclusion of this paper, we briefly discuss the perturbations of blocking in response to climate change, which is currently a hot topic, as suggested in~\citep{woollings2018blocking,kautz2022atmospheric}.  It is worth noting that numerical climate models have always faced challenges in accurately representing blocking events, since these models tend to underestimate both the occurrence and persistence of blocking events, as suggested in~\citep{tibaldi1990operational, d1998northern, davini2016northern}. It has also been observed that apparent improvements of blocking representation in a numerical model can sometimes occur through compensation of errors, as mentioned in~\citep{davini2017improved}. Additionally, increasing the horizontal resolution of a numerical model can enhance the transient eddy forcing of blocks, as highlighted in~\citep{matsueda2009blocking, schiemann2017resolution}. These findings align with our observations made in this paper that the perturbations of the preexisting synoptic-scale eddies are prone to result in unstable and chaotic behavior in the evolution of blocking events. It is also a valuable topic to discuss the climatological seasonal impact of blocking, as mentioned in~\citep{newman1998impact}.

\section*{Acknowledgments}
{\small This work was supported by Grant No.12241105 of NSFC and Grant No.YSBR-034 of CAS.}

\appendix
\section{Derivation of the NMI model}
\label{sec: derivation-nmi}

In this section, we complement the details of the NMI model's derivation in~\Cref{subsec: nmi-model}.

\subsection{The wave-superposition form of the preexisting synoptic-scale eddies streamfunction $\psi_1'$ and its phase velocities}
\label{subsec: synoptic-o32}

Let us first put the asymptotic expansions of both the planetary-scale blocking anomaly and synoptic-scale eddies streamfunctions,~\eqref{eqn: planetary-expansion} and~\eqref{eqn: synoptic-expansion}, into the characteristic equation of synoptic-scale eddies~\eqref{eqn: synoptic-scale}. Taking the lowest approximation, we obtain that $\psi_1'$ satisfies the $O(\epsilon^{\frac32})$-order approximating equation of synoptic-scale eddies as 
\begin{equation}
\label{eqn: synoptic-order-32}
\left( \frac{\partial}{\partial t} + U \frac{\partial}{\partial x} \right)\left( \nabla^2\psi'_1 - F\psi'_1 \right) + PV_y \frac{\partial \psi'_1}{\partial x} = 0. 
\end{equation}   
Then, according to the observation, we may assume that the synoptic-scale eddies streamfunction takes the following form as
\begin{equation}
\label{eqn: synoptic-wave} 
\psi_1'(x,t;X_1)  = \frac{2f_0(X_1)}{\epsilon^{\frac32}}\big( \cos(k_1x-\omega_1t) - \cos(k_2x - \omega_2t) \big)\sin\left( \frac{my}{2} - \frac{\pi}{8}\right), 
\end{equation}
where $f_0(X_1) = a_0  \exp\left[ - \mu (X_1 + \epsilon x_T)^2 \right]$. Some simple substitution of variables tells us that the synoptic-scale streamfunction $\psi'_1$ takes the wave-superposition form of~\eqref{eqn: synoptic-1}. Taking the wave-superposition form~\eqref{eqn: synoptic-1} into the $O(\epsilon^{\frac32})$-order approximating equation of synoptic-scale eddies~\eqref{eqn: synoptic-order-32}, the phase velocities in~\eqref{eqn: s-phase-velocity} are derived. 

\subsection{The single-wave form of the blocking wavy anomaly stramfunction $\psi_1$ and its phase velocity}
\label{subsec: planetary-o1}
   
Putting the asymptotic expansions of both the planetary-scale blocking anomaly and synoptic-scale eddies streamfunctions,~\eqref{eqn: planetary-expansion} and~\eqref{eqn: synoptic-expansion}, into the characteristic equation of planetary-scale blocking anomaly~\eqref{eqn: planetary-scale}, we take the $O(\epsilon)$-order approximation and obtain that $\psi_1$ satisfies
\begin{equation}
\label{eqn: plantary-order-1}
\left( \frac{\partial}{\partial t} + U \frac{\partial}{\partial x} \right)\left( \nabla^2\psi_1 - F\psi_1 \right) + PV_y \frac{\partial \psi_1}{\partial x} = 0. 
\end{equation}   
The blocking wavy anomaly streamfunction $\psi_{1}$ is assumed with the form~\eqref{eqn: blocking-wave-1},
where $B$ is the complex amplitude only dependent on the slow-varying variables, $X_1, X_2, \ldots$ and $T_1,T_2,\ldots$. Taking the single-wave form of the blocking wavy anomaly streamfunction~\eqref{eqn: blocking-wave-1} into the $O(\epsilon)$-order approximating equation of the planetary-scale blocking anomaly streamfunction~\eqref{eqn: planetary-scale}, we derive that the phase velocity satisfies~\eqref{eqn: p-phase-velocity}. 


\subsection{The linear relationship of the complex blocking amplitude $B$ and the group velocity of the blocking wavy anomaly $\psi_1$}
\label{subsec: planetary-o2}

Putting the asymptotic expansions of both the planetary-scale blocking anomaly and synoptic-scale eddies streamfunctions,~\eqref{eqn: planetary-expansion} and~\eqref{eqn: synoptic-expansion}, into the characteristic equation of planetary-scale blocking anomaly~\eqref{eqn: planetary-scale}, we take the $O(\epsilon^2)$-order approximation and obtain that $\psi_2$ satisfies
\begin{align}
   & \left( \frac{\partial}{\partial x} + U \frac{\partial}{\partial x} \right)\left( \nabla^2\psi_2 - F\psi_2 \right) + PV_y \frac{\partial \psi_2}{\partial x} \nonumber \\
=  & - 2 \left( \frac{\partial}{\partial t} + U \frac{\partial}{\partial x} \right) \frac{\partial^2\psi_1}{\partial x\partial X_1} -  \left( \frac{\partial}{\partial T_1} + U \frac{\partial}{\partial X_1} \right)\left( \nabla^2\psi_1 - F\psi_1 \right) - PV_y \frac{\partial \psi_1}{\partial X_1}  - J(\psi_1, \nabla^2\psi_1). \label{eqn: plantary-order-2}
\end{align}   
Let the associated zonal-mean anomaly $\psi_2$ also satisfy the linear barotropic quasi-geostrophic equation. Then, we can obtain 
\[
2 \left( \frac{\partial}{\partial t} + U \frac{\partial}{\partial x} \right) \frac{\partial^2\psi_1}{\partial x\partial X_1} +  \left( \frac{\partial}{\partial T_1} + U \frac{\partial}{\partial X_1} \right)\left( \nabla^2\psi_1 - F\psi_1 \right) + PV_y \frac{\partial \psi_1}{\partial X_1}  - J(\psi_1, \nabla^2\psi_1) = 0.
\]
Putting the single-wave form of the blocking wavy anomaly streamfunction~\eqref{eqn: blocking-wave-1} into it, we obtain the group velocity~\eqref{eqn: velocity} and the linear relationship as
\begin{equation}
\label{eqn: linear-cg}
\frac{\partial B}{\partial T_1} + c_g \frac{\partial B}{\partial X_1} = 0.
\end{equation}

\subsection{The static and sinusoidal form of the associated zonal-mean anomaly $\psi_2$}
\label{subsec: planetary-o3-1}
Putting the asymptotic expansions of both the planetary-scale blocking anomaly and synoptic-scale eddies streamfunctions,~\eqref{eqn: planetary-expansion} and~\eqref{eqn: synoptic-expansion}, into the characteristic equation of planetary-scale blocking anomaly~\eqref{eqn: planetary-scale}, we take the $O(\epsilon^3)$-order approximation and obtain that $\psi_3$ satisfies
\begin{align}
     \left( \frac{\partial}{\partial t} + U \frac{\partial}{\partial x} \right)&\left( \nabla^2\psi_3 - F\psi_3 \right) + PV_y \frac{\partial \psi_3}{\partial x} \nonumber \\
   & =- 2 \left( \frac{\partial}{\partial t} + U \frac{\partial}{\partial x} \right) \frac{\partial^2\psi_2}{\partial x\partial X_1} -  \left( \frac{\partial}{\partial T_1} + U \frac{\partial}{\partial X_1} \right)\left( \nabla^2\psi_2 - F\psi_2 \right) - PV_y \frac{\partial \psi_2}{\partial X_1} \nonumber \\
    & \mathrel{\phantom{=}} - \underbrace{2  \left( \frac{\partial}{\partial t} + U \frac{\partial}{\partial x} \right) \frac{\partial^2\psi_1}{\partial x \partial X_2}}_{\mathbf{I}_1}  - \underbrace{\left( \frac{\partial}{\partial T_2} + U \frac{\partial}{\partial X_2} \right) \left( \nabla^2\psi_1 - F\psi_1 \right)}_{\mathbf{I}_2} - \underbrace{PV_y \frac{\partial \psi_1}{\partial X_2}}_{\mathbf{I}_3} \nonumber \\
    & \mathrel{\phantom{=}} - \underbrace{2 \left( \frac{\partial}{\partial T_1} + U \frac{\partial}{\partial X_1} \right) \frac{\partial^2\psi_1}{\partial x\partial X_1}}_{\mathbf{II}_1} - \underbrace{\left( \frac{\partial}{\partial t} + U \frac{\partial}{\partial x} \right) \frac{\partial^2 \psi_1}{\partial X_1^2}}_{\mathbf{II}_2} \nonumber \\
    & \mathrel{\phantom{=}}- \underbrace{2J\left(\psi_1, \frac{\partial^2\psi_1}{\partial x \partial X_1}\right)}_{\mathbf{III}_1} - \underbrace{\left( \frac{\partial \psi_1}{\partial X_1}\frac{\partial \nabla^2\psi_1}{\partial y} - \frac{\partial \psi_1}{\partial y}\frac{\partial \nabla^2\psi_1}{\partial X_1} \right)}_{\mathbf{III}_2} \nonumber \\
    & \mathrel{\phantom{=}}- \underbrace{J(\psi_1, \nabla^2\psi_2)}_{\mathbf{IV}_1} - \underbrace{J(\psi_2, \nabla^2\psi_1)}_{\mathbf{IV}_2} \nonumber \\
    & \mathrel{\phantom{=}}- \underbrace{J(\psi_1',\nabla^2\psi'_1)_P}_{\mathbf{V}} \label{eqn: plantary-order-3}
\end{align}   
Let $\psi_3$ also satisfy the linear barotropic quasi-geostrophic equation. With the single-wave form of the blocking wavy anomaly streamfunction~\eqref{eqn: blocking-wave-1}, we know that $\nabla^2\psi_1$ is proportional to~$\psi_1$. Hence, taking the zonal average of~\eqref{eqn: plantary-order-3}, we obtain that the associated zonal-mean anomaly streamfunction $\psi_2$ satisfies 
\begin{equation}
\label{eqn: zonal-average-psi2}
\left( \frac{\partial}{\partial T_1} + U \frac{\partial}{\partial X_1} \right)\left( \frac{\partial^2\psi_2}{\partial y^2} - F\psi_2 \right) + PV_y \frac{\partial \psi_2}{\partial X_1}  =  \frac{4mk^2}{\epsilon^2L_y} \frac{\partial |B|^2}{\partial X_1} \cos(2my)
\end{equation}
Obviously, the assumption of the associated zonal-mean anomaly with the form $\psi_{2} = -\epsilon^{-2}g|B|^2\cos(2my)$ is reasonable, since 
\[
\left.\frac{\partial\psi_2}{\partial y}\right|_{y=0}=\left.\frac{\partial\psi_2}{\partial y}\right|_{y=L_y} = 0
\] satisfy the boundary condition, thus we obtain that the associated zonal-mean anomaly $\psi_1$ satisfies~\eqref{eqn: associate-2}. With the equality of zonal average~\eqref{eqn: zonal-average-psi2}, the coefficient is obtained as 
\[
g = \frac{4m k^2(m^2+k^2+F)^2}{ PV_yL_y\left[(4m^2+F)(m^2+F-k^2) - (m^2+k^2+F)^2 \right]}.
\]

\subsection{Nonlinear time-evolution behavior of the complex amplitude $B$}
\label{subsec: planetary-o3-2}

Since the nondimensional blocking wavy anomaly streamfunction $\psi_1$ is the superposition of the single wave $\frac{1}{\epsilon}\sqrt{\frac{2}{L_y}} \sin\left(my - \frac{\pi}{4}\right)Be^{i(kx-\omega t)}$ and its conjugate $\frac{1}{\epsilon}\sqrt{\frac{2}{L_y}} \sin\left(my - \frac{\pi}{4}\right)\overline{B}e^{-i(kx-\omega t)}$, we only use the part $\frac{1}{\epsilon}\sqrt{\frac{2}{L_y}} \sin\left(my - \frac{\pi}{4}\right)Be^{i(kx-\omega t)}$ for convenience. Then, we show the right-hand side of~\eqref{eqn: plantary-order-3} by five parts as 

\begin{itemize}
\item \textbf{Part-I}: Putting $\psi_1=\frac{1}{\epsilon}\sqrt{\frac{2}{L_y}} \sin\left(my - \frac{\pi}{4}\right)Be^{i(kx-\omega t)}$ into Part-I, we obtain that
      \begin{align*}
      \mathbf{I}_1 + \mathbf{I}_2 + \mathbf{I}_3                    =  &2  \left( \frac{\partial}{\partial t} + U \frac{\partial}{\partial x} \right) \frac{\partial^2\psi_1}{\partial x \partial X_2}  + \left( \frac{\partial}{\partial T_2} + U \frac{\partial}{\partial X_2} \right) \left( \nabla^2\psi_1 - F\psi_1 \right) + PV_y \frac{\partial \psi_1}{\partial X_2} \\ 
                                                                    = & - \frac{1}{\epsilon}\sqrt{\frac{2}{L_y}} \left( m^2+k^2+F \right) \left( \frac{\partial B}{\partial T_2} + c_g \frac{\partial B}{\partial X_2}\right) \sin\left(my - \frac{\pi}{4}\right)Be^{i(kx-\omega t)}.
      \end{align*}   

\item \textbf{Part-II}: With the phase velocity~\eqref{eqn: p-phase-velocity} and the linear relationship~\eqref{eqn: linear-cg}, we take $\psi_1=\frac{1}{\epsilon}\sqrt{\frac{2}{L_y}} \sin\left(my - \frac{\pi}{4}\right)Be^{i(kx-\omega t)}$ into Part-II as
      \begin{align*}
      \mathbf{II}_1 + \mathbf{II}_2 & =  2 \left( \frac{\partial}{\partial T_1} + U \frac{\partial}{\partial X_1} \right) \frac{\partial^2\psi_1}{\partial x\partial X_1} + \left( \frac{\partial}{\partial t} + U \frac{\partial}{\partial x} \right) \frac{\partial^2 \psi_1}{\partial X_1^2} \\
                                    & = \frac{i}{\epsilon}\left[-\omega + kU+ \frac{2kPV_y(m^2-k^2+F)}{(m^2+k^2+F)^2}  \right] \frac{\partial^2 \psi_1}{\partial X_1^2} \\
                                    & = \frac{i}{\epsilon}\sqrt{\frac{2}{L_y}}\frac{kPV_y\left[3(m^2+F)-k^2\right]}{(m^2+k^2+F)^2}\frac{\partial^2 B}{\partial X_1^2} \sin\left(my - \frac{\pi}{4}\right)Be^{i(kx-\omega t)}
      \end{align*} 

\item \textbf{Part-III}: With the property that $\nabla^2\psi_1$ is proportional to~$\psi_1$, we known $\mathbf{III}_2=0$. With the property of Jacobians, we obtain $\mathbf{III}_1$ is proportional to $\cos(2my)$, thus
      \[
      \mathbf{III}_1 + \mathbf{III}_2 \propto \cos(2my).   
      \]  

\item \textbf{Part-IV}: Putting $\psi_1=\frac{1}{\epsilon}\sqrt{\frac{2}{L_y}} \sin\left(my - \frac{\pi}{4}\right)Be^{i(kx-\omega t)}$ and $\psi_2=-\epsilon^{-2}g|B|^2\cos(2my)$ into Part-IV, we obtain that
      \begin{align*}
      \mathbf{IV}_1 + \mathbf{IV}_2 & = J(\psi_1, \nabla^2\psi_2) + J(\psi_2, \nabla^2\psi_1) \\
                                    & = -\frac{2i}{\epsilon^3} \sqrt{\frac{L_y}{2}}  gkm(3m^2-k^2) |B|^2B \sin\left(my - \frac{\pi}{4}\right) \sin(2my) e^{i(kx-\omega t)}
      \end{align*}

\item \textbf{Part-V}: Here, we only consider the coefficient of the wave with wavenumber $k_2-k_1$. Hence, we take the superposition form of $\psi_1'$~\eqref{eqn: synoptic-wave} into Part-V and obtain that
        \[
        \frac{1}{2L_x}\int_{-L_x}^{L_x}\mathbf{V}e^{-i(k_2-k_1)x}dx = \frac{if_{0}(X_1)}{\epsilon^3}  \frac{m(k_1+k_2)(k_1^2 - k_2^2)}{4} \cdot \frac{L_y}{2} e^{-i(\omega_2-\omega_1)t}
        \]
\end{itemize}
Taking some basic calculations, we obtain the following two equalities as
\[
\int_{0}^{L_y}\sin^2\left(my - \frac{\pi}{4} \right)dy=\frac{L_y}{2}, \quad \text{and} \quad \int_{0}^{L_y}\sin^2\left(my - \frac{\pi}{4}\right)\sin(2my) dy=-\frac{L_y}{4}.
\] 
Filtering out the wave with the zonal wavenumber $k=2k_0$ and the meridional wavenumber $m$ of the right-hand side of the $O(\epsilon^3)$-order expansion~\eqref{eqn: plantary-order-3}, we obtain the forced NLS equation of the complex blocking amplitude $B$ with the periodic boundary condition~\eqref{eqn: ampli-nls}.

{\small
\bibliographystyle{abbrvnat}
\bibliography{sigproc}}

\end{document}